\documentclass[10pt,aps,prd,twocolumn,nofootinbib,noshowkeys,noshowpacs,longbibliography,
superscriptaddress,tightenlines,floatfix]{revtex4-2}
\usepackage[colorlinks=true,linktocpage=true,linkcolor=darkblue,citecolor=red,urlcolor=darkblue]{hyperref}
\usepackage{amsmath,amsfonts,amsthm,amssymb}
\usepackage{pifont}
\usepackage{graphics,graphicx}
\usepackage{color}
\usepackage{orcidlink}
\definecolor{darkblue}{RGB}{0,0,196}
\definecolor{darkgreen}{RGB}{0,120,0}

\def\beq{\begin{eqnarray}}
\def\eeq{\end{eqnarray}}
\def\half{\frac{1}{2}}
\def\f{\frac}
\def\nn{\nonumber}

\def\bm{\boldsymbol}

\def\pd{\partial}
\def\d{{\rm d}}
\def\e{{\rm e}}
\newcommand{\E}{{\rm e}}
\newcommand{\D}{{\rm d}}
\def\codevmu{{\stackrel{\leftrightarrow}{\partial}}\!\,^\mu}
\def\codevnu{{\stackrel{\leftrightarrow}{\partial}}\!\,^\nu}
\def\mL{m_\Lambda}
\def\mT{m_T}
\def\ap{\alpha_p}

\usepackage{xcolor}
\usepackage{ulem}
\definecolor{purple}{rgb}{0.8,0,0.6}

\begin{document}
\preprint{}
 \title{Modeling \texorpdfstring{$\Lambda$}{Lambda} polarization in Au\texorpdfstring{$+$}{-}Au collisions at \texorpdfstring{$\sqrt{s_{\rm NN}}=200$}{200} GeV using relativistic spin hydrodynamics}
    \author{Matteo Buzzegoli\orcidlink{0000-0002-2114-5431}}
    \email{matteo.buzzegoli@e-uvt.ro}
    \affiliation{Department of Physics, West University of Timisoara, Bulevardul Vasile P\^arvan 4, Timisoara 300223, Romania}
    \affiliation{Institute of Nuclear Physics, Polish Academy of Sciences, Krakow, 31-342, Poland}
    \author{Aleksandar Geci\'{c}\orcidlink{0009-0002-3032-9360}}
    \email{aleksandar.gecic10@e-uvt.ro}
    \affiliation{Department of Physics, West University of Timisoara, Bulevardul Vasile P\^arvan 4, Timisoara 300223, Romania}
    \author{Rajeev Singh\orcidlink{0000-0001-5855-4039}}
    \email{rajeev.singh@e-uvt.ro}
    \email{rajeevofficial24@gmail.com}
    \affiliation{Department of Physics, West University of Timisoara, Bulevardul Vasile P\^arvan 4, Timisoara 300223, Romania}
    \affiliation{Institut f$\ddot{u}$r Theoretische Physik, Goethe University, Max-von-Laue-Straße 1, D-60438 Frankfurt am Main, Germany}
\date{\today} 
\bigskip
\begin{abstract}
We investigate spin polarization dynamics in relativistic heavy-ion collisions using ideal relativistic spin hydrodynamics, employing non-boost-invariant longitudinal solutions as the hydrodynamic background. Operating in the small-polarization regime, where spin evolves perturbatively on top of the bulk expansion, we first analyze a $(1+1)$D setup with transverse homogeneity. In this framework, symmetry-constrained initial conditions for the spin potential lead to non-trivial evolution and generate both local and global $\Lambda$ hyperon polarization consistent with qualitative experimental trends, though they fail to reproduce observed azimuthal structures. To address this limitation, we extend the framework by incorporating transverse flow and spatial anisotropy at freeze-out, constructing a \textit{novel} $(1+1+2)$D model that preserves the longitudinal dynamics. We demonstrate that the inclusion of a longitudinal spin acceleration component, coupled with transverse expansion, results in the emergence of a quadrupole pattern in the longitudinal polarization. The resulting momentum-dependent and integrated observables exhibit qualitative and, after fitting the spin potential initial intensity, reasonably good quantitative agreement with experimental data for Au+Au collisions at $\sqrt{s_{\rm NN}}=200$\,GeV. Finally, we provide predictions for the in-plane transverse spin polarization, an observable that, to our knowledge, has not yet been experimentally measured.
\end{abstract}
\maketitle
\section{Introduction and physical motivation}
\label{sec:intro}
Relativistic heavy-ion collisions create strongly interacting matter under extreme conditions of temperature, energy density, and angular momentum~\cite{Heinz:2013th,Gale:2013da,Braun-Munzinger:2015hba,Arslandok:2023utm}. Over the past two decades, soft hadronic observables have established that the quark-gluon plasma (QGP) produced in these collisions behaves as an almost perfect fluid, whose evolution is accurately described by relativistic hydrodynamics~\cite{Teaney:2009qa,Schafer:2009dj,Martinoia:2024hip}. In non-central collisions, however, the system is characterized not only by strong collective flow but also by a very large initial orbital angular momentum. A major open question is how this macroscopic angular momentum is redistributed during the evolution, and to what extent it is converted into microscopic spin polarization of the produced hadrons~\cite{Becattini:2007sr,Becattini:2024uha,Niida:2024ntm}. The observation of non-zero hyperon polarization has made this question experimentally accessible and has turned spin observables into a new probe of the space-time structure of the QGP~\cite{Becattini:2024uha,Niida:2024ntm}.

A decisive milestone was the STAR measurement of global $\Lambda$ and $\bar{\Lambda}$ polarization in non-central Au$+$Au collisions~\cite{STAR:2017ckg}, which provided direct evidence that the matter created in heavy-ion collisions is vortical and that a fraction of the initial orbital angular momentum is transferred to particle spin~\cite{Liang:2004ph}. Subsequent measurements established the beam-energy, centrality, and rapidity dependence of global polarization and confirmed that polarization observables carry non-trivial information about the collective dynamics of the medium. Beyond the global signal, STAR also measured the polarization component along the beam direction and found a characteristic azimuthal modulation with a quadrupole structure in momentum space~\cite{STAR:2019erd}. This result was especially important because it showed that spin polarization in heavy-ion collisions is not exhausted by a simple global alignment mechanism but is sensitive to local flow gradients, anisotropic expansion, and the detailed space-time structure of the freeze-out hypersurface. Measurements by ALICE in Pb$+$Pb collisions at $\sqrt{s_{\rm NN}}=5.02$ TeV~\cite{ALICE:2019onw,ALICE:2021pzu} and, more recently, the observation of local $\Lambda$ polarization in p$+$Pb collisions by CMS~\cite{CMS:2025nqr} have further broadened the phenomenological landscape, indicating that spin polarization phenomena persist across collision systems and energies.

These experimental developments have stimulated intense theoretical activity~\cite{Becattini:2024uha}. In the hydrodynamic description, relating the spin degrees of freedom to the local thermodynamic state of the medium requires a spin potential $\omega_{\mu\nu}$, whose evolution must be studied consistently together with the bulk flow. More generally, the problem is embedded in the broader framework of relativistic hydrodynamics with spin, where the fundamental conserved quantities include not only energy and momentum but also total angular momentum~\cite{Becattini:2011zz,Florkowski:2017ruc,Florkowski:2018fap,
Montenegro:2017rbu,Hattori:2019lfp,Hongo:2021ona,Gallegos:2022jow,Das:2022azr,Singh:2025hnb,Cao:2022aku,Weickgenannt:2022zxs,Weickgenannt:2022qvh,Becattini:2023ouz,Drogosz:2024gzv,Daher:2025pfq,Abboud:2025shb,Bhadury:2025wuh,Sapna:2025yss,Matsuda:2025whj}. A natural consequence is that, besides the energy-momentum tensor, one must specify a spin tensor and its dynamics. Several formulations have been developed from kinetic theory and quantum-statistical considerations, and an important issue in this context is the role of the pseudogauge choice~\cite{Hehl:1976vr,Buzzegoli:2024mra,Becattini:2025oyi}.

Among the available choices, we will focus on the de Groot-van Leeuwen-van Weert (GLW) decomposition~\cite{DeGroot:1980dk}, which we find particularly useful for phenomenological applications because it is associated to a symmetric energy-momentum tensor and a separately conserved spin tensor in the ideal limit~\cite{Florkowski:2018ahw}, and simplifies the first-order non-dissipative contributions to the spin polarization to only the spin potential one~\cite{Buzzegoli:2021wlg}.

A widely used simplification is the regime of small spin polarization, in which the spin potential $\omega^{\mu\nu}$ is treated perturbatively. In this limit, the hydrodynamic background can be solved independently of the spin sector, while the spin degrees of freedom evolve on top of that background according to the spin equations of motion~\cite{Florkowski:2019qdp}. This approximation has made it possible to formulate tractable models of spin dynamics and connect them to hyperon polarization observables. In particular, earlier studies demonstrated how relativistic spin hydrodynamics can be used to study spin polarization and compute measurable quantities~\cite{Florkowski:2021wvk,Singh:2024cub}. At the same time, some studies also made it clear that highly symmetric backgrounds impose stringent restrictions on the allowed structure of the polarization signal~\cite{Singh:2020rht,Singh:2026wvf}.

The tension between experimental data and simple theoretical expectations is most visible in the case of local longitudinal polarization. Early thermal and hydrodynamic calculations based on a direct identification of spin polarization with thermal vorticity reproduced a quadrupole structure in the longitudinal component but often obtained the opposite sign or an incorrect magnitude compared with the RHIC data~\cite{Karpenko:2016jyx,Xie:2017upb,Wu:2019eyi,Ivanov:2020udj,Fu:2020oxj,Li:2017slc,Wei:2018zfb}. This discrepancy has motivated a broader re-examination of the mechanisms responsible for polarization. In particular, it has been argued that shear-induced contributions~\cite{Becattini:2021suc,Becattini:2021iol,Liu:2021uhn,Fu:2021pok} are essential for local longitudinal polarization and dominate the conventional thermal-vorticity contribution. The polarization signal may also be sensitive to the temperature at the freeze-out, the equation of states, and to shear and bulk viscosity~\cite{Becattini:2021iol,Yi:2021ryh,Palermo:2024tza}. Recent work on in-plane transverse polarization has emphasized that additional spin observables, beyond the standard out-of-plane and beam direction components, can provide a more complete characterization of spin dynamics in heavy-ion collisions~\cite{Karpenko:2016jyx,Florkowski:2021wvk,Arslan:2025tan}. However, its measurement is technically difficult due to complex acceptance corrections~\cite{Niida:2024ntm}.

To address these issues within spin hydrodynamics, it is essential to go beyond Bjorken symmetry. Although boost-invariant flow provides a valuable baseline, it suppresses the longitudinal structure that is expected to be important for realistic spin polarization phenomena~\cite{Florkowski:2019qdp}. Non-boost-invariant longitudinal expansion generates non-trivial rapidity dependence in temperature and flow velocity, and thus provides a more flexible environment in which spin degrees of freedom can evolve~\cite{Florkowski:2021wvk,Singh:2022uyy}. A particularly useful development in this direction was the construction of a family of exact longitudinal solutions of ideal relativistic hydrodynamics by Shi, Jeon, and Gale~\cite{Shi:2022iyb}\footnote{The solution in~\cite{Shi:2022iyb} has been used for various other physical analyses~\cite{Chen:2023vrk,Chen:2024pez,Chen:2024grb,Shao:2025ygy}.}. These solutions describe longitudinally expanding fluids with finite rapidity plateaus and allow both symmetric and asymmetric rapidity profiles while still retaining analytic control. They have already been shown to reproduce charge-particle pseudo-rapidity distributions reasonably well in several collision systems~\cite{Shi:2022iyb}, making them attractive as analytically controlled backgrounds for spin-hydrodynamic studies.

The purpose of the present work is to investigate spin polarization dynamics in such non-boost-invariant longitudinally expanding backgrounds and to assess how much of the observed phenomenology can already be understood within ideal relativistic spin hydrodynamics. We work in the GLW formulation and in the small-polarization regime, so that the bulk evolution and spin evolution can be treated sequentially. We first consider a $(1+1)$ dimensional setup, where the system is homogeneous in the transverse plane but exhibits a non-trivial rapidity structure. This provides a clean framework in which the effect of longitudinal gradients on the spin degrees of freedom can be isolated, the symmetry properties of different spin components can be identified explicitly, and the minimal set of spin initial conditions required by the geometry of non-central collisions can be analyzed in detail. In this setting, we study the coupled evolution of the spin potential components and compute both local and global polarization observables for $\Lambda$ hyperons.

At the same time, a purely $(1+1)$-dimensional description is expected to be too restrictive for a realistic account of the full polarization pattern, especially for observables that are known experimentally to be tied to the anisotropy of the transverse expansion. For this reason, we construct an extended $(1+1+2)$-dimensional model (hereafter referred to as \textit{1-1-2 model}),
in which the longitudinal dynamics remain governed by the exact non-boost-invariant solution, while transverse flow and spatial anisotropy are incorporated phenomenologically at freeze-out. This construction does not replace a fully dynamical $(3+1)$D simulation, such as \cite{Singh:2024cub}, but it allows one to test in a controlled manner how transverse expansion, elliptic deformation, and additional spin potential components affect the momentum dependence and azimuthal structure of the polarization signal. In particular, it offers a setting in which the role of spin acceleration components can be assessed quantitatively.

The present analysis is \textit{motivated} by two related questions. First, to what extent can non-boost-invariant longitudinal dynamics alone improve the description of spin observables relative to boost-invariant baselines? Second, what additional structures become accessible once transverse freeze-out geometry is included, even if the underlying bulk evolution is still analytically controlled and effectively lower dimensional? 

Our results show that the $(1+1)$D setup already captures several qualitative features of global and local polarization, while the $1-1-2$ extension is capable of generating a quadrupole structure in the longitudinal polarization through the combined effect of transverse flow and a longitudinal spin acceleration component. In this sense, the present work identifies a \textit{minimal mechanism} by which acceleration and anisotropic freeze-out geometry can contribute to the observed polarization pattern while keeping the hydrodynamic background sufficiently simple to allow transparent analytical and numerical analysis.

This paper is organized as follows. Sections~\ref{subsec:conservationsLaws}-\ref{subsec:SpinEOMs} briefly review the formalism of perfect-fluid relativistic spin hydrodynamics in the GLW pseudogauge and specify the approximation scheme adopted in this work, followed by the analysis of the initialization of spin potential components in Section~\ref{subsec:SpinInit}. Expressions to compute the momentum-dependent and momentum-integrated spin polarization are mentioned in Section~\ref{subsec:mean-spin-1plus1}. In Section~\ref{sec:SJGFlow}, we introduce the non-boost-invariant longitudinal hydrodynamic background and benchmark it against bulk hadronic observables. In Section~\ref{sec:Spin_1_plus_1}, we study the evolution of spin degrees of freedom and the resulting $\Lambda$-hyperon polarization in $(1+1)$D setup. In Section~\ref{sec:Spin_1_plus_3}, we extend the analysis to a $(1+1+2)$D freeze-out model and investigate the corresponding local and global polarization observables, where Section~\ref{subsec:VarParam} shows how the results are sensitive to the parameters of the model. Finally, in Section~\ref{sec:summary}, we summarize our findings and discuss possible future directions. Appendix~\ref{sec:GeneralFormOmegaP} provides a detailed analysis of the spin potential initialization, and Appendix~\ref{sec:FOIntegral} contains the explicit expressions of the freeze-out integral in the $1-1-2$ model.\\

\noindent \textbf{Notations and conventions}: In this work, we employ the mostly minus Minkowski metric, that is $\eta_{\mu\nu} = {\rm diag}\left(1,-1,-1,-1\right)$. The scalar product of two four-vectors is denoted as $a \cdot b\equiv a_\mu b^\mu$. The Levi-Civita tensor follows the convention $\epsilon^{0123} = -\epsilon_{0123} = 1$ throughout the work, and we also adopt natural units $c = \hbar = k_B = 1$. We denote symmetrization and anti-symmetrization by
\beq
A_{(\mu\nu)} = \frac{1}{2}\left(A_{\mu\nu} + A_{\nu\mu}\right),\quad
A_{[\mu\nu]} = \frac{1}{2}\left(A_{\mu\nu} - A_{\nu\mu}\right),\nn
\eeq
respectively. We define the Lorentz-invariant momentum measure as $\d P = d^3p/\left(E_p(2 \pi )^3\right)$, where $E_p = \sqrt{m^2 + \bm{p}^2}$ is the on-mass-shell particle energy and $p^\mu = \left(E_p, \bm{p}\right)$ is the particle four-momentum.
\section{Perfect-fluid relativistic spin hydrodynamics}
\label{sec:SpinHydro}
In this section, we give a brief review of the relativistic perfect-fluid spin hydrodynamics formalism for massive particles with spin-$\half$ adopting the de Groot-van Leeuwen-van Weert (GLW) form of the spin tensor~\cite{DeGroot:1980dk,Florkowski:2018ahw,Florkowski:2019qdp}.

We work in the regime of small spin polarization, where the spin potential $(\omega^{\mu\nu})$ is treated perturbatively. In this approximation, the energy-momentum tensor $T^{\mu\nu}$ is evaluated at zeroth-order in $\omega^{\mu\nu}$, while the spin tensor $S^{\lambda,\mu\nu}$ is retained to linear order. As a consequence, the conservation equations for energy and momentum decouple from the spin dynamics, allowing one to first solve for the hydrodynamic background and subsequently evolve the spin degrees of freedom on top of it. This approximation neglects the back-reaction of spin on the bulk evolution, which would arise at higher orders in $\omega^{\mu\nu}$. For simplicity and phenomenological relevance to mid-rapidity heavy-ion collisions at high energies, we further assume a vanishing chemical potential.
\subsection{Conservation laws}
\label{subsec:conservationsLaws}
The conservation equation for the ideal fluid energy-momentum tensor is written as
\beq
\pd_\mu T^{\mu\nu}=0\,, \quad \text{with} \quad T^{\mu\nu} = \varepsilon \, u^\mu u^\nu - P\, \Delta^{\mu\nu}\,,
\label{eq:energy-momentum-tensor}
\eeq
where $u^\mu$ is the fluid velocity and $\Delta^{\mu\nu} = \eta^{\mu\nu}-u^\mu u^\nu$ is the spatial projection operator orthogonal to the fluid velocity, with $\varepsilon$ and $P$ being the energy density and pressure, respectively. For an ideal relativistic gas of classical massive particles, one finds~\cite{Florkowski:2019qdp}
\beq
\varepsilon = \f{2\, T^4 z^2}{\pi^2}\left[z K_1 (z) + 3 K_2 (z)\right] , \quad P = \f{2\,T^4 z^2}{\pi^2} K_2(z)\,,
\eeq
where, owing to the small polarization limit, we neglected the contribution of the spin potential.
Here, $K_1, K_2$ represent the modified Bessel functions of 2$^{\rm nd}$ kind, and $z \equiv m/T$ is the ratio of the particle mass and temperature.

As the energy-momentum tensor, Eq.~\eqref{eq:energy-momentum-tensor}, is symmetric in its indices, the conservation of total angular momentum
\beq
\pd_\lambda J^{\lambda,\mu\nu} \equiv T^{\mu\nu} - T^{\nu\mu} + \pd_\lambda S^{\lambda,\mu\nu} = 0\,,
\eeq
forces the spin to be conserved separately, that is
\beq
\pd_\lambda S^{\lambda,\mu\nu} = 0\,.
\label{eq:spin-conservation}
\eeq
The spin tensor operator of a free Dirac field in the GLW pseudogauge is defined as~\cite{DeGroot:1980dk}
\begin{equation}
\label{eq:SGLW}
\begin{split}
\widehat{S}^{\lambda,\mu\nu}_{\rm GLW}(x)=&
\frac{i}{8}\widehat{\bar\Psi}(x)\left\{\gamma^\lambda,\left[\gamma^\mu,\,\gamma^\nu \right] \right\}\widehat{\Psi}(x)\\
&-\frac{i}{4m}\widehat{\bar\Psi}(x)\left(\sigma^{\lambda\mu}\codevnu-\sigma^{\lambda\nu}\codevmu\right)\widehat{\Psi}(x),
\end{split}
\end{equation}
where
\begin{equation*}
\codevmu=\left({\stackrel{\rightarrow}{\partial}}\!\,^\mu-{\stackrel{\leftarrow}{\partial}}\!\,^\mu\right),\quad
\sigma^{\mu\nu}=\frac{i}{2}\left[\gamma^\mu,\gamma^\nu\right].
\end{equation*}
For an ideal relativistic fluid, the GLW spin tensor at first order in spin potential is given by~\cite{Florkowski:2019qdp}
\beq
S^{\lambda, \mu\nu} &=& u^\lambda \left(S_1 \omega^{\mu\nu} + S_2 u^{[\mu} a^{\nu]}\right)\nn\\
&& \qquad \qquad +\, S_3 \left(u^{[\mu}\omega^{\nu]\lambda} + \eta^{\lambda[\mu} a^{\nu]}\right),
\label{eq:spin_tensor}
\eeq
where the coefficients are
\beq
S_1 &=& \f{T^3 z^2}{2 \pi^2}\left[K_2(z) \left(1 + \f{8}{z^2} \right) + \f{2\, K_1 (z)}{z}\right],\label{eq:S1}\\
S_2 &=& \f{T^3 z^2}{ 2\pi^2}\left[K_2(z)\left(2 + \f{48}{z^2}\right)   + \f{12 \, K_1 (z)}{z}\right],\label{eq:S2}\\
S_3 &=& - \f{T^3}{ \pi^2}\left[z K_1 (z) + 4 K_2 (z)\right] = -\f{T^3}{\pi^2} z K_3 (z),\nn\\
&\equiv&\half\left(S_1 - \f{S_2}{2}\right).\label{eq:S3}
\eeq
These coefficients depend on the thermodynamic properties of the medium and determine the relative importance of different spin modes during evolution.

It is important to emphasize that, in the assumptions of ideal fluid and small polarization, there is no direct back-reaction of spin dynamics on the hydrodynamic background; thus, the spin degrees of freedom evolve independently. 
In this framework, the mass $m$ entering the equations above represents an effective parameter describing the interaction of the spin polarized particles with the medium and as such it does not need to be matched with the equation of state of the background hydrodynamics, see also~\cite{Singh:2020rht,Singh:2024cub,Singh:2026wvf}. The impact of $m$ on the evolution of spin potential is studied below; for the main results we take $m=0.3$ GeV.
\subsection{Spin degrees of freedom}
\label{subsec:Spin degrees of freedom}
The spin potential, by definition, is a second-rank tensor that is antisymmetric in its indices and can be decomposed as
\beq
\omega_{\mu\nu} = 2\,a_{[\mu} u_{\nu]}  + \epsilon_{\mu\nu\alpha\beta}\, u^\alpha \, \omega^\beta\,,
\label{eq:spin_potential}
\eeq
which satisfies $a \cdot u = 0 = \omega \cdot u$, and the four-vectors $a_\mu$ and $\omega_\mu$ can be obtained as
\begin{equation}    
a_\mu = \omega_{\mu\alpha} u^\alpha , \qquad \omega_\mu = \half \epsilon_{\mu\alpha\beta\gamma}\, \omega^{\alpha \beta} u^\gamma\,.
\end{equation}
Furthermore, $a^\mu$ and $\omega^\mu$ can be expressed in terms of scalar spin components along the basis vectors orthogonal to the fluid velocity, that is $(x^\mu,y^\mu,z^\mu)$, as
\beq
\label{eq:spin_potential_a}
a^\mu &=& a_x \, x^\mu + a_y \, y^\mu + a_z \, z^\mu \,,\\
\label{eq:spin_potential_omega}
\omega^\mu &=& \omega_x \, x^\mu + \omega_y \, y^\mu + \omega_z \, z^\mu\,,
\eeq
see Eqs.~(\ref{eq:basis_u})-(\ref{eq:basis_z}) below. In analogy with the thermal vorticity, the four-vector $a^\mu$ denotes the acceleration part of the spin potential, and $\omega^\mu$ denotes the rotation part.
\subsection{Flow and basis vectors}
\label{subsec:flow-basis-vetors}
We perform our computations using the projection method~\cite{Florkowski:2011jg}, which is a numerical technique that simplifies complex multi-dimensional equations by projecting tensor equations onto specific directions (e.g., orthogonal to the four-velocity) allowing us to identify key degrees of freedom in spin hydrodynamics.

In addition to the fluid velocity, we introduce four-vectors $x^\mu, y^\mu$ that represent the  directions transverse to the beam and $z^\mu$ representing the longitudinal direction, playing an important role because of the collision's initial geometry.
It is important to note that $u^\mu$ is time-like, whereas the other three four-vectors are space-like and are chosen orthogonal to each other. Hence, they satisfy the following relations
\beq
u \cdot u \,\,=\,\, 1\,, && x\cdot x \,\,=\,\, y \cdot y \,\,=\,\, z\cdot z \,\,=\,\, -1\,,\nn\\
u \cdot x &=& u \cdot y \,\,=\,\, u \cdot z \,\,=\,\, 0\,,  \nn \\ 
x \cdot y &=&  x \cdot z \,\,=\,\, y \cdot z \,\,=\,\, 0\,.  \label{XYYZZX}
\eeq
To describe a system that is homogeneous in the transverse plane but has a non-trivial rapidity structure, the four-vector basis used in this work is chosen to be the following
\beq
\label{eq:basis_u}
u^\alpha &=& \left(u^0,0,0,u^3\right)\,,\\
x^\alpha &=& \left(0,1,0,0\right)\,,\\
y^\alpha &=& \left(0,0,1,0\right)\,,\\
\label{eq:basis_z}
z^\alpha &=& \left(u^3,0,0,u^0\right)\,.
\eeq
\subsection{Spin hydrodynamic equations of motion}
\label{subsec:SpinEOMs}
Projecting the conservation equation~(\ref{eq:energy-momentum-tensor}) for the energy -momentum tensor with fluid velocity $(u^\mu)$ and spatial projector $(\Delta^{\mu\nu})$ gives 
\beq
u \cdot \pd \, \varepsilon + \left(\varepsilon + P\right) \pd \cdot u &=& 0\,,\nn\\
\left(\varepsilon + P\right) u \cdot \pd \, u^\mu - \nabla^\mu P &=& 0 \,,
\label{eq:EoM-of-energy-momentum-conservation}
\eeq
which are the energy and momentum equations, respectively,
where $\nabla^\mu \equiv \pd^\mu - u^\mu u^\alpha \pd_\alpha$. Solving Eqs.~\eqref{eq:EoM-of-energy-momentum-conservation} gives the evolution of temperature and fluid velocity.

To obtain the equations of motion for the spin degrees of freedom, we use the spin tensor~\eqref{eq:spin_tensor} in the conservation law~\eqref{eq:spin-conservation} as input, and apply the projection method by projecting the resulting equation with $u_\mu x_\nu$, $u_\mu y_\nu$, $u_\mu z_\nu$, $y_\mu z_\nu$, $x_\mu z_\nu$, and $x_\mu y_\nu$. Since $\omega^{\mu\nu}$ is antisymmetric, it contains six independent components. Correspondingly, we obtain six independent equations of motion. This yields a closed system governing the evolution of the spin degrees of freedom:
\begin{widetext}
    \begin{subequations}\label{eq:spinEoM}\beq
    2\, \pd_\mu \left(S_3\,a_x\,u^\mu\right) - \pd_\mu \left(S_3\, \omega_y\, z^\mu\right) + S_3\, a_x \left(u^\mu z\cdot \pd z_\mu\right) + 2\, S_1\, \omega_y \left(u^\mu u \cdot \pd z_\mu\right)&=& 0\,,\label{eq:UXspinEoM}\\
    2\, \pd_\mu \left(S_3\,a_y\,u^\mu\right) + \pd_\mu \left(S_3\, \omega_x\, z^\mu\right) + S_3\, a_y \left(u^\mu z\cdot \pd z_\mu\right) - 2\, S_1\, \omega_x \left(u^\mu u \cdot \pd z_\mu\right)&=& 0\,,\label{eq:UYspinEoM}\\
    \pd_\mu \left(S_3\,a_z\,u^\mu\right) &=& 0\,,\label{eq:UZspinEoM}\\
    2\,\pd_\mu \left(S_1\, \omega_x\, u^\mu\right) - \pd_\mu \left(S_3 \,a_y\,z^\mu\right) + S_3 \,\omega_x \left(z^\mu z\cdot \pd u_\mu \right) + 2\,S_3\, a_y \left(z^\mu u\cdot \pd u_\mu\right)  &=& 0\,,
    \label{eq:YZspinEoM}\\
    2\,\pd_\mu \left(S_1\, \omega_y\, u^\mu\right) + \pd_\mu \left(S_3 \,a_x\,z^\mu\right) + S_3 \,\omega_y \left(z^\mu z\cdot \pd u_\mu \right) - 2\,S_3\, a_x \left(z^\mu u\cdot \pd u_\mu\right)  &=& 0\,,
    \label{eq:XZspinEoM}\\
    \pd_\mu \left(S_1\,\omega_z\,u^\mu\right) &=& 0\,.\label{eq:XYspinEoM}
    \eeq\end{subequations}
\end{widetext}
It is important to highlight some properties of the spin equations of motion:
\begin{enumerate}
    \item Note that the transverse spin acceleration component $a_x(a_y)$ is coupled with the spin vorticity components $\omega_y(\omega_x)$, see Eqs.~\eqref{eq:UXspinEoM} \& \eqref{eq:XZspinEoM} and Eqs.~\eqref{eq:UYspinEoM} \& \eqref{eq:YZspinEoM}.
    \item The longitudinal spin acceleration $a_z$ and spin vorticity component $\omega_z$ evolve independently and do not couple with any other spin degree of freedom, see Eqs.~\eqref{eq:UZspinEoM} and \eqref{eq:XYspinEoM}.
\end{enumerate}
This simplification emerges because we are working in $(1+1)$D and with systems that are transversely homogeneous but non-boost invariant. In a general $(3+1)$D evolution, all spin components couple with each other~\cite{Singh:2022uyy}.

The Eqs.~\eqref{eq:UXspinEoM}-\eqref{eq:XYspinEoM} will reduce to the equations of motion presented in~\cite{Florkowski:2019qdp} if we assume boost-invariance, where all spin components evolve independently of each other.
\subsection{Spin initialization}
\label{subsec:SpinInit}
\begin{figure}[ht!]
    \centering
    \includegraphics[width=1\linewidth]{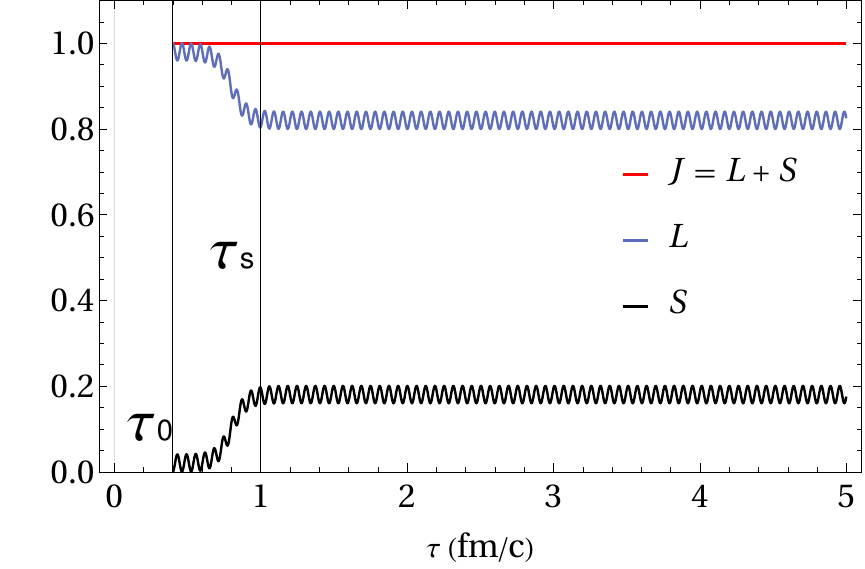}
    \caption{Representation of the general physical picture of orbital and spin angular momentum evolution in heavy-ion collisions.}
    \label{fig:SpinOrbit}
\end{figure}
Initialization of spin potential components $a_i$ and $\omega_i$ is chosen to reflect the physical conditions realized in non-central relativistic heavy-ion collisions. In such collisions, the total angular momentum $\boldsymbol{J}$ of the system, evaluated in the center-of-mass frame, is initially purely orbital, $\boldsymbol{L}$, oriented perpendicular to the reaction plane and pointing in the negative direction $-\hat{y}$~\cite{STAR:2017ckg,STAR:2019erd}. During the subsequent evolution, interactions in the fluid transfer part of this orbital angular momentum into intrinsic spin, $\boldsymbol{S}$, as shown in Fig.~\ref{fig:SpinOrbit}. Owing to the conservation of angular momentum, this redistribution can be expressed as
\beq
\boldsymbol{J}_{\rm initial} = \boldsymbol{L}_{\rm initial}
= \boldsymbol{J}_{\rm final}
= \boldsymbol{L}_{\rm final} + \boldsymbol{S}_{\rm final},
\label{eq:L}
\eeq
implying that, on average, the generated spin aligns with the direction of the initial total angular momentum, as shown by the measurements of global spin polarization~\cite{STAR:2017ckg,STAR:2019erd}.

For Au$+$Au collisions at top RHIC energies, conventional $(3+1)$D hydrodynamic models, which lack an explicit spin tensor, have successfully provided a quantitative and qualitative description of most spin polarization measurements~\cite{Becattini:2021iol,Fu:2021pok,Palermo:2024tza}. These analyses suggest that spin degrees of freedom reach equilibrium at a characteristic time prior to decoupling. A primary objective of spin hydrodynamics is to describe this equilibration process. 
Naturally, such an analysis necessitates the inclusion of interactions and the dynamical conversion between orbital and spin angular momentum, as studied recently in~\cite{Sapna:2025yss}.

However, ideal spin hydrodynamics remains a powerful tool if one assumes that the spin degrees of freedom are independently conserved starting from a time $\tau_s$, see Fig.~\ref{fig:SpinOrbit}. In this framework, the spin tensor is initialized at $\tau_s>\tau_i$ and evolved independently. A distinct advantage of employing ideal spin hydrodynamics within the GLW pseudogauge, as opposed to conventional hydrodynamics, is the ability to incorporate acceleration and rotation within the spin sector, even in a simplified $(1+1)$D model. This approach allows us to infer the necessary structure of the spin potential required to reproduce experimental data, thereby testing the hypothesis of spin conservation and providing some estimate for $\tau_s$. In what follows, we demonstrate how the spin potential can be initialized to provide a realistic description of spin polarization, even in flows where the orbital angular momentum is vanishing.

First, we show that the flow solutions used in this work do not generate orbital angular momentum. We remind that the spatial components of the angular momentum vector are related to the antisymmetric angular momentum tensor via
\begin{equation}
L^k = -\frac{1}{2} \epsilon^{kij} L^{ij},
\end{equation}
so that a non-zero $y$-component of $L$ corresponds to a finite $xz$-component of the angular momentum tensor. The global orbital angular momentum can be obtained by integrating its density, which is given by the energy momentum tensor as
\beq
L^{\mu\nu} = \int_\Sigma \d\Sigma_\lambda\, L^{\lambda,\mu\nu} = \int_\Sigma \d\Sigma_\lambda \left(x^\mu T^{\lambda\nu} - x^\nu T^{\lambda\mu}\right).
\eeq
Since the orbital angular momentum is a conserved quantity, its value does not depend on the hypersurface $\Sigma$. A convenient choice is the hypersurface of constant $\tau$, e.g., $\tau=\bar{\tau}$, described by $\d\Sigma_\lambda = \bar{\tau} n_\lambda \d x\,\d y\,\d\eta_s$ with $n_\lambda = (\cosh \eta_s,0,0,\sinh \eta_s)$. Substituting the constitutive relation for the energy-momentum tensor, Eq.~\eqref{eq:energy-momentum-tensor}, one can write
\beq
L^{\mu\nu} &=& A_\perp \bar{\tau} \int \!\! \d\eta_s\, n_\lambda  \Big[\left(\varepsilon + P\right) u^\lambda \left(x^\mu u^\nu - x^\nu u^\mu \right)\nn\\
&& \qquad - P \left(x^\mu \eta^{\lambda\nu} - x^\nu \eta^{\lambda\mu}\right)\Big]\,,
\eeq
where $A_\perp$ constitutes the factors coming from the integral over the transverse area.
In this work, we use the analytical flow given in Eq.~\eqref{eq:SJGflow} with $a=1$, from which one finds that $L^{\mu\nu} = 0$, see also~\cite{Florkowski:2021wvk}. This result follows from the combination of longitudinal symmetry and the low dimensionality of the model and should be understood as a property of the idealized model rather than a general statement about realistic heavy-ion collisions. As we will explicitly show later, the spin polarization in the GLW pseudogauge can only be induced by the spin potential; therefore, a vanishing orbital angular momentum does not forbid or hinder the study of spin polarization in this setup.

Moving to the initialization of the spin components, following the previous discussion, the physics and the measurements of non-central collisions dictate that the only non-vanishing spin vector at the decoupling is the $\hat{y}$ one, i.e. $S^{13}_{\rm FO} \neq 0$~\cite{Florkowski:2021wvk}.
As the spin tensor is conserved, requiring this condition at the decoupling or at time $\tau_s$ is equivalent: $S^{\mu\nu}_{\rm FO}=S^{\mu\nu}(\tau_s)=S^{\mu\nu}$.
This requirement allows us to physically interpret which spin component is important for the current analysis and how to initialize them (see Appendix~\ref{sec:GeneralFormOmegaP} for more details).
From the definition of the spin angular momentum
\begin{equation}
S^{\mu\nu} = \int_\Sigma \d\Sigma_\lambda\, S^{\lambda,\mu\nu}\,,
\end{equation}
substituting Eq.~\eqref{eq:spin_tensor},
using the decomposition of spin potential~\eqref{eq:spin_potential}, and a $(1+1)$D flow $u^\mu(\tau,\eta_s)=(u^0,$ $0,\,0,\,u^3)$ along with the normal vector $n_\mu(\eta_s)=(\cosh \eta_s$ $,0,0,\sinh \eta_s)$ evaluated at $\tau=\tau_s$, one can obtain all the components of $S^{\mu\nu}$, such as
\beq
S^{01} &=& \f{A_\perp\tau_s}{2}  \int\!\! \d\eta_s\, S_3 a_x \Big[\cosh \eta_s \left( 1 - 3 \left(u^0\right)^2 \right)\nn\\
&+& 3 \sinh \eta_s \, u^0 u^3\Big] - \omega_y \Big[S_3 \sinh \eta_s \left(u^0\right)^2 \label{eq:SFO-01}\\
&+& \left(2 S_1 - S_3\right) \cosh \eta_s \,u^0 u^3 
- 2\, S_1 \sinh \eta_s\, \left(u^3\right)^2\Big],\nn
\eeq
\beq
S^{02} &=& \f{A_\perp\tau_s}{2}  \int \!\! \d\eta_s\, S_3 a_y \Big[\cosh \eta_s \left(1- 3 \left(u^0\right)^2 \right)\nn\\
&+& 3 \sinh \eta_s \, u^0 u^3\Big] + \omega_x \Big[S_3 \sinh \eta_s \left(u^0\right)^2 \label{eq:SFO-02}\\
&+& \left(2 S_1 - S_3\right) \cosh \eta_s \,u^0 u^3 
- 2\, S_1 \sinh \eta_s\, \left(u^3\right)^2\Big],\nn
\eeq
\beq
S^{03} &=& A_\perp \tau_s \int \!\! \d\eta_s\, S_3 a_z \left(\sinh\eta_s\, u^3 - \cosh \eta_s \,u^0\right)\,,\label{eq:SFO-03}
\eeq
\beq
S^{12 } &=& A_\perp \tau_s \int \!\! \d\eta_s\, S_1 \omega_z \left( \sinh\eta_s\, u^3 - \cosh \eta_s \,u^0\right)\,,\label{eq:SFO-12}
\eeq
\beq
S^{13} &=& \f{A_\perp\tau_s}{2}  \int\!\! \d\eta_s\, S_3 a_x \Big[\sinh \eta_s - 3 \cosh\eta_s\, u^0 \, u^3\nn\\
&+& 3 \sinh \eta_s  \left(u^3\right)^2\Big] + \omega_y \Big[2 S_1 \cosh\eta_s \left(u^0\right)^2 \label{eq:SFO-13}\\
&-& \left(2 S_1 - S_3\right) \sinh \eta_s \,u^0 u^3 
- S_3 \cosh \eta_s\, \left(u^3\right)^2\Big],\nn
\eeq
\beq
S^{23} &=& \f{A_\perp\tau_s}{2}  \int\!\! \d\eta_s\, S_3 a_y \Big[\sinh \eta_s - 3 \cosh\eta_s\, u^0 \, u^3\nn\\
&+& 3 \sinh \eta_s  \left(u^3\right)^2\Big] - \omega_x \Big[2 S_1 \cosh\eta_s \left(u^0\right)^2 \label{eq:SFO-23}\\
&-& \left(2 S_1 - S_3\right) \sinh \eta_s \,u^0 u^3 
- S_3 \cosh \eta_s\, \left(u^3\right)^2\Big]\nn,
\eeq
from which we observe that to describe the physical situation mentioned before
\begin{itemize}
    \item $\omega_y$ and $a_y$ need to be $\eta_s$-even functions;
    \item $\omega_x$, $\omega_z$, $a_x$, and $a_z$ need to be $\eta_s$-odd functions.
\end{itemize}
Imposing these symmetry constraints guaranties that the momentum averaged spin polarization is directed along $\hat{y}$, that is, parallel to the total angular momentum of the system.
For consistency and to reduce the degrees of freedom in the initialization, the shape of the functions giving the spin components is built using the flow components in Eq.~\eqref{eq:SJGflow}, $u^0$ being an $\eta_s$-even function and $u^3$ being odd (for $a=1$), ensuring that they remain finite and go to zero at large rapidity.

As mentioned, a complete and realistic physical picture should start from an initial orbital angular momentum and a vanishing spin potential, which is later generated dynamically. Alternatively, the use of $(3+1)$D dynamics allows for the initialization of spin at time $\tau_s$ with other hydrodynamic fields such as vorticity and shear~\cite{Singh:2024cub}. On the other hand, the use of a simplified model proves the applicability of spin hydrodynamics, and it is effective in understanding how the spin potential should look to reproduce the data.
\subsection{Spin polarization of \texorpdfstring{$\Lambda$}{Lambda} hyperons}
\label{subsec:mean-spin-1plus1}
Spin polarization is obtained by computing the mean Pauli-Lubanski (PL) vector for particles having momentum $p$ that are emitted from the freeze-out hypersurface $\Sigma_{\rm FO}$~\cite{Becattini:2024uha}.
Using the GLW pseudogauge, the spin polarization vector is given by~\cite{Buzzegoli:2021wlg}
\begin{subequations}\label{meanspin}\beq
P_{\mu} (p) &=& \frac{\int_{\Sigma_{\rm FO}} \!\!
\d^3 \Sigma\cdot p\, \mathcal{S}_\mu(x,p)}{\int_{\Sigma_{\rm FO}} \!\! \d^3  \Sigma\cdot p\, f(x,p) },\\
\mathcal{S}_{\mu} (x,p) &=& -\frac{1}{4m_\Lambda}\tilde{\omega }_{\mu \beta }p^{\beta}
    f(x,p)\left(1-f(x,p)\right),
\eeq
\end{subequations}
where $\tilde{\omega }_{\mu \beta } = \half \epsilon_{\mu\beta\alpha\gamma}\, \omega^{\alpha\gamma}$ is the dual of the spin potential, $m_\Lambda$ is the Lambda hyperon mass,
we denote the momentum as
\begin{equation}\label{eq:Momentum_p_rapidity}
p^\mu = \left( m_T\cosh y_p,\, p_T \cos\phi,\,
    p_T\sin\phi,\, m_T \sinh y_p\right),    
\end{equation}
with $p_T$ and $m_T = \sqrt{m_\Lambda^2+p_T^2}$, respectively, representing the transverse momentum and the transverse mass. The scalar product of particle four-momentum and the freeze-out element is given in Eq.~\eqref{eq:Sigma-cdot-momentum} for the $(1+1)$D model and in Eqs.~\eqref{eq:dSigmaP112}-\eqref{eq:dSigmaPxY} for the $1-1-2$ model. 
In the numerical calculations, we neglect quantum corrections to the Dirac thermal statistics and we use
\beq
f(x,p)\left(1-f(x,p)\right)\simeq f(x,p)\simeq\exp(-\beta(x)\cdot p)\,,
\eeq
with $\beta \cdot p = \f{u \cdot p}{T}$. 

To compare with the experiments, we obtain the spin polarization vector in the rest frame of the particle, denoted by $P_\mu^*$, which is obtained with the boost:
\begin{subequations}\label{meanspinBoost}\beq
P_0^*(p) &=& \frac{E_p}{m_\Lambda}P_0-\frac{\bm{P}\cdot \bm{p}}{m_\Lambda},\\
\bm{P}^* (p) &=& \bm{P} - \frac{\bm{p}}{E_p(E_p+m_\Lambda)}\bm{P}\cdot \bm{p},
\eeq \end{subequations}
where $E_p$ is the energy of the particle.
Hereafter, the spin polarization of the $\Lambda$ hyperon, in comparison with the experimental data, is always intended in the rest frame of the particle, even if the symbol $^*$ is omitted from $P$. Furthermore, when needed, we corrected the experimental data with the updated $\Lambda$ decay constant $\alpha_\Lambda = 0.732$ \cite{ParticleDataGroup:2020ssz}.
It is important to note that, unlike the Belinfante and the Canonical pseudogauge, the thermal vorticity and the thermal shear contributions to the spin polarization in the GLW pseudogauge~\eqref{meanspin} are vanishing~\cite{Buzzegoli:2021wlg}. 
Thanks to this feature, in the GLW pseudogauge, it is possible to obtain an accurate description of spin polarization even if the fluid is $1+1$ dimensional and has no vorticity. Furthermore, since the gradients of temperature do not contribute, the isothermal equilibrium evaluation of spin polarization yields the same results as the more generic one; see the discussion in~\cite{Becattini:2021iol}.

A detailed characterization of spin polarization in relativistic heavy-ion collisions requires resolving its dependence on the transverse momentum $(p_T)$ and the azimuthal angle $(\phi)$. Such a differential analysis probes the interplay between local vorticity, flow anisotropies, and non-equilibrium effects in the QGP, thereby providing access to the microscopic mechanisms responsible for spin polarization.

To compare with the experimental data, we compute the spin polarization of $\Lambda$s starting from Eq.~\eqref{meanspin}. The global, or integrated over momentum, spin polarization at a fixed value of rapidity is obtained by:
\beq
\langle P_\mu \rangle = \left.\frac{\int\d\phi\!\int\!\! p_T\d p_T\!\int_{\Sigma_{\rm FO}} \!\!\!
\d^3 \Sigma\cdot p\, \mathcal{S}^*_\mu(x,p)}{\int\d\phi\!\int\!\! p_T\d p_T\!\int_{\Sigma_{\rm FO}} \!\! \d^3  \Sigma\cdot p\, f(x,p) }\right|_{y_p},
\label{eq:momentum-integrated-polarization}
\eeq
and it is integrated over the $p_T$ range $\{0.5,\, 6\}$ GeV to match the experimental setup.
In our analysis, it follows from the choice of spin initialization that the only non-vanishing global spin polarization is the transverse one, i.e. $\langle P_y \rangle$.
Polarization as a function of the azimuthal angle $P_\mu(\phi)$ are obtained at mid-rapidity $y_p=0$ as
\beq
P_{\mu} (\phi)=\left.\frac{\int p_T\d p_T\int_{\Sigma_{\rm FO}} \!\!
\d^3 \Sigma\cdot p\, \mathcal{S}^*_\mu(x,p)}{\int p_T\d p_T\int_{\Sigma_{\rm FO}} \!\! \d^3  \Sigma\cdot p\, f(x,p) }\right|_{y_p=0},
\eeq
and is integrated over the $p_T$ range $\{0,\, 6.0\}$ GeV$/c$,
whereas the second Fourier harmonic as a function of transverse momentum $p_T$ at mid-rapidity is defined as~\cite{Becattini:2017gcx,STAR:2019erd,ALICE:2021pzu}
\begin{equation*}
P_\mu(\bm{p}_T,\,y_p=0) = \langle P_\mu \rangle
    + \sum_{k=1}^\infty 2 \langle P_\mu \sin(2k\phi)\rangle\sin(2k\phi) \,,
\end{equation*}
and is computed using
\begin{equation}
\langle P_\mu \sin(2\phi)\rangle = \left.\frac{\int\d\phi\int_{\Sigma_{\rm FO}} \!\!
\d^3 \Sigma\cdot p\, \mathcal{S}^*_\mu(x,p)\sin(2\phi)}{\int \d\phi\int_{\Sigma_{\rm FO}} \!\! \d^3  \Sigma\cdot p\, f(x,p) }\right|_{y_p=0},
\end{equation}
and similarly $\langle P_\mu \sin(2\phi)\rangle$ averaged in transverse momentum over the range $\{0,\, 6.0\}$ GeV$/c$ is
\begin{equation}
\left.\frac{\int p_T\d p_T\int\d\phi\int_{\Sigma_{\rm FO}} \!\!
\d^3 \Sigma\cdot p\, \mathcal{S}^*_\mu(x,p)\sin(2\phi)}{\int p_T\d p_T\int \d\phi\int_{\Sigma_{\rm FO}} \!\! \d^3  \Sigma\cdot p\, f(x,p) }\right|_{y_p=0}.
\end{equation}
\section{Longitudinally expanding relativistic ideal hydrodynamics}
\label{sec:SJGFlow}
\begin{figure*}[t!bh]
    \centering
    \includegraphics[width=8.9cm]{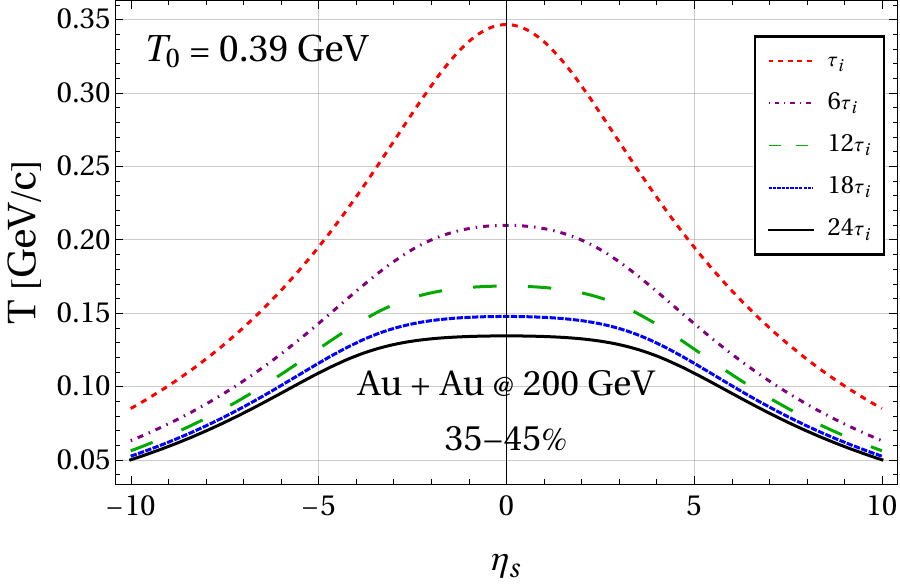}
    \includegraphics[width=8.9cm]{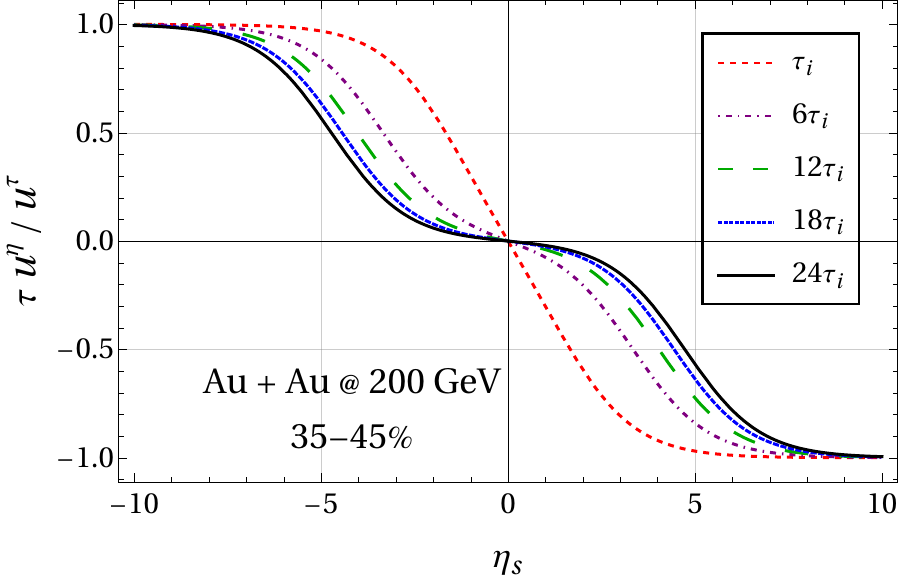}
    \caption{The temperature (left) and the flow (right) evolution with spacetime rapidity for the $(1+1)$D symmetric ($a=1$) SJG flow with hydrodynamic evolution starting at $\tau_i=0.4$~fm$/c$ in the $35-45$\% centrality class. See Table~\ref{tab:AuAu200GeV_parameters} for the hydrodynamic parameters used.}
    \label{fig:Tand Flow profiles}
\end{figure*}
In this section, we introduce the $(1+1)$D analytic solutions to the ideal energy-momentum equations of motion \eqref{eq:EoM-of-energy-momentum-conservation} that were first presented in Ref.~\cite{Shi:2022iyb} (hereafter referred to as `SJG flow') and which we use for our analysis. These solutions are applicable for both symmetric and asymmetric collisions, e.g., Au$+$Au and p$+$Pb collisions, respectively.

Taking $\varepsilon=3P$ as the equation of state (EOS),
the flow components of this solution are expressed as~\cite{Shi:2022iyb}
\beq
&& u^0 (\tau,\eta_s) = u^\tau \cosh \eta_s + \tau u^{\eta_s} \sinh \eta_s \,,\nn\\
&& u^3 (\tau,\eta_s) = u^\tau \sinh \eta_s + \tau u^{\eta_s} \cosh \eta_s\,,\label{eq:SJGflow}
\eeq
with
\begin{align*}
u^\tau &=& \half \left[\left(\f{t_0\, \e^{-\eta_s} + \tau  a}{t_0\, \e^{\eta_s} + \f{\tau}{a}}\right)^{\half} + \left(\f{t_0 \,\e^{\eta_s} + \f{\tau}{a}}{t_0\, \e^{-\eta_s} + \tau  a}\right)^{\half}\right],\nn\\
\tau u^{\eta_s} &=& \f{1}{2} \left[\left(\f{t_0 \,\e^{-\eta_s} + \tau  a}{t_0\, \e^{\eta_s} + \f{\tau}{a}}\right)^{\half} - \left(\f{t_0\, \e^{\eta_s} + \f{\tau}{a}}{t_0\, \e^{-\eta_s} + \tau  a}\right)^{\half}\right],\nn
\end{align*}
whereas the temperature evolution is given by~\cite{Shi:2022iyb}
\beq
T(\tau,\eta_s)\! =\! T_0\! \left[\!\f{1}{\tau_0^2}\!\left[\!\tau^2\! +\! t_0^2 \!+\! \tau t_0 \left(a \e^{\eta_s} \!+\! \f{\e^{-\eta_s}}{a} \right)\right]\right]^{\f{1}{6 a^2}-\f{1}{3}}
\label{eq:SJGTemperature}
\eeq
with $T_0$ being the initial temperature.
In the above 
\begin{itemize}
    \item $\tau$ is the proper time.
    \item $\eta_s$ is the longitudinal spacetime rapidity.
    \item $t_0$ is a positive constant controlling the rapidity structure, specifically its width.
    \item $\tau_0$ is a positive number giving the time scale of the evolution. Even though it may differ from the initial value of proper time, we have kept $\tau_0=\tau_i$ throughout this work. It represents the time required for the colliding nuclei to pass through each other in relativistic heavy-ion collisions.
    \item $a$ is the positive dimensionless parameter quantifying asymmetry in rapidity. $a=1$ represents symmetric collisions, such as Au$+$Au, and $a\neq 1$ denotes asymmetric collisions, such as p$+$Pb.
\end{itemize}
The solutions Eqs.~\eqref{eq:SJGflow}-\eqref{eq:SJGTemperature} reduce to the Bjorken solutions when $t_0 = 0$ and $a=1$. To ensure the convergence of energy density, $a$ must remain in the range $\left\{\sqrt{\frac{1-c_s^2}{1+c_s^2}},\,\sqrt{\frac{1+c_s^2}{1-c_s^2}}\right\}$; in this work, we always have $a=1$ and $c_s^2=1/3$.

In Fig.~\ref{fig:Tand Flow profiles}, we show the evolution of temperature (left) and fluid velocity (right) with respect to spatial rapidity $\eta_s$ for various proper times for symmetric Au$+$Au collisions at the center of mass energy 200 GeV. The hydrodynamic evolution starts at the initial proper time $\tau_i = 0.4$ fm$/$c. 
As expected in relativistic heavy-ion collisions, the temperature decreases with the increase in proper time and rapidity. The flow gradients also start to develop with rapidity.

Now, to highlight the applicability of SJG flow to QCD matter created in relativistic heavy-ion collisions, we compute the charged particle pseudo-rapidity distribution for Au$+$Au collisions at $\sqrt{s_{\rm NN}}=200$ GeV using Eqs.~\eqref{eq:SJGflow}-\eqref{eq:SJGTemperature}.

We employ the Cooper-Frye freeze-out procedure to compute the momentum distribution of particles
\beq
\f{\d N_{\rm ch}}{p_T \d p_T \,\d \phi \, \d y_p} = \int_{\Sigma_{\rm FO}} \f{\d^3\Sigma_\mu \, p^\mu}{(2\pi)^3} \f{\Theta(u\cdot p)}{\e^{\beta \cdot p}\pm 1}\,,
\label{eq:Cooper-Frye-formula}
\eeq
where $\beta=u/T$ is the inverse four-temperature, $u \cdot p$ is the particle's energy in the rest frame of the fluid, $\Theta (u \cdot p)$ is the step function, $p_T$ is the transverse momentum, $\phi$ is the azimuthal angle of momentum, $y_p$ is the rapidity, see Eq.~\eqref{eq:Momentum_p_rapidity}.
The freeze-out hypersurface $\Sigma_{\rm FO}$ is defined as the constant energy density hypersurface, which in our case is equivalent to the constant temperature hypersurface obtained from
\begin{equation}
T(\tau,\,\eta_s)=T_{\rm FO}.
\end{equation}
In this work we set $T_{\rm FO}=155$ MeV.
The freeze-out hypersurface element $\d^3 \Sigma_\mu$  is obtained from the previous equation and the temperature profile in Eq.~\eqref{eq:SJGTemperature} using
\beq\label{eq:VolumeElement}
\d^3 \Sigma_\mu = \epsilon_{\mu\alpha\beta\gamma} \f{\pd x^\alpha \, \pd x^\beta \, \pd x^\gamma}{\pd \zeta \, \pd \zeta^\prime \, \pd \zeta^{\prime \prime}}\, \d \zeta \, \d \zeta^\prime \, \d \zeta^{\prime \prime}\,,
\eeq
with $\zeta$, $\zeta^\prime$, and $\zeta^{\prime\prime}$ denoting the hypersurface coordinates.
Given Eq.~(\ref{eq:SJGflow}), a convenient choice is~\cite{Shi:2022iyb}
\beq
\zeta &=& \half \ln\left(1+\f{\tau a }{t_0} \e^{\eta_s}\right) - \half \ln\left(1+\f{\tau}{t_0 a} \e^{-\eta_s}\right)\,,\\
\zeta^\prime &=& x\,,\\
\zeta^{\prime\prime} &=& y\,.
\eeq
In Eq.~\eqref{eq:Cooper-Frye-formula}, $u \cdot p = m_T \cosh(y_p - \zeta)$ where $m_T \equiv \sqrt{m^2 + p_T^2}$ is the transverse mass, and the scalar product of the four-momentum and the freeze-out element, at freeze-out proper-time $\tau(\zeta) \geq \tau_i$, is given by
\beq
\d^3 \Sigma_\mu p^\mu &=& 2 \,t_0\, m_T\, \d\zeta \,\d x\, \d y \f{\sqrt{(2-a^2)(2-a^{-2})}}{4-a^2-a^{-2}} \e^{\f{q_1+q_2}{2}}\nn\\
&\times& \cosh\left(y_p - \zeta + \half \ln\f{2 a^2-1}{2-a^2}\right),
\label{eq:Sigma-cdot-momentum}
\eeq
where
\beq
q_1 &\equiv& \ln \left(1+\f{\tau a}{t_0} \e^{\eta_s}\right), \nn\\
q_2 &\equiv& \ln \left(1+\f{\tau}{t_0 a} \e^{-\eta_s}\right).
\label{eq:q1-q2}
\eeq
Thus, the pseudo-rapidity $(\eta)$ distribution of charged particle multiplicity $(N_{\rm ch})$ is expressed as~\cite{Shi:2022iyb}
\beq
\f{\d N_{\rm ch}}{\d\eta} &=& A_\perp \int^{\f{\ln\left(\f{T_0}{T_{\rm FO}}\right)^3 + \ln \left(\f{\tau_i}{t_0}\right)}{2-a^{-2}}}_{-\f{\ln\left(\f{T_0}{T_{\rm FO}}\right)^3 + \ln \left(\f{\tau_i}{t_0}\right)}{2-a^{2}}} \d\zeta \, \e^{\f{\left(a^2-a^{-2}\right)\zeta}{4-a^2-a^{-2}}} \Theta (\tau-\tau_i)\nn\\
&\times& \int^\infty_0 \f{p_T^2\,\d p_T\,\cosh{\eta}}{\e^{\f{\sqrt{m^2+p_T^2 \cosh^2\eta}}{T_{\rm FO}}\cosh{\zeta} - \f{p_T}{T_{\rm FO}}\sinh{\eta}\sinh{\zeta}}\pm 1}\nn\\
&\times& \Bigg[ \cosh\left(\zeta - \half \ln \f{2a^2 - 1}{2-a^2}\right)\nn\\
&-& \f{p_T\sinh\eta \sinh \left(\zeta - \half \ln \f{2a^2 - 1}{2-a^2}\right)}{\sqrt{m^2 + p_T^2 \cosh^2 \eta}}\Bigg].
\label{eq:pseudo-rapidity-distribution1+1}
\eeq
In the above expression, transverse dynamics is effectively factorized into the overall normalization $A_\perp$, which is fixed to reproduce the total number of charged particles, while longitudinal dynamics is governed by the SJG solution.
\begin{table}[ht!]
\centering
\begin{tabular}{c|c|c|c|c}
Centrality (\%) & $a$ & $\tau_0$(fm$/c$) & $T_0$(GeV) & $t_0$(fm$/c$)\\
\hline
$0-6$ & 1.00 & 0.4 & 0.42 & 0.28\\
$6-15$ & 1.00 & 0.4 & 0.42 & 0.26\\
$15-25$ & 1.00 & 0.4 & 0.40 & 0.21\\
$25-35$ & 1.00 & 0.4 & 0.39 & 0.18\\
$35-45$ & 1.00 & 0.4 & 0.39 & 0.17\\
$45-55$ & 1.00 & 0.4 & 0.39 & 0.16\\
$55-65$ & 1.00 & 0.4 & 0.38 & 0.15\\
$65-75$ & 1.00 & 0.4 & 0.37 & 0.14\\
\end{tabular}
\caption{The hydrodynamic parameters of the symmetric $(1+1)$D SJG flow for Au$+$Au collisions at $\sqrt{s_{\rm NN}}=200$ GeV; furthermore we used the gaussian smearing width $\sigma_p=1$ and the freeze-out temperature $T_{\rm FO}= 155$ MeV.}
\label{tab:AuAu200GeV_parameters}
\end{table}
It is also important to take into account the hadron scattering effects and resonance decays, for which we perform a rough analysis by combining Eq.~\eqref{eq:Cooper-Frye-formula} with a Gaussian smearing~\cite{Shi:2022iyb}
\beq
\f{\d N^\prime_{\rm ch}}{\d\eta^\prime} = \int \!\! \d \eta \,\f{1}{\sqrt{2\pi}\sigma_p} \, \e^{-\f{\left(\eta^\prime - \eta \right)^2}{2 \sigma_p^2}}\, \f{\d N_{\rm ch}}{\d\eta} \,,
\label{eq:pseudo-rapidity-distribution1+1-Gaussian}
\eeq
where $\sigma_p$ is a phenomenological parameter controlling the width of the rapidity smearing, mimicking the effects of hadronic re-scattering and resonance decays. In our analysis, $\sigma_p$ is treated as an effective parameter, set to $\sigma_p=1.0$.

\begin{figure}[ht!]
    \centering
    \includegraphics[width=1\linewidth]{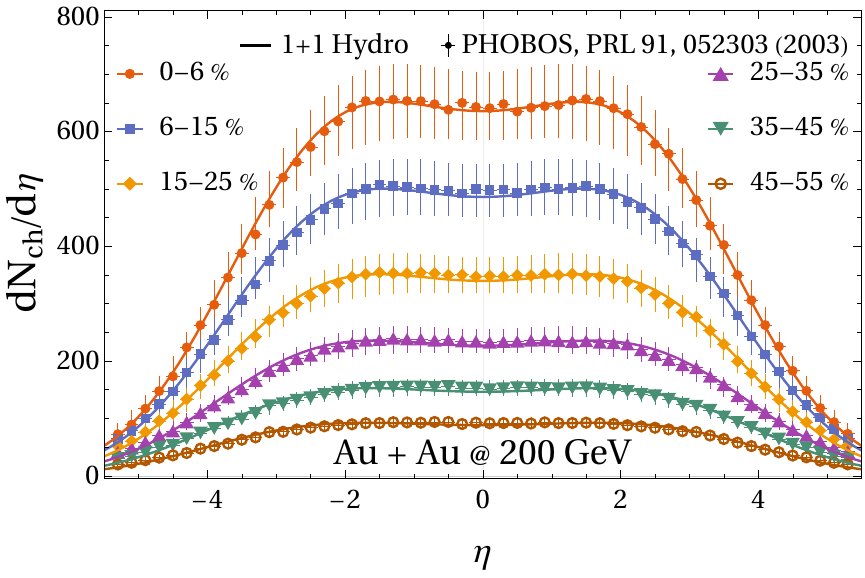}
    \caption{The charged particle pseudo-rapidity distribution obtained with the $(1+1)$D symmetric ($a=1$) SJG flow for Au$+$Au collisions at $\sqrt{s_{\rm NN}}=200$ GeV for various centrality classes. The hydrodynamic parameters used in our computations are reported in Table~\ref{tab:AuAu200GeV_parameters}. The experimental data is taken from Ref.~\cite{Back:2002wb}.} 
    \label{fig:Gold-Gold_ChargedParticlePseudorapidity_1_plus_1}
\end{figure}
Figure~\ref{fig:Gold-Gold_ChargedParticlePseudorapidity_1_plus_1} shows the pseudo-rapidity distribution of charged particle multiplicity (with Gaussian smearing) as the sum of pions $(\pi^\pm)$, kaons $(K^\pm)$, and protons-antiprotons $p(\bar{p})$ for various centrality classes. The hydrodynamic parameters we used in our analysis are reported in Table~\ref{tab:AuAu200GeV_parameters}. We observe that $(1+1)$D hydrodynamic calculations reproduce the experimental data~\cite{Back:2002wb} reasonably well for all centrality classes. This agreement indicates that SJG flow could be useful for phenomenological modeling of the spin polarization of $\Lambda$ hyperons observed in relativistic heavy-ion collisions~\cite{STAR:2017ckg,STAR:2018gyt,STAR:2019erd,STAR:2021beb,STAR:2023eck}.
\section{Spin polarization in \texorpdfstring{$1+1$}{1+1}D}
\label{sec:Spin_1_plus_1}
%
\begin{figure*}[t!hb]
    \centering
    \includegraphics[width=8.9cm]{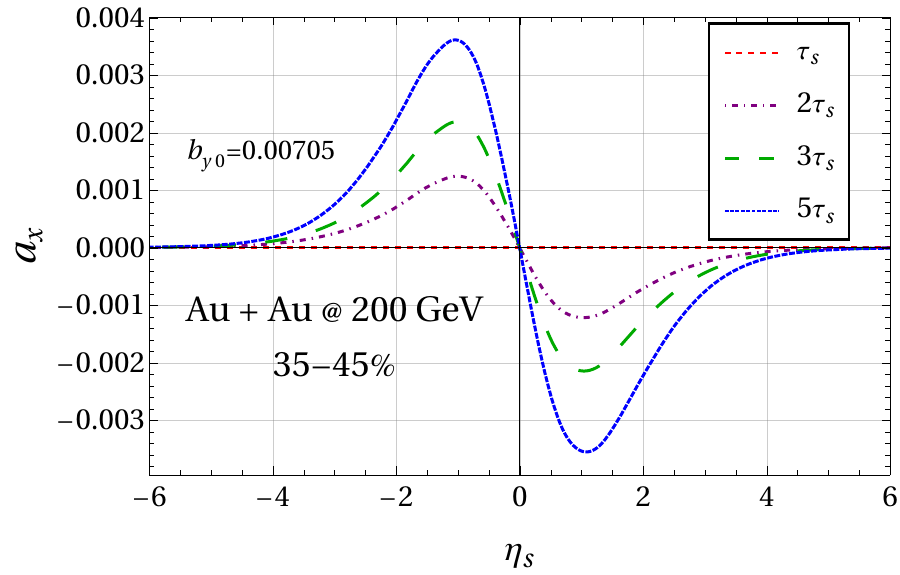}
    \includegraphics[width=8.9cm]{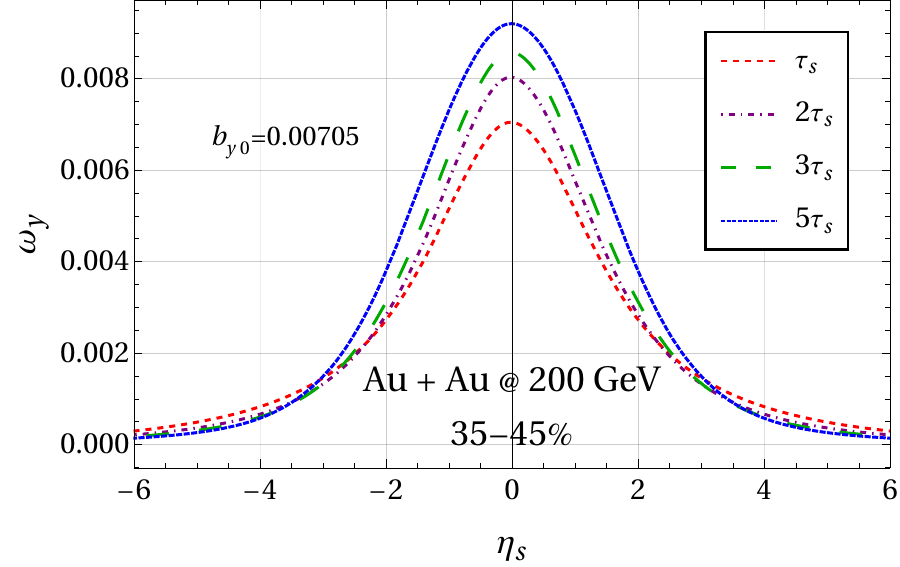}
    \includegraphics[width=8.9cm]{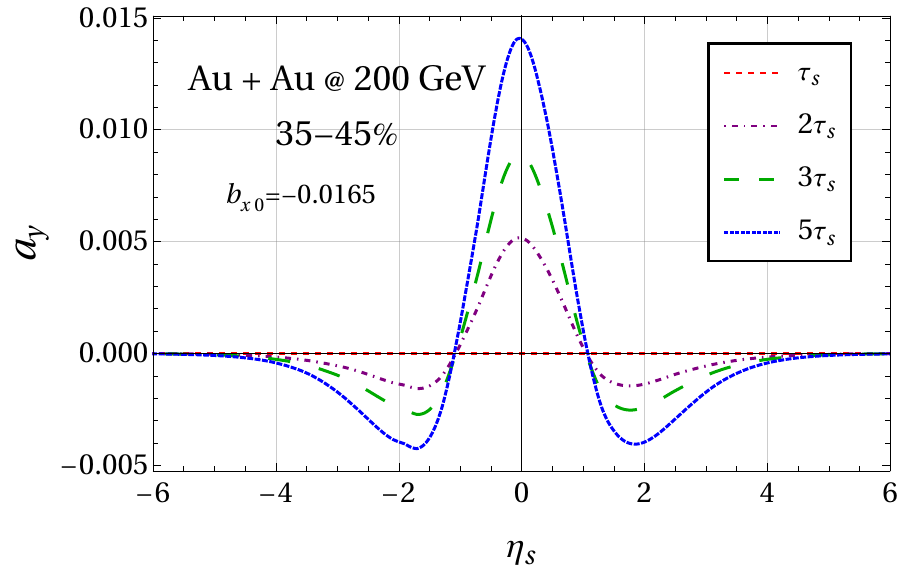}
    \includegraphics[width=8.9cm]{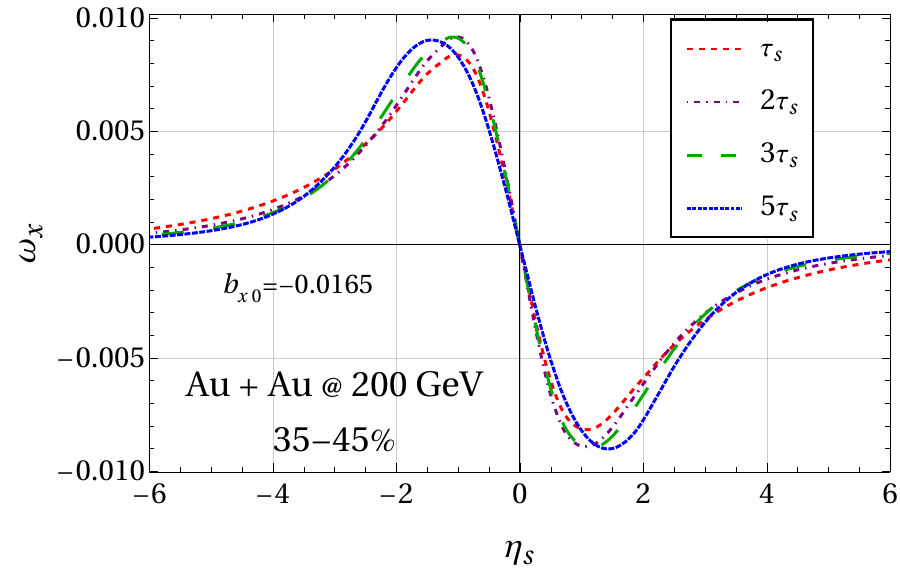}
    \caption{The spin potential components $a_x$ (top left), $\omega_y$ (top right), $a_y$ (bottom left), and $\omega_x$ (bottom right) resulting from ideal spin hydrodynamic evolution with $(1+1)$D symmetric ($a=1$) SJG flow for a Au$+$Au collision at $\sqrt{s_{\rm NN}}=200$ GeV within centrality class $35-45$\% and with spin evolution starting at $\tau_s=1$ fm$/c$.}
    \label{fig:spin-components-1plus1-evolution}
\end{figure*}
In this section, we obtain the evolution of spin degrees of freedom, quantified by the spin potential $\omega_{\mu\nu}$, using $(1+1)$D perfect-fluid spin hydrodynamics on top of the $(1+1)$D perfect-fluid hydrodynamic background using SJG flow. Then, using the obtained dynamics of spin potential, we obtain information on spin polarization observables and compare it with the experimental data for Au$+$Au symmetric collisions at the center of mass energy of $\sqrt{s_{\rm NN}}=200$ GeV.

From the analysis of Sec.~\ref{subsec:SpinInit}, we concluded that the initialization of spin components must be done using a function consistent with the flow and compatible with a global spin polarization directed only along the $\hat{y}$ direction, with local spin polarization possible in all three directions. The global spin polarization requirement is satisfied by choosing the symmetries of spin potential components as given in Sec.~\ref{subsec:SpinInit}.  To identify a minimal set of components, we compute the local spin polarization \eqref{meanspinBoost} at mid-rapidity $y_p=0$ with a $(1+1)$D flow $u^\mu(\tau,\, \eta_s)=(u^0,\,0,\,0,\,u^3)$, see Appendix~\ref{sec:GeneralFormOmegaP}.

\begin{figure}[ht!]
\centering
\includegraphics[width=1\linewidth]{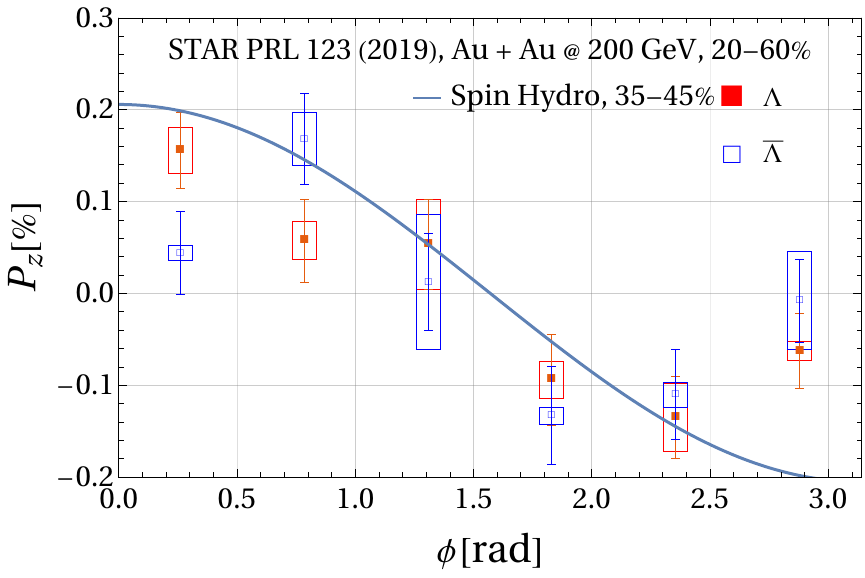}
\caption{The local longitudinal spin polarization $P_z$ as a function of momentum azimuthal angle $\phi$ for Au$+$Au collision at $\sqrt{s_{\rm NN}}=200$ GeV within centrality class $35-45$\% for a $(1+1)$D freeze-out. The experimental data for the comparison is taken from~\cite{STAR:2019erd}.}
\label{fig:Pz_2d}
\end{figure}

It is straightforward to see that $P_x$ and $P_y$ can receive contributions only from $a_x$ and $\omega_y$, while $P_z$ can receive contributions only from $a_y$ and $\omega_x$, see Eqs.~\eqref{eq:PxExp11}-\eqref{eq:PzExp11}. For these reasons, we initialize the following two spin components and keep all other spin components initially vanishing
\beq
{\omega_x}(\tau_s,\, \eta_s) &=& \frac{b_{x0}\,u^3(\tau_s,\,\eta_s)}{ u^0(\tau_s,\,\eta_s)^2}, \quad [\eta_s-{\rm odd}]
\label{eq:MinimalLongitudinalSpinIC}\\
{\omega_y}(\tau_s,\, \eta_s) &=& \frac{b_{y0}}{ u^0(\tau_s,\,\eta_s)}, \quad [\eta_s-{\rm even}]
\label{eq:MinimalSpinIC}
\eeq
where $b_{x0}$ and $b_{y0}$ are amplitude parameters, and $\tau_s > \tau_i$ is the proper-time when we start the spin evolution. The parameter $b_{y0}$ is fitted to the experimental global spin polarization value, that is, $\langle P_J\rangle=0.243\%$ for the 20-50\% centrality class, and $b_{x0}$ is fitted to the local longitudinal spin polarization $P_z$, specifically the data point $P_z(\phi=1.31)=0.053\%$ for the 20-60\% centrality class, see Fig.~\ref{fig:Pz_2d}; furthermore, in this work, we used $\tau_s=1$ fm$/c$.

Figure~\ref{fig:spin-components-1plus1-evolution} shows the evolution of $\omega_y$ (top right) and $\omega_x$ (bottom right) which are $\eta_s$-even and $\eta_s$-odd functions, respectively, resulting from the Eqs.~\eqref{eq:spinEoM}, the initialization~\eqref{eq:MinimalLongitudinalSpinIC}-\eqref{eq:MinimalSpinIC} and the background flow \eqref{eq:SJGflow} with the parameters in table~\ref{tab:AuAu200GeV_parameters} for $35-45\%$ centrality class. We observe that even though we do not initialize the acceleration spin components, they are also generated because $a_x$ (top left) and $a_y$ (bottom left) are coupled with $\omega_y$ and $\omega_x$, respectively. 
The symmetry properties of all spin components with respect to $\eta_s$ remain intact throughout the evolution, which is a result of the spin equations of motion and initial conditions, \eqref{eq:MinimalLongitudinalSpinIC}-\eqref{eq:MinimalSpinIC}.

\begin{figure}[ht!]
    \centering
    \includegraphics[width=\linewidth]{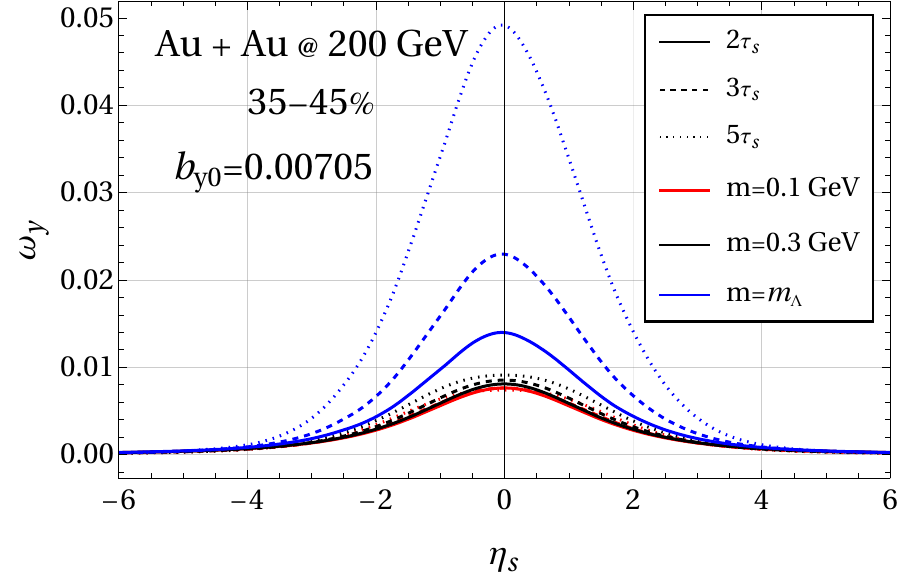}
    \caption{The spin potential component $\omega_y$ resulting from ideal spin hydrodynamic evolution as in Fig.~\ref{fig:spin-components-1plus1-evolution} with different effective masses; the initial intensity parameter $b_{y0}$ has been kept the same for all the effective masses.}
\label{fig:SpinPot_var_mass_no_rescale}
\end{figure}
Figure~\ref{fig:SpinPot_var_mass_no_rescale} shows the impact of effective mass $m$, appearing inside the thermodynamic coefficients \eqref{eq:S1}-\eqref{eq:S3} of the equations of motion~\eqref{eq:spinEoM}, on the evolution of $\omega_y$ obtained for different values of $m$ (100 MeV, 300 MeV and $m_\Lambda=1116$ MeV) starting with the same initial condition with fixed value $b_{y0}=0.00705$. It seems that the effective mass controls how fast the peak of the spin potential component grows with time $\tau$. The same behavior is observed for the other spin potential components with larger effective masses resulting in higher values for the spin potential (see Fig.~\ref{fig:SpinPot_az_var_mass_no_rescale} for the behavior of $a_z$ for different masses). As the initial intensities $b_{x0}$, $b_{y0}$, and later $a_{z0}$, are fitted to the spin polarization data, the final predictions are basically insensitive to the effective mass $m$, but they result in different initial values for the spin potential.

It is important to stress an observation that the spin component evolution satisfies the constraint for the convergence of spin potential~\cite{Abboud:2025shb} 
\beq
\frac{m}{T} > \f{\sqrt{3}}{2} \sqrt{\boldsymbol{a}^2 + \boldsymbol{\omega}^2}\,,
\label{eq:spin-potential-constraint}
\eeq
which makes the spin evolution equations non-linearly causal and symmetric-hyperbolic. As long as the evolution of the spin components satisfies the condition~\eqref{eq:spin-potential-constraint}, the solutions will be applicable to relativistic heavy-ion collisions. Here, $\boldsymbol{a}^2$ and $\boldsymbol{\omega}^2$ denote the magnitudes of the spin potential components. In our numerical implementation, this condition is verified throughout the evolution to ensure that the solutions remain within the regime of validity of the theory.

\begin{figure*}[ht!]
    \centering
    \includegraphics[width=5.91cm]{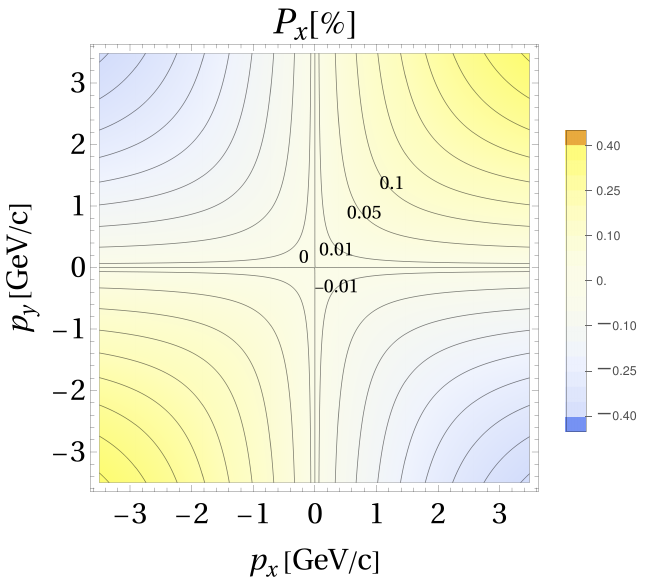}
    \includegraphics[width=5.91cm]{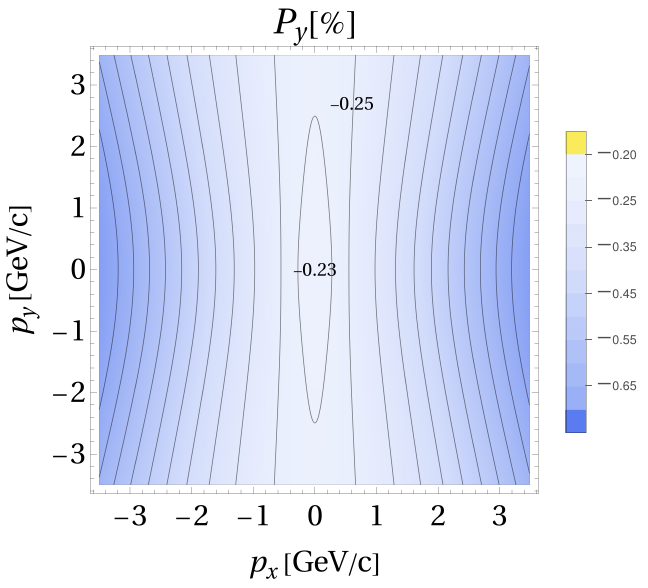}
    \includegraphics[width=5.91cm]{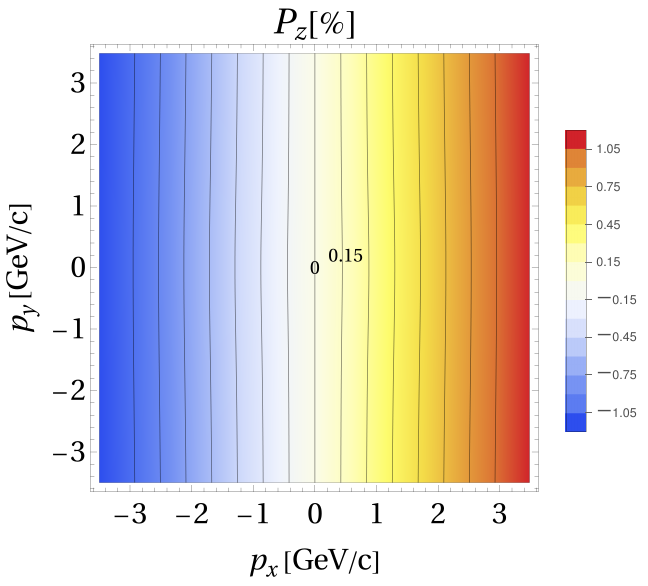}
    \caption{The local spin polarization vector components $(P_x, P_y, P_z)$ at mid-rapidity $(y_p = 0)$ as a function of transverse momentum $(p_x,p_y)$ for Au$+$Au collisions at $\sqrt{s_{\rm NN}}=200$ GeV within centrality class $35-45$\% for a $(1+1)$D freeze-out.}
\label{fig:density_plot_2d}
\end{figure*}

\begin{figure*}[ht!b]
\centering
\includegraphics[width=8.9cm]{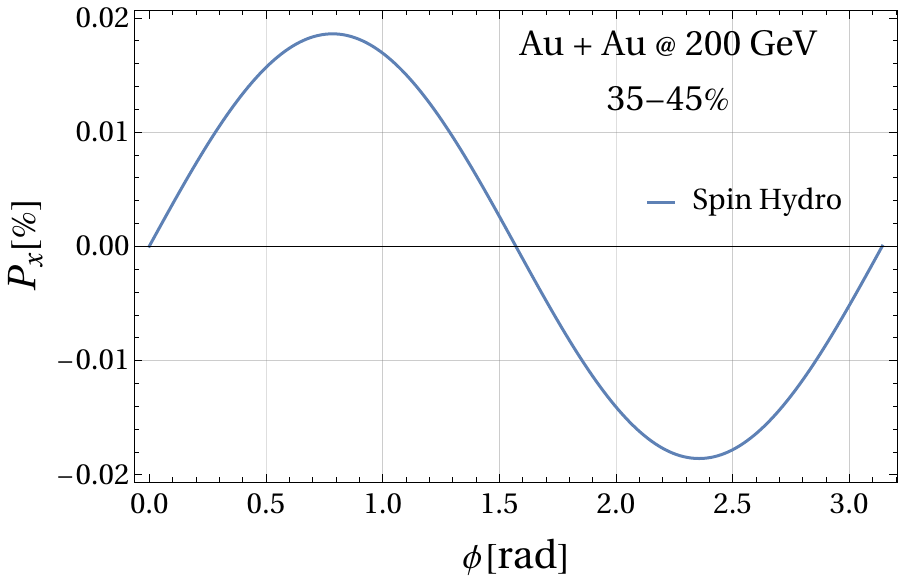}
\includegraphics[width=8.9cm]{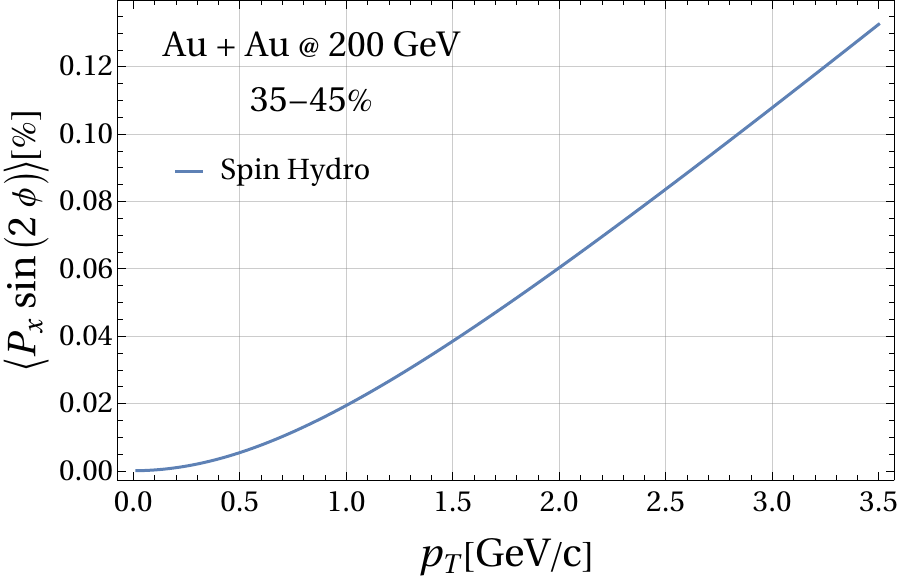}
\caption{The in-plane transverse spin polarization $P_x$ as a function of momentum azimuthal angle $\phi$ (left), and its second Fourier coefficient with the transverse momentum $p_T$  (right) for Au$+$Au collision at $\sqrt{s_{\rm NN}}=200$ GeV within centrality class $35-45$\% for $(1+1)$D freeze-out.}
\label{fig:Px_2d}
\end{figure*}

Knowing the evolution of spin potential, we can now compute the spin polarization observables for Au$+$Au symmetric collisions through Eqs.~\eqref{meanspin}-\eqref{eq:momentum-integrated-polarization}.
Figure~\ref{fig:density_plot_2d} shows the behavior of $P_\mu$ with respect to momentum $p_x$ and $p_y$ at mid-rapidity $y_p =0$ for Au$+$Au collisions at $\sqrt{s_{\rm NN}}=200$ GeV within centrality class $35-45$\%.
From Fig.~\ref{fig:density_plot_2d} (left panel), the component $P_x$ (henceforth referred to as \textit{in-plane}) exhibits a clear quadrupole pattern, with its sign alternating across successive quadrants, which is motivated by the directed flow.

Figure~\ref{fig:density_plot_2d} (middle panel) shows the behavior of $P_y$. In agreement with the chosen initialization of spin, this component remains negative, indicating that the spin angular momentum vector points opposite to the positive $y$-direction in heavy-ion collisions.

An observable of particular experimental relevance is the  spin polarization along the beam direction~\cite{STAR:2019erd,ALICE:2021pzu}, henceforth referred to as \textit{longitudinal}, see fig.~\ref{fig:density_plot_2d} (right panel). Insights into its behavior can be obtained from symmetry considerations of the chosen spin initializations (Eqs.~\eqref{eq:MinimalLongitudinalSpinIC} and \eqref{eq:MinimalSpinIC}) and Eq.~\eqref{eq:PzExp11}. Due to the assumed homogeneity in the transverse plane, the longitudinal spin polarization cannot achieve the experimentally observed quadrupole structure. However, such a structure may arise from elliptic flow generated by the initial spatial anisotropy in the transverse plane~\cite{Voloshin:2017kqp}. Motivated by this, in the next section, we extend our $(1+1)$D flow to the transverse region, modeling the transverse direction $x$ and $y$, and we observe quantitative agreement of our computations with the experimental data.

\begin{figure*}[ht!]
    \centering
    \includegraphics[width=8.9cm]{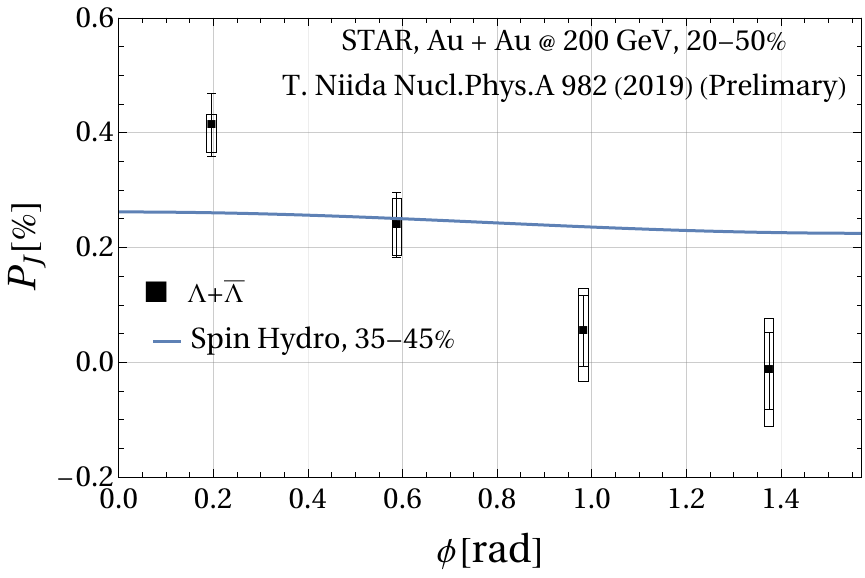}
    \includegraphics[width=8.9cm]{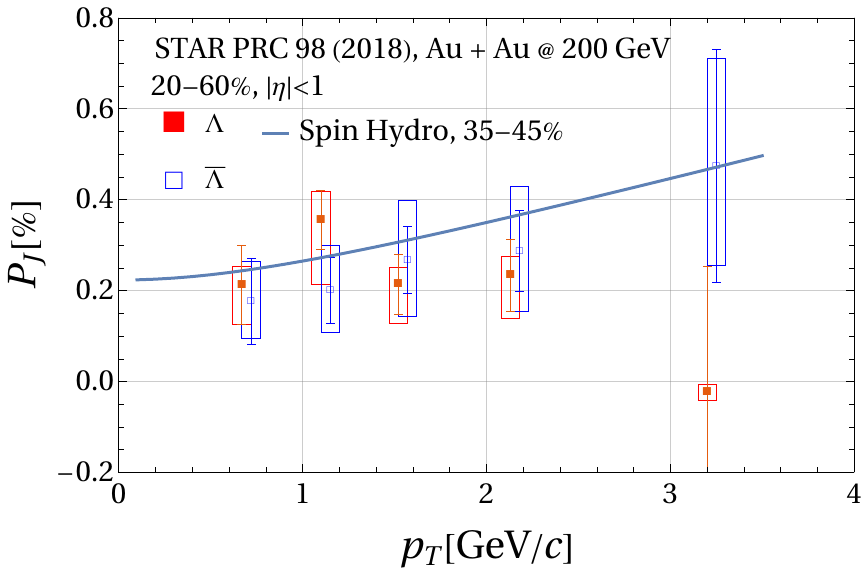}
    \caption{Spin polarization along the direction of total angular momentum of the system $J\,(-\hat{y})$ as a function of momentum azimuthal angle $\phi$ (left) and the transverse momentum $p_T$ (right) for Au$+$Au collision at $\sqrt{s_{\rm NN}}=200$ GeV within centrality class $35-45$\% for a $(1+1)$D freeze-out. The data for the left plot is taken from~\cite{Niida:2018hfw} and for the right plot is taken from~\cite{STAR:2018gyt}.}
\label{fig:PJ_2d}
\end{figure*}

The behavior of the in-plane transverse polarization $P_x$ as a function of $\phi$ (left) and $p_T$ (right) is shown in Figure~\ref{fig:Px_2d} for Au$+$Au collision at $\sqrt{s_{\rm NN}}=200$ GeV within the centrality class $35-45$\%.
The clear quadrupole pattern in Figure~\ref{fig:density_plot_2d} (left panel) results in a non-vanishing second Fourier harmonic which increases linearly with $p_T$, Fig.~\ref{fig:Px_2d} (right panel).
This prediction, if measured in experiments, will contribute to a complete understanding of spin polarization dynamics in heavy-ion collisions. We note that the trend of $P_x$ has the same qualitative behavior of longitudinal local polarization seen in experiments.
\begin{figure*}[hbt!]
    \centering
    \includegraphics[width=8.9cm]{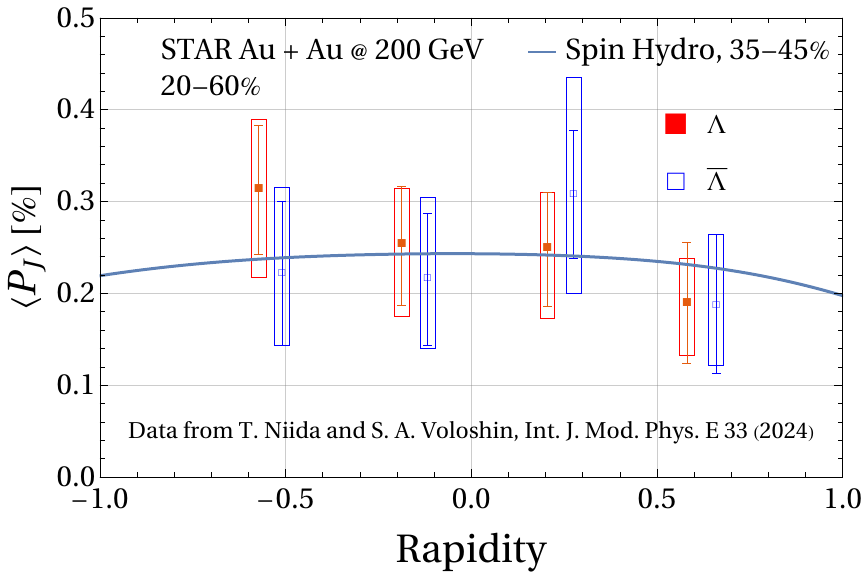}
    \includegraphics[width=8.9cm]{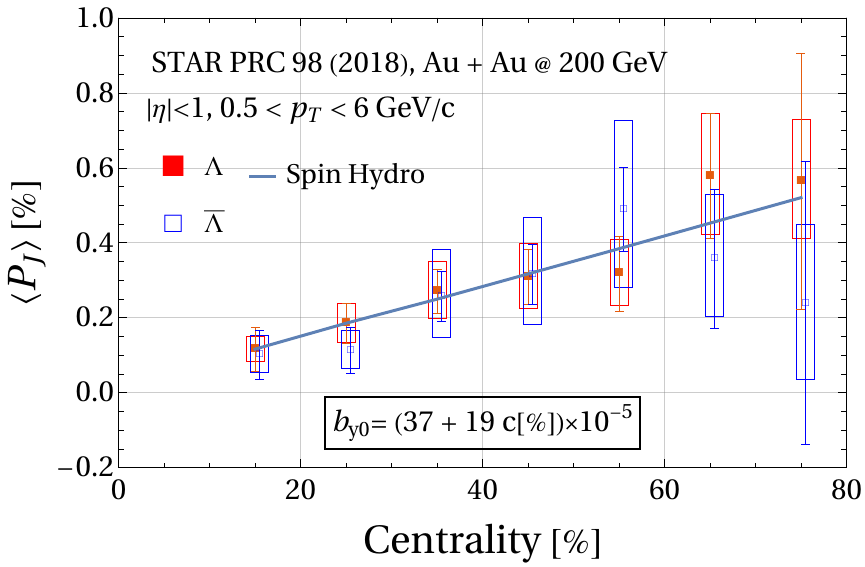}
    \caption{The rapidity (left) and centrality (right) dependence of the global spin polarization along $J$ for Au$+$Au collision at $\sqrt{s_{\rm NN}}=200$ GeV for a $(1+1)$D freeze-out.
    The data for the left plot is taken from~\cite{Niida:2024ntm} and for the right plot is taken from~\cite{STAR:2018gyt}.}
\label{fig:PJ_centrality_2d}
\end{figure*}

We then analyze the spin polarization along the direction of total angular momentum $J$, i.e., $P_J$, as a function of azimuthal angle and transverse momentum in Figure~\ref{fig:PJ_2d}. We observe that, unlike the preliminary experimental data~\cite{Niida:2018hfw}, $P_J$ is only slightly changing with $\phi$, in accordance with other predictions~\cite{Palermo:2024tza}. However, in the right panel of Figure~\ref{fig:PJ_2d}, one can see that, within the error bars, the $p_T$ dependence of $P_J$ is consistent with the experimental observations~\cite{STAR:2018gyt}. These results show that the assumptions of spin conservation and spin initialization, \eqref{eq:MinimalLongitudinalSpinIC}-\eqref{eq:MinimalSpinIC}, are capable of describing the spin polarization along $J$.

The local longitudinal spin polarization $P_z$ as a function of azimuthal angle $\phi$ is reported in Figure~\ref{fig:Pz_2d}. From Figs.~\ref{fig:density_plot_2d} and~\ref{fig:Pz_2d} it is clear that the second harmonic $\langle P_z \sin(2\phi\rangle$ is vanishing, which we refrain from showing here. This is expected due to the assumed homogeneity in the transverse plane in our $(1+1)$D setup.

Figure~\ref{fig:PJ_centrality_2d} shows the behavior of momentum integrated $\langle P_J \rangle$ (opposite to the $y$-component of $\langle P_\mu \rangle$) with respect to rapidity $(y_p)$ (left panel) and centrality (right panel) computed from Eq.~\eqref{eq:momentum-integrated-polarization}. To obtain predictions for the centrality dependence of $\langle P_J\rangle$, we fitted the parameter $b_{y0}$ in the initialization \eqref{eq:MinimalSpinIC} to the data points of the first two centrality classes and assumed a linear increase with centrality $c$: $b_{y0}(c)=b_{y0}^{(0)}+b_{y0}^{(1)}\cdot c$. Both the rapidity and centrality dependence of $\langle P_J \rangle$ are consistent with the experimental observations~\cite{STAR:2018gyt,Niida:2024ntm} within the error bars.

In summary, we highlight that ideal spin hydrodynamics in $(1+1)$D has been studied and used to compute the local and global polarization of Lambda hyperons, which has reasonable qualitative and quantitative agreement with all the experimental data. Even though there are six independent spin components, it was enough to consider only two of them to qualitatively describe the physics of polarization in heavy-ion collisions. This analysis suggests that dissipation and gradient corrections of spin potential seem to be subleading in high energy Au$+$Au collisions, in accordance with previous findings that those corrections could be important in small systems~\cite{CMS:2025nqr} and low energies~\cite{HADES:2014ttv,STAR:2021beb,HADES:2022enx}. 

We also found that a $(1+1)$D analysis is too restrictive, especially in describing the $\phi$-dependence of local longitudinal spin polarization. Hence, in the next section, we extend the $(1+1)$D to include transverse expansion at the freeze-out, which we refer to as the \textit{1-1-2 model}.
Then, we demonstrate that the inclusion of transverse freeze-out geometry, together with a non-vanishing longitudinal spin acceleration component $a_z$, is capable of explaining the data.
\section{Spin polarization with \texorpdfstring{$3$}{3}D freeze-out}
\label{sec:Spin_1_plus_3}
In this section, we extend the $(1+1)$D longitudinal hydrodynamic solution by introducing a phenomenological model for transverse expansion at freeze-out. Importantly, the dynamical evolution remains governed by the $(1+1)$D SJG solution, while transverse flow and spatial deformation are incorporated only at the level of the freeze-out hypersurface. This construction allows us to capture essential transverse effects without solving the full $(3+1)$D spin hydrodynamic equations.
\subsection{\texorpdfstring{$1-1-2$}{112} model}
To extend the flow to the transverse direction, we change the fluid velocity at later times to
\begin{equation}\label{eq:u_1_plus_3}
u^\mu = \left(\gamma_\perp\, u^0, \bm{u}_\perp, \gamma_\perp\,u^3 \right),
\end{equation}
with $\gamma_\perp = \sqrt{1+|\bm{u}_\perp|^2}$ such that $u^2=1$.
We model the flow components in the transverse plane according to an elliptic shape with axes $R_x$ and $R_y$~\cite{Heiselberg:1998es}
\begin{equation}\label{eq:uperp}
\bm{u}_\perp = u_\perp\frac{\left( R_y^2\cos\varphi,\, R_x^2\sin\varphi\right)}{\sqrt{R_y^4\cos^2\varphi + R_x^4\sin^2\varphi}}\,,
\end{equation}
with $u_\perp$ being a constant parameter giving the strength of transverse flow.
We consider the plasma at late times to be an ellipsoidal region with coordinates as
\begin{equation}
\label{eq:EllipticCoord}
x = \frac{r}{\sqrt{1+\delta}}\cos\varphi\,,\quad
y = \frac{r}{\sqrt{1-\delta}}\sin\varphi\,,
\end{equation}
where $\varphi\in[0,2\pi]$ and $r\in[0,R_{\rm FO}]$. The parameters $R_{\rm FO}$ and $\delta$ control, respectively, the overall system size at freeze-out and the elliptic deformation of the system.
\begin{figure*}[ht!]
\centering
\includegraphics[width=5.91cm]{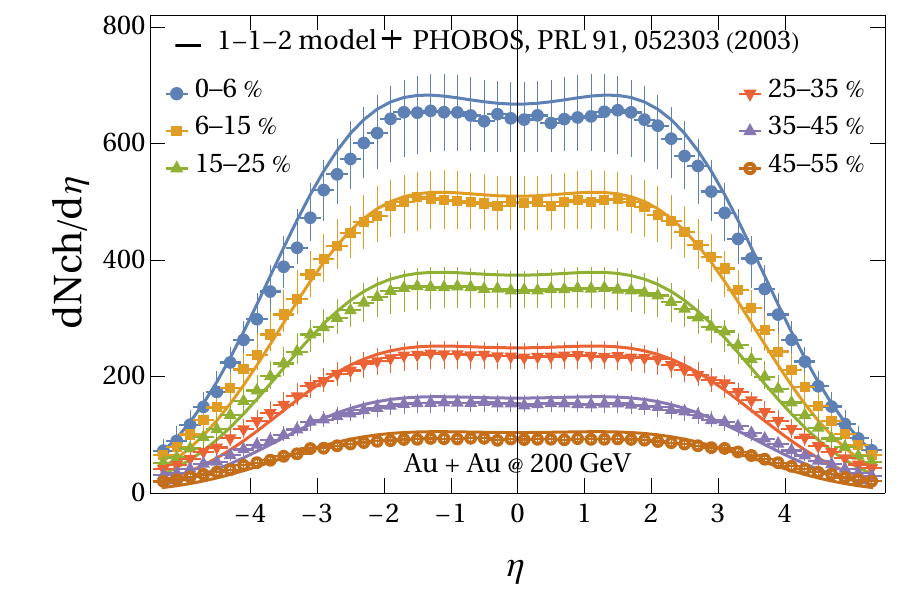}
\includegraphics[width=5.91cm]{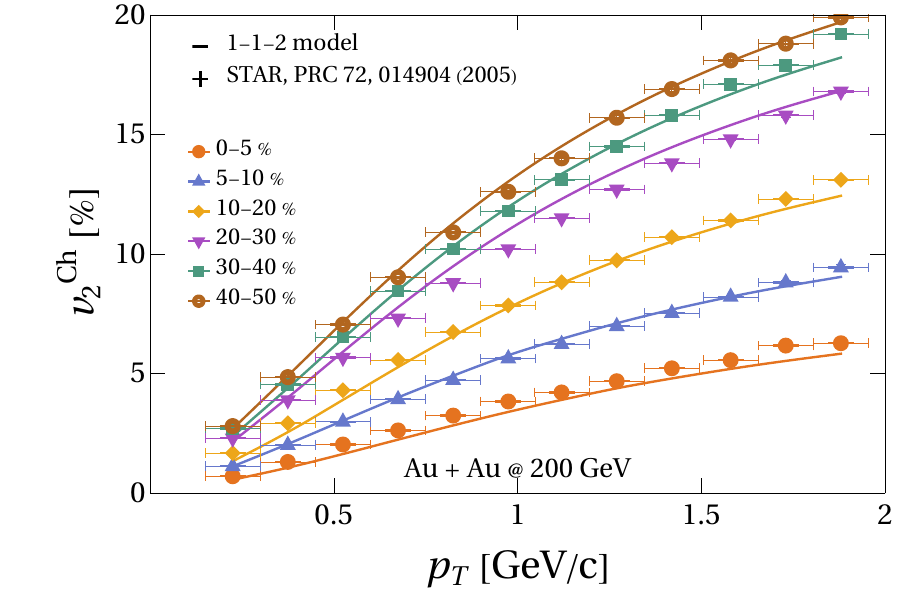}
\includegraphics[width=5.91cm]{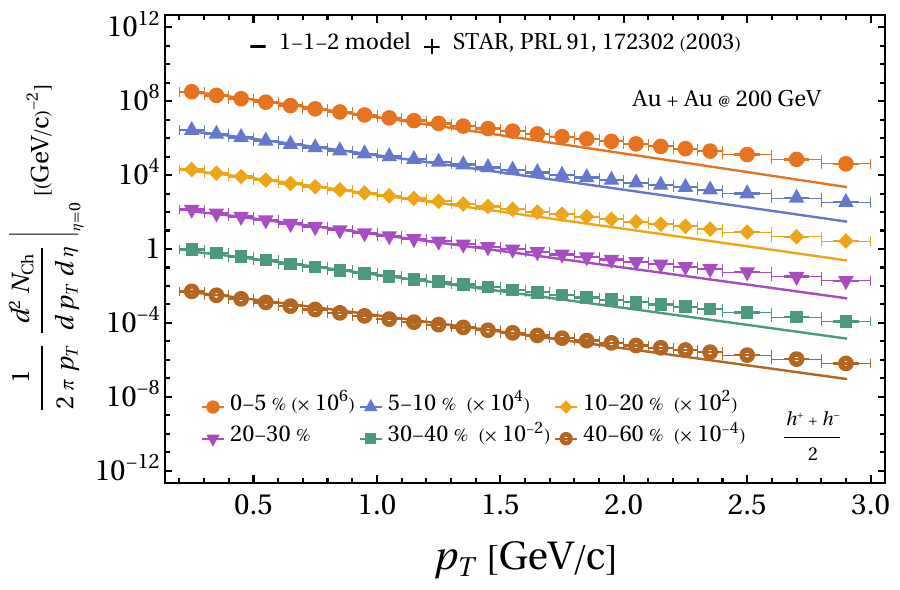}
\caption{The hadron spectra distribution obtained (using Eqs.~\eqref{eq:charged particle multiplicity-1-1-2}, \eqref{eq:v2-1-1-2}, and \eqref{eq:transverse momentum spectrum-1-1-2}) with the $1-1-2$ model for Au$+$Au collisions at $\sqrt{s_{\rm NN}}=200$ GeV for various centrality classes. The hydrodynamic parameters used are reported in Table~\ref{tab:AuAu200GeV_parameters} and~\ref{tab:AuAu200GeV_plus_2}. The data for the left plot is taken from~\cite{Back:2002wb}, the data for the middle plot is taken from~\cite{STAR:2004jwm}, and the data for the right plot is taken from~\cite{STAR:2003fka}.
}
\label{fig:hadronspectra_4d}
\end{figure*}
Those parameters are related to the axis of the ellipse as
\begin{align}
\delta = \frac{R_y^2 - R_x^2}{R_y^2 + R_x^2}\,,\quad
R_x =  \frac{R_{\rm FO}}{\sqrt{1+\delta}}\,,\quad R_y = \frac{R_{\rm FO}}{\sqrt{1-\delta}}\,.
\end{align}
In addition to the transverse flow, we have to include the elliptic shape of the freeze-out hypersurface. To achieve this goal, we deform the temperature profile such that it has a Gaussian distribution in $x$ and $y$ directions according to an ellipse. For the $1-1-2$ model, we then extend the temperature profile $T(\tau,\, \eta_s)$ of the SJG solutions, given in Eq.~\eqref{eq:SJGTemperature}, to the following temperature profile
\begin{equation}\label{eq:EllipticTemp}
\begin{split}
T(\tau,\, x,\, y,\, \eta_s) = T(\tau,\, \eta_s) \exp\left(-\frac{x^2}{2\sigma_x^2} - \frac{y^2}{2\sigma_y^2} \right).
\end{split}
\end{equation}
Defining $\sigma$ as
\begin{eqnarray}
\sigma_x^2 = \frac{\sigma^2}{1+\delta}\,,\quad
\sigma_y^2 = \frac{\sigma^2}{1-\delta}\,,
\end{eqnarray}
and using the elliptic coordinates (\ref{eq:EllipticCoord}), the exponential in Eq.~\eqref{eq:EllipticTemp} becomes
\begin{equation}
\exp\left(-\frac{x^2}{2\sigma_x^2} -\frac{y^2}{2\sigma_y^2} \right)
    = \exp\left(-\frac{r^2}{2\sigma^2}\right).
\end{equation}
The variance $\sigma$ can be related to the radius at the freeze-out $R_{\rm FO}$ and a small elliptic temperature correction parameter $\epsilon_T$. We take
\begin{equation}
\sigma = \lambda\, R_{FO},
\end{equation}
then for large $\sigma$, that is $\lambda \gg 1$, we have
\begin{align}
\exp\left(-\frac{r^2}{2\sigma^2}\right) &\simeq 1 - \frac{r^2}{2\sigma^2}
    =  1 - \frac{1}{2\lambda^2}\frac{r^2}{R_{\rm FO}^2}\,,\nn\\
    &\equiv 1 - \epsilon_T \frac{r^2}{R_{\rm FO}^2}\,,\label{eq:expansion}
\end{align}
with
\begin{equation*}
\epsilon_T \equiv \frac{1}{2\lambda^2} \ll 1\,.
\end{equation*}
The expansion in Eq.~\eqref{eq:expansion} assumes that the transverse width $\sigma$ is sufficiently large compared to the system size, such that $r^2/\sigma^2 \ll 1$. For the parameter values used in our analysis, this condition is always satisfied.
In total, we have 4 new parameters to fit to the hadron spectra: $u_\perp$, $R_{\rm FO}$, $\delta$, and $\sigma/R_{\rm FO}$.

Unlike the $(1+1)$D case, where the transverse components of the hypersurface element vanish, the inclusion of transverse flow and deformation leads to non-vanishing $\d\Sigma_x$ and $\d\Sigma_y$, see Appendix~\ref{sec:FOIntegral}. These contributions play a crucial role in modifying the resulting spin polarization observables.
With the generalized freeze-out hypersurface and the transverse flow \eqref{eq:uperp}, we use the Cooper-Frye formula \eqref{eq:Cooper-Frye-formula} to fit the parameters to the pseudo-rapidity distribution of charged particle multiplicity, see Figure~\ref{fig:hadronspectra_4d} (left panel), the $p_T$ dependence of elliptic flow $v_2$ of the charged particles, see Figure~\ref{fig:hadronspectra_4d} (middle panel), and the transverse momentum spectrum of charged particles, see Figure~\ref{fig:hadronspectra_4d} (right panel). See Appendix~\ref{sec:FOIntegral} for more details. We observe reasonably good agreement with the experimental data in all centrality classes within uncertainties. The hydrodynamic transverse plane parameters used in our analysis are given in Table~\ref{tab:AuAu200GeV_plus_2}. The generalized freeze-out hypersurface and the transverse flow \eqref{eq:uperp} are also included in Eqs.~\eqref{meanspin}-\eqref{eq:momentum-integrated-polarization} to obtain the spin polarization of $\Lambda$ hyperons.

\begin{table}[ht!]
\centering
\begin{tabular}{c|c|c|c|c|c|c}
Centrality (\%)& $N_{\rm ch}^{\rm exp}$ & $u_\perp$ & $R_{\rm FO}$(fm) & $\delta$ & ${\sigma}/{R_{\rm FO}}$ & $N_{\rm ch}^{\rm th}$\\
\hline
$0-6$ & $4941^{+525}_{-525}$ & 0.31 & 14.4 & 0.1 & 2.0 & 5201 \\
\hline
$6-15$ & $3851^{+407}_{-407}$ & 0.36 & 13.4 & 0.14 & 1.5 & 3805 \\
\hline
$15-25$ & $2737^{+288}_{-288}$ & 0.37 & 12.8 & 0.195 & 1.5 & 2791 \\
\hline
$25-35$ & $1862^{+194}_{-194}$ & 0.4 & 11.1 & 0.25 & 1.5 & 1852 \\
\hline
$35-45$ & $1219^{+117}_{-117}$  & 0.4 & 9.0 & 0.27 & 1.6 & 1222 \\
\hline
$45-55$ & $745^{+77}_{-77}$ & 0.4 & 7.3 & 0.29 & 1.6 & 775 \\
\hline
$55-65$ & - & 0.36 & 5.6 & 0.3 & 1.5 & - \\
\hline
$65-75$ & - & 0.32 & 4.0 & 0.31 & 1.5 & -
\end{tabular}
\caption{The hydrodynamic transverse plane parameters for Au$+$Au collisions at $\sqrt{s_{\rm NN}}=200$ GeV used in this work.}
\label{tab:AuAu200GeV_plus_2}
\end{table}
We note that although the longitudinal flow is derived from an ideal, non-dissipative hydrodynamic solution, the transverse components are incorporated (at the freeze-out) via a phenomenological model with parameters tuned to experimental data. While the amplitude of the pseudo-rapidity distribution is primarily sensitive to the freeze-out radius $R_{\rm FO}$ and the width $\sigma$ (parameters that in the previous $(1+1)$D model were accounted for by the pre-factor $A_\perp$ in Eq.~\eqref{eq:pseudo-rapidity-distribution1+1}), the elliptic flow and inverse transverse mass spectra are highly sensitive to transverse expansion. In a full $(3+1)$D hydrodynamic framework, these observables would necessitate viscous corrections. By fitting these transverse parameters, we effectively incorporate dissipative effects into the background flow, even though the evolution of the spin potential remains governed strictly by the ideal $(1+1)$D dynamics.
\subsection{Spin polarization of \texorpdfstring{$\Lambda$}{Lambda} hyperons in \texorpdfstring{$(1+1+2)$}{1+1+2}D}
\label{subsec:mean-spin-1plus3}
The inclusion of the transverse flow  drastically changes the behavior of local spin polarization, requiring a re-evaluation of the spin potential initialization. Since the freeze-out hypersurface remains $\eta_s$-even, the requirement for global spin polarization still mandates that the spin potential components must have the properties listed in Sec.~\ref{subsec:SpinInit}. For the $(1+1)$D analysis, we found that $\omega_y$ and $a_x$ contribute to the local spin polarization components $P_x$ and $P_y$, while $\omega_x$ and $a_y$ contribute to $P_z$ without creating a quadrupole structure. These conclusions still hold for the $1-1-2$ model. 

However, local spin polarization now receives contributions also from $u^x$, $u^y$, $\d\Sigma_x$, and $\d\Sigma_y$. In particular, from the analytical expression of $(\tilde{\omega}_{3\lambda}p^\lambda)^*$, we see that the spin component $a_z$ gives a non-vanishing contribution to $P_z$ that is proportional to $p_x\,p_y$ (see Eq.~\eqref{eq:Pz-even-1-1-2} in Appendix~\ref{sec:GeneralFormOmegaP}), which is needed to reproduce the experimental data. In order to reduce the number of parameters in the $(1-1-2)$D model, we turned off the $\omega_x$ and $a_y$ components and included $a_z$ instead. Furthermore, we see that at mid-rapidity $a_z$ does not contribute to $P_x$ and $P_y$, and conversely, $\omega_y$ and $a_x$ do not contribute to $P_z$. 

\begin{figure*}[t!hb]
\centering
\includegraphics[width=8.9cm]{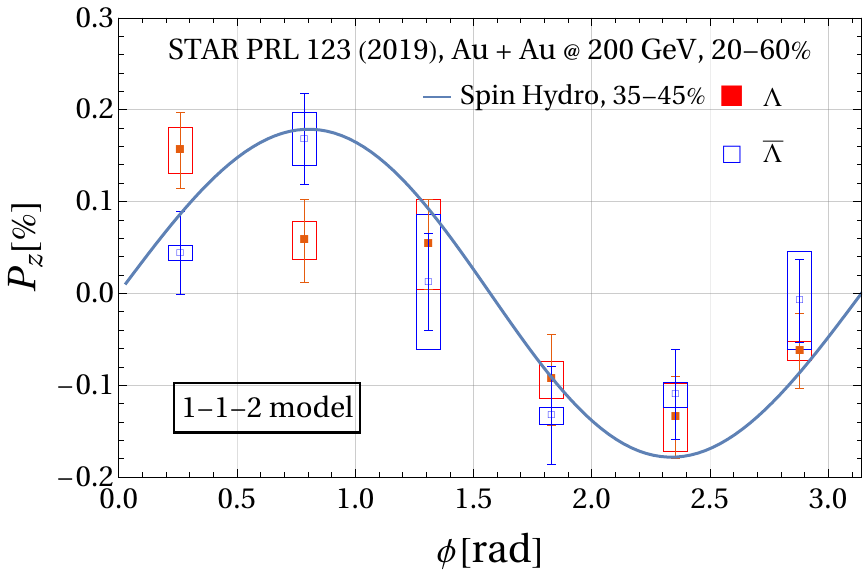}
\includegraphics[width=8.9cm]{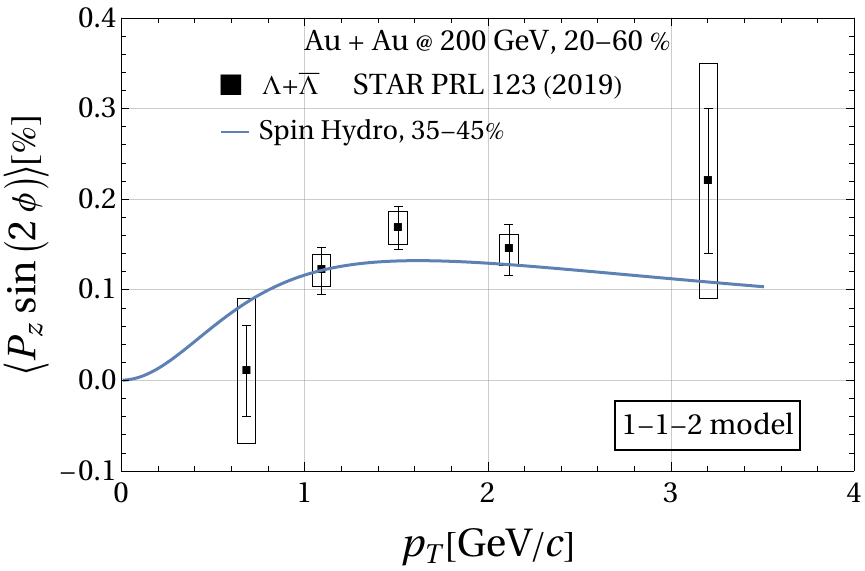}
\caption{The local longitudinal spin polarization $P_z$ as a function of momentum azimuthal angle $\phi$ (left) and its second Fourier coefficient with the transverse momentum $p_T$  (right) for Au$+$Au collision at $\sqrt{s_{\rm NN}}=200$ GeV within centrality class $35-45$\% for a $1+1+2$D freeze-out. The experimental data for the comparison is taken from~\cite{STAR:2019erd}.}
\label{fig:Pz_4d}
\end{figure*}

In conclusion, we found that the \textit{minimal} set of components is obtained by initializing the following spin potential components
\beq
{a_z}(\tau_s,\, \eta_s) &=& \frac{a_{z0}\,u^3(\tau_s,\,\eta_s)}{ u^0(\tau_s,\,\eta_s)^2}\,, \quad [\eta_s-{\rm odd}]
\label{eq:MinimalLongitudinalSpinIC1+3}\\
{\omega_y}(\tau_s,\, \eta_s) &=& \frac{b_{y0}}{ u^0(\tau_s,\,\eta_s)}\,. \quad [\eta_s-{\rm even}]
\label{eq:MinimalSpinIC1+3}
\eeq
The parameter $b_{y0}$ is again fitted to the experimental global spin polarization value, and $a_{z0}$ is fitted to the local longitudinal spin polarization $P_z$, specifically $\langle P_z\sin(2\phi)\rangle(p_T=1.09\text{ GeV}/c)=0.121\%$ for 20-60\% centrality class, see Fig.~\ref{fig:Pz_4d} (right panel).

\begin{figure}[t!h]
\centering
\includegraphics[width=1\linewidth]{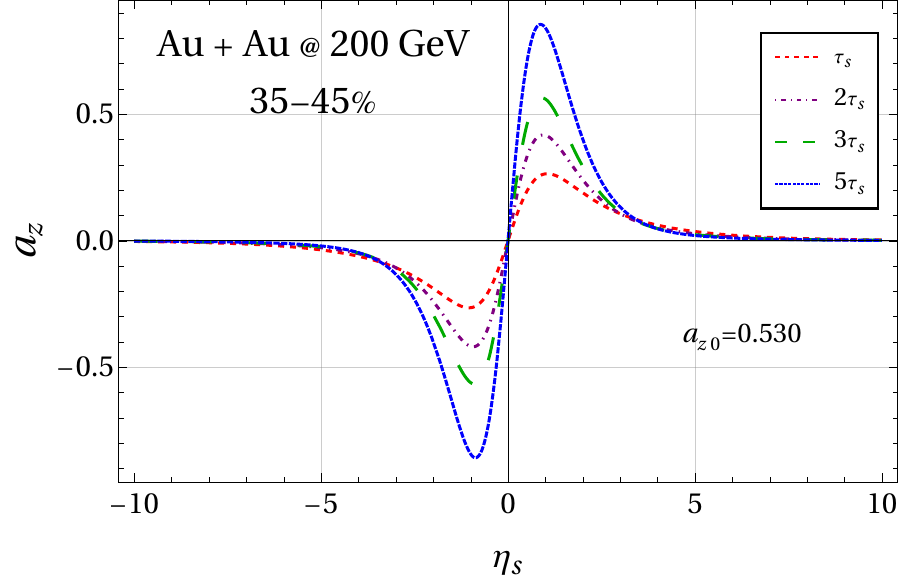}
\caption{The spin potential component $a_z$ resulting from ideal spin hydrodynamic with $(1+1)$D symmetric ($a=1$) SJG flow for a Au$+$Au collision at $\sqrt{s_{\rm NN}}=200$ GeV within centrality class $35-45$\% and with spin evolution starting at $\tau_s=1$ fm$/c$.}
\label{fig:SpinPot_az}
\end{figure}

\begin{figure}[ht!]
    \centering
    \includegraphics[width=\linewidth]{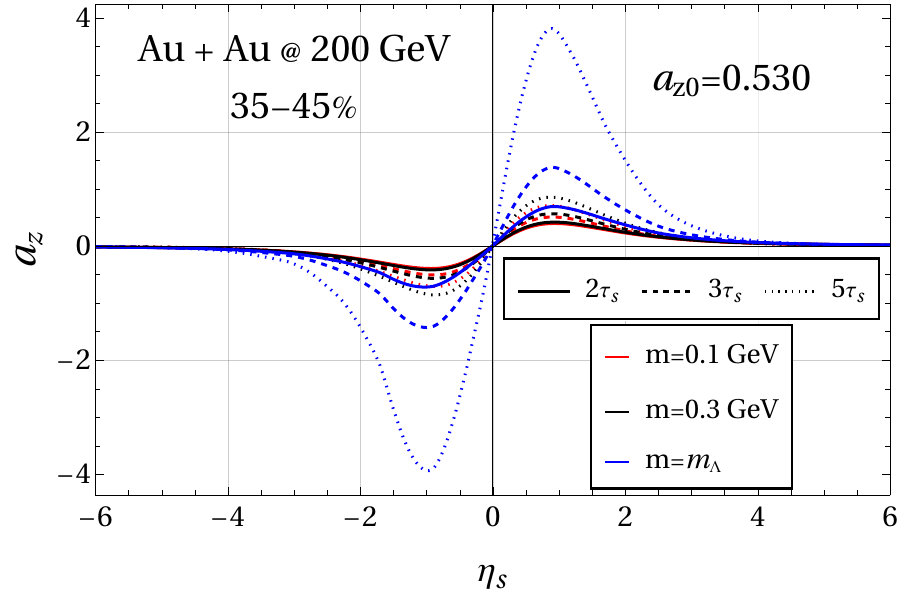}
    \caption{The spin potential component $a_z$ resulting from ideal spin hydrodynamic evolution as in Fig.~\ref{fig:SpinPot_az} with different effective masses; the initial intensity parameter $a_{z0}$ has been kept the same for all the effective masses.}
\label{fig:SpinPot_az_var_mass_no_rescale}
\end{figure}

From Eqs.~\eqref{eq:UXspinEoM} and \eqref{eq:XZspinEoM}, we note that the spin component $\omega_y$ couples with the $a_x$ component. Their evolution is the same as depicted in Figure~\ref{fig:spin-components-1plus1-evolution} (top left and top right); hence, we do not show it in this section again. However, $a_z$ evolves independently (see Eq.~\eqref{eq:UZspinEoM}). Its behavior with spacetime rapidity for various proper-times is shown in Figure~\ref{fig:SpinPot_az} for Au$+$Au collision at $\sqrt{s_{\rm NN}}=200$ GeV within centrality class $35-45$\%. As before, we kept the spin evolution starting time as $\tau_s = 1$ fm$/$c. We observe that it increases with $\eta_s$ and then quickly vanishes in forward rapidity. The symmetry of $a_z$ is dictated by its equation of motion and initialization.
The impact of the effective mass $m$ on the evolution of $a_z$ is shown in Fig.~\ref{fig:SpinPot_az_var_mass_no_rescale} and it is similar to what observed for $\omega_y$ in Fig.~\ref{fig:SpinPot_var_mass_no_rescale}.

\begin{figure}[ht!]
\centering
\includegraphics[width=1\linewidth]{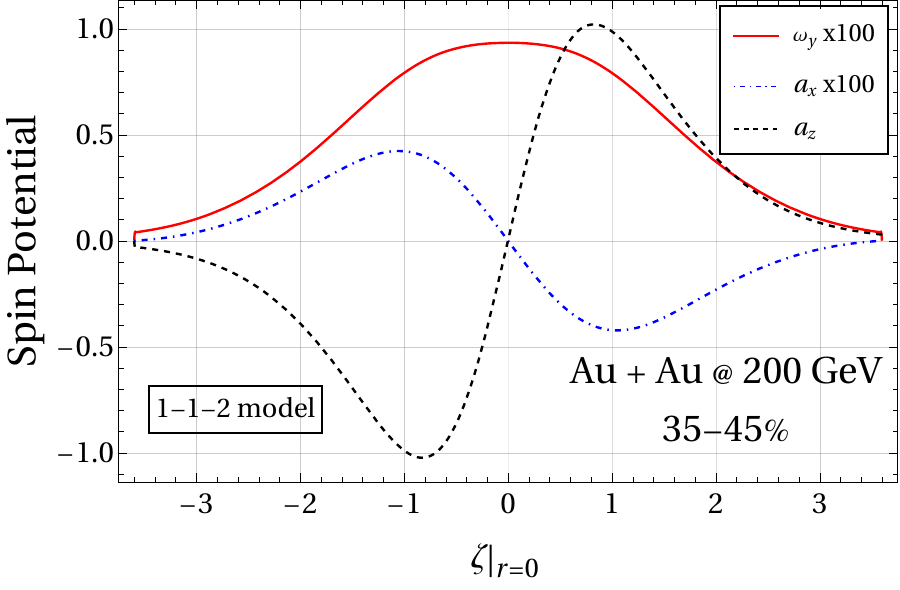}
\caption{The spin potential components at the center ($r=0$) of the 3D freeze-out as functions of the freeze-out variable $\zeta$ for Au$+$Au collision at $\sqrt{s_{\rm NN}}=200$ GeV within centrality class $35-45$\%.}
\label{fig:SpinPot_FO}
\end{figure}

To understand how the longitudinal spin acceleration component $a_z$ can help elucidate the physics of local longitudinal spin polarization, we plot the contributions of our non-vanishing spin components $\omega_y$, $a_x$, and $a_z$ at the freeze-out with respect to the freeze-out parameter $\zeta$ at the center $(r=0)$ in Figure~\ref{fig:SpinPot_FO}.

The region $-1<\zeta<1$ corresponds to the mid-rapidity regime, with the dominant contribution to particle emission at freeze-out, making the enhancement of $a_z$ in this region particularly relevant for polarization observables.

Indeed, we shall see that this is the case, confirming that the spin acceleration component can contribute to spin polarization~\cite{Palermo:2024tza}. This indicates that for the $1-1-2$ model analysis, switching on $\omega_y$ and $a_z$ while keeping all other spin components initially vanishing is enough to explain the spin polarization observables. We also observe that $a_z$ is the dominant component, reaching or even exceeding unity, which formally violates the assumption of small spin polarization. However, $a_z$ remains small throughout the majority of the evolution and across most of the rapidity range. This suggests that corrections arising from treatments beyond the linear polarization approximation would be negligible.

\begin{figure*}[b!ht]
\centering
\includegraphics[width=5.91cm]{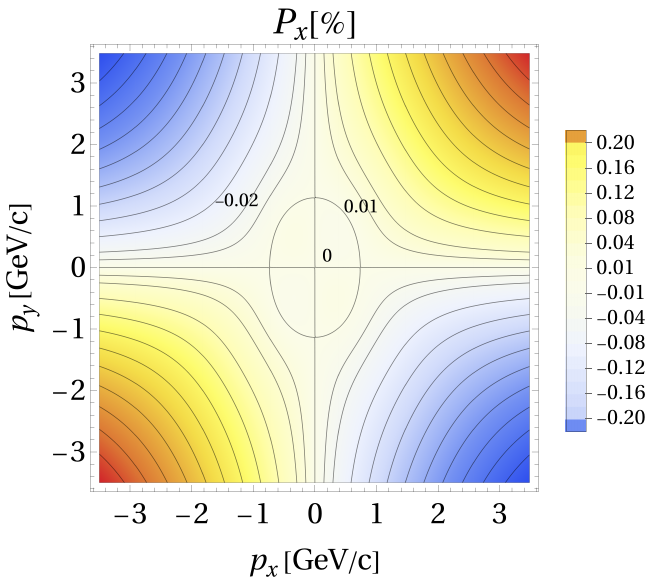}
\includegraphics[width=5.91cm]{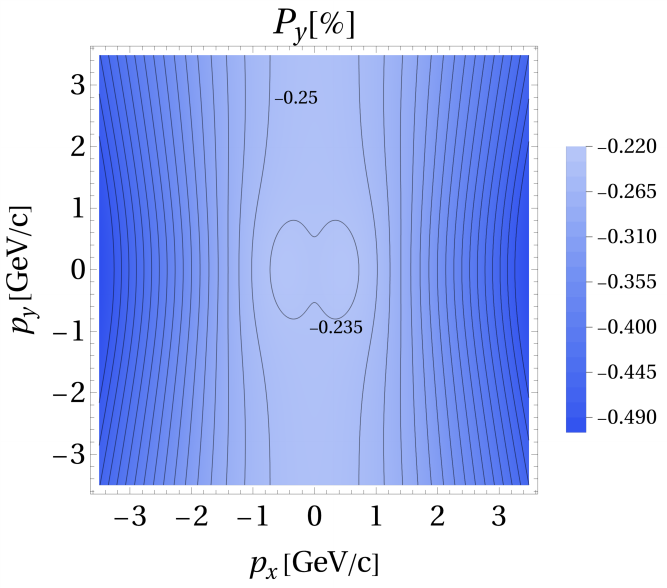}
\includegraphics[width=5.91cm]{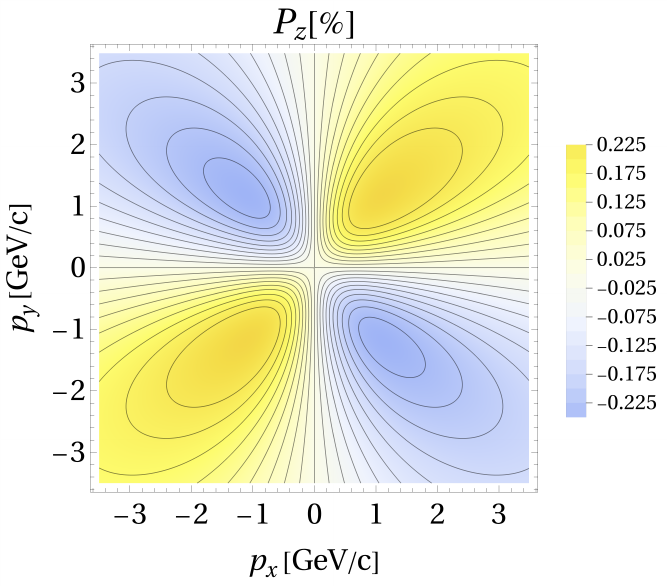}
\caption{The local spin polarization vector components at mid-rapidity as a functions of transverse momentum components for Au$+$Au collisions at $\sqrt{s_{\rm NN}}=200$ GeV within centrality class $35-45$\% for $1+1+2$D freeze-out.}
\label{fig:density_plot_4d}
\end{figure*}
We first show the computed spin polarization $P_\mu$ as a function of $p_x$ and $p_y$ in Figure~\ref{fig:density_plot_4d} at mid-rapidity. In comparison with Figure~\ref{fig:density_plot_2d}, we observe that the quadrupole structure in $P_x$ is still intact; however, there is a decrease in magnitude. The same conclusion also holds for the $P_y$ component of the spin polarization vector, where the magnitude is decreasing, but the qualitative behavior remains the same. To emphasize again, negative sign in $P_y$ means that the spin angular momentum vector points opposite to the positive $y$-direction in heavy-ion collisions, as chosen by fitting $b_{y0}$.

Interestingly, as we expected, the non-vanishing $a_z$ evolution and transverse flow components incorporated in the $1-1-2$ model give rise to the quadrupole pattern in the local longitudinal spin polarization $P_z$, similar to $P_x$, with a higher magnitude, which is driven by the elliptic flow $v_2$ in Figure~\ref{fig:hadronspectra_4d}. 
\begin{figure*}[ht!]
\centering
\includegraphics[width=8.9cm]{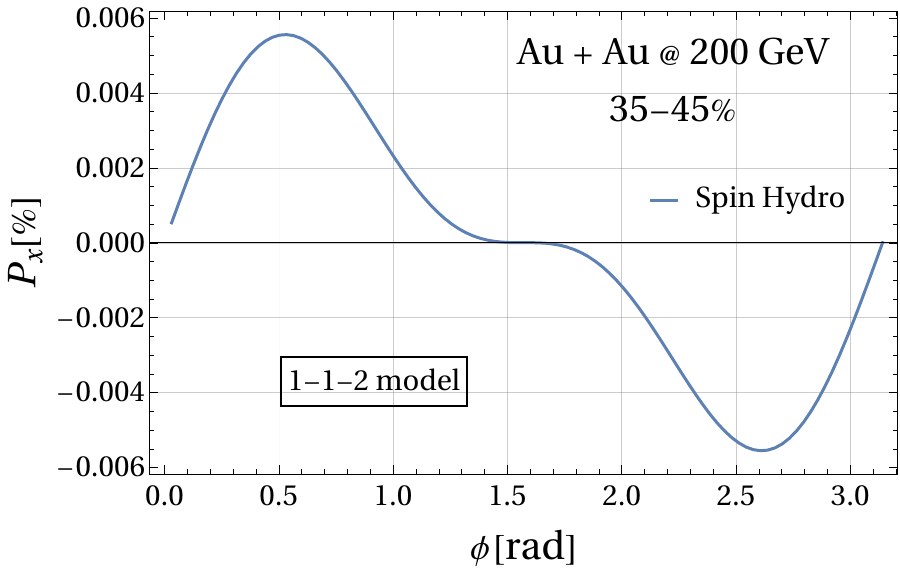}
\includegraphics[width=8.9cm]{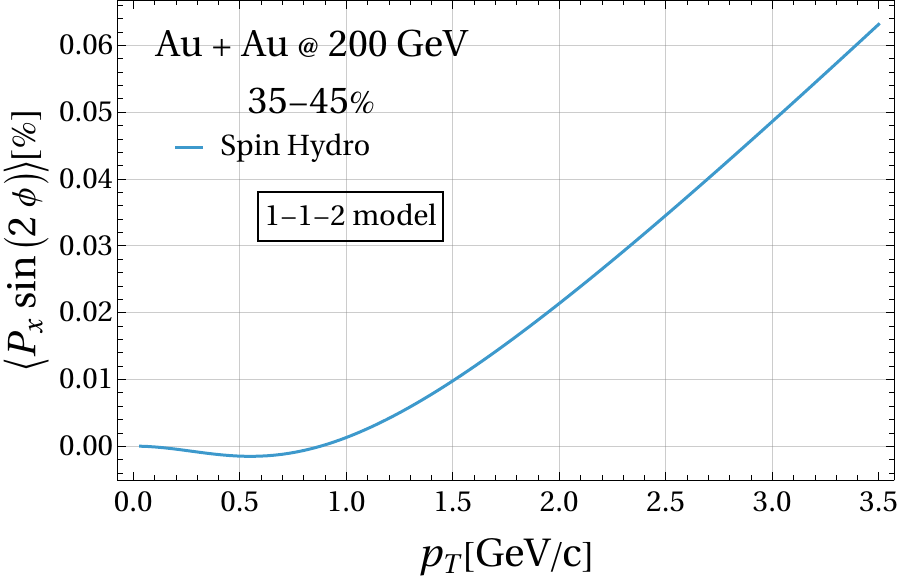}
\caption{The in-plane transverse spin polarization $P_x$ as a function of momentum azimuthal angle $\phi$ (left), and its second Fourier coefficient with the transverse momentum $p_T$  (right) for Au$+$Au collision at $200$ GeV within centrality class $35-45$\% for $1+1+2$D freeze-out.}
\label{fig:Px_4d}
\end{figure*}

Next, we plot the behavior of in-plane transverse spin polarization $P_x$ with respect to the azimuthal angle $\phi$ and transverse momentum $p_T$ in Figure~\ref{fig:Px_4d}. We observe that, in comparison to Figure~\ref{fig:Px_2d} obtained with $(1+1)$D flow, the qualitative trend remains the same despite the decrease in magnitude. However, there is a dip in the magnitude of the second Fourier coefficient of $P_x$ with the transverse momentum $p_T$  (right panel) around $p_T = 0.6$ GeV$/$c, which is due to the transverse flow expansion introduced in our modeling (see Eq.~\eqref{eq:Px_112_model}).
The observed $\phi$ dependence of in-plane transverse polarization has also been seen in a recent blast-wave model analysis~\cite{Arslan:2025tan} (see Refs.~\cite{Karpenko:2016jyx,Xia:2018tes,Sun:2021nsg,Florkowski:2021wvk} for earlier studies on in-plane spin polarization). This prediction, if measured in experiments, will contribute to a complete understanding of spin polarization dynamics in heavy-ion collisions.

\begin{figure*}[ht!]
\centering
\includegraphics[width=8.9cm]{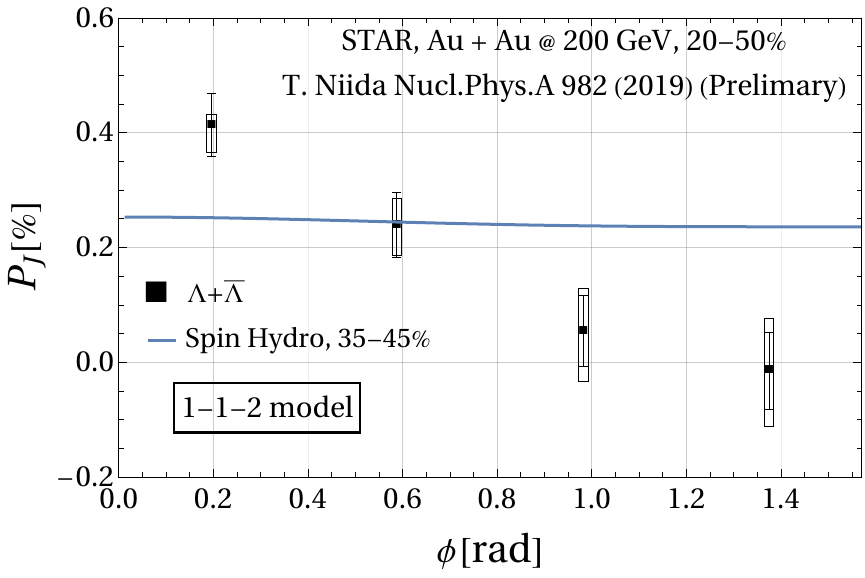}
\includegraphics[width=8.9cm]{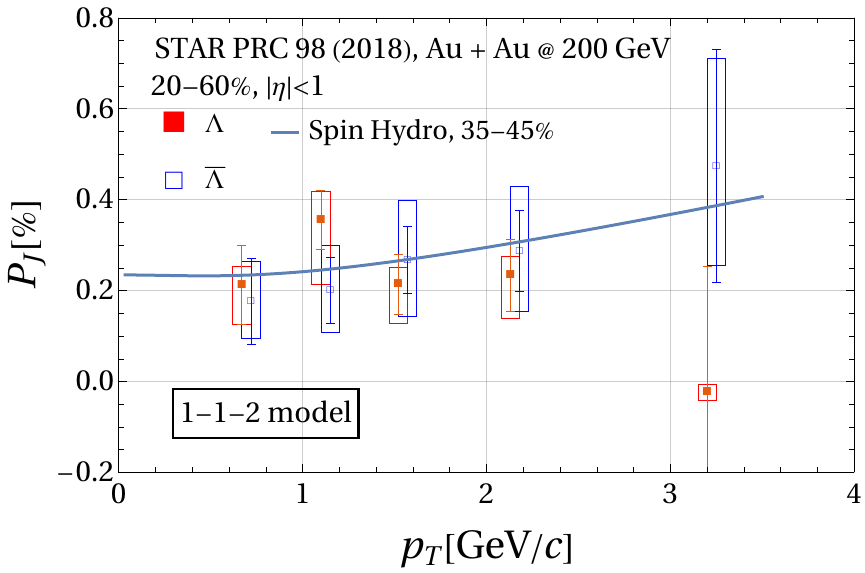}
\caption{Spin polarization along the total angular momentum of the system $J\,(-\hat{y})$ as a function of momentum azimuthal angle $\phi$ (left) and the transverse momentum $p_T$ (right) for Au$+$Au collision at $\sqrt{s_{\rm NN}}=200$ GeV within centrality class $35-45$\% for $1+1+2$D freeze-out. The data for the left plot is taken from~\cite{Niida:2018hfw} and for the right plot is taken from~\cite{STAR:2018gyt}.}
\label{fig:PJ_4d}
\end{figure*}

We then show $P_J$, the spin polarization component along the direction of total angular momentum $J$, as a function of $\phi$ and $p_T$ in Figure~\ref{fig:PJ_4d}. In the $1-1-2$ model, the qualitative behavior remains the same as in Figure~\ref{fig:PJ_2d}.

\begin{figure*}[t!hb]
\centering
\includegraphics[width=8.9cm]{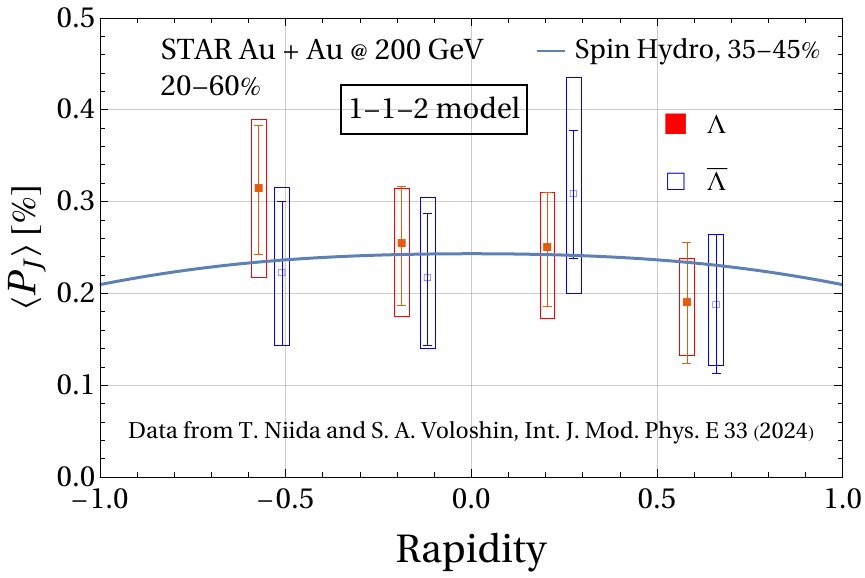}
\includegraphics[width=8.9cm]{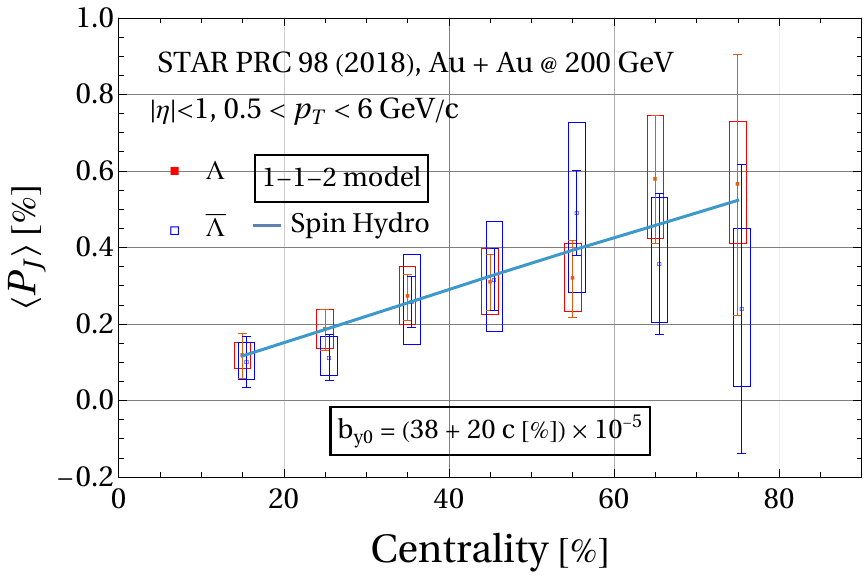}
\caption{The rapidity (left) and centrality (right) dependence of the global spin polarization along $J$ for Au$+$Au collision at $\sqrt{s_{\rm NN}}=200$ GeV for a $1+1+2$D freeze-out. The data for the left plot is taken from~\cite{Niida:2024ntm} and for the right plot is taken from~\cite{STAR:2018gyt}.}
\label{fig:PJ_centrality_4d}
\end{figure*}

Figure~\ref{fig:PJ_centrality_4d} shows the behavior of momentum integrated $\langle P_J \rangle$ (the $y$-component of $\langle P_\mu \rangle$) with respect to rapidity $(y_p)$ (left panel) and centrality (right panel) computed from Eq.~\eqref{eq:momentum-integrated-polarization}. Both the rapidity and centrality dependence of $\langle P_J \rangle$ are consistent with the experimental observations~\cite{STAR:2018gyt,Niida:2024ntm} within the uncertainties. As before, the centrality dependence of $\langle P_J\rangle$ was obtained by fitting the parameter $b_{y0}$ in the initialization \eqref{eq:MinimalSpinIC1+3} to the data points of the first two centrality classes and assuming a linear increase with centrality $c$: $b_{y0}(c)=b_{y0}^{(0)}+b_{y0}^{(1)}\cdot c$. In comparison with Figure~\ref{fig:PJ_centrality_2d}, both the magnitude and trend remain the same, indicating that the transverse flow has almost no contribution to global polarization and that a linear increase assumption of the rotational spin component with centrality is compatible with the data.

\begin{figure}[ht!]
\centering
\includegraphics[width=1\linewidth]{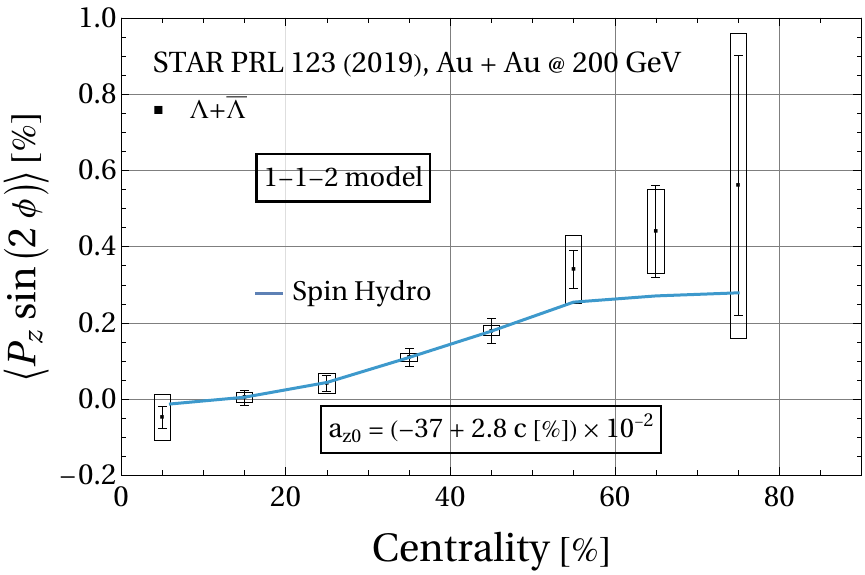}
\caption{The centrality dependence of the second Fourier coefficient of the longitudinal local spin polarization for Au$+$Au collision at $\sqrt{s_{\rm NN}}=200$ GeV. Results are obtained with $1+1+2$D freeze-out. The experimental data for the comparison is taken from~\cite{STAR:2019erd}.}
\label{fig:Pz_centrality_4d}
\end{figure}

Figures~\ref{fig:Pz_4d} and \ref{fig:Pz_centrality_4d} show the behavior of local longitudinal spin polarization with respect to $\phi$, $p_T$, and centrality. With the inclusion of the longitudinal spin acceleration component $a_z$ in $(1+1+2)$D, we were able to explain the $\phi$-dependence of $P_z$, which strongly suggests the importance of the spin acceleration component. It is interesting to note that, differently from previous numerical calculations~\cite{Palermo:2024tza,Singh:2024cub}, the second harmonic of the longitudinal polarization does not increase linearly with $p_T$, but instead increases at low $p_T$, reaches a maximum, and then decreases, Fig.~\ref{fig:Pz_4d} (right panel). 

This differ from the transverse-momentum dependence of the in-plane transverse polarization (Fig.~\ref{fig:Px_4d}), which grows linearly. This difference can be understood directly from the mid-rapidity structure of the Pauli-Lubanski vector given in Appendix~\ref{sec:GeneralFormOmegaP}. The second harmonic of $P_x$ receives a direct contribution proportional to $p_xp_y=(p_T^2/2)\sin(2\phi)$, multiplied by the spin components $a_x$ and $\omega_y$. The main contribution of this term is obtained when multiplied by the volume element $p^\tau\d\Sigma_\tau + p^\eta\d\Sigma_\eta$ in Eq.~\eqref{eq:dSigmaPtauEta} which is proportional to $m_T=\sqrt{m_\Lambda^2 + p_T^2}$. These factors combined, $p_xp_y$ and $m_T$, make $\langle P_x\sin(2\phi)\rangle$ grow approximately as $p_T^2$ at small transverse momentum and approximately linearly at larger transverse momentum, explaining Fig.~\ref{fig:Px_4d}.

By contrast, as discussed in Appendix~\ref{sec:GeneralFormOmegaP}, in the $1-1-2$ model the quadrupole structure of $P_z$  is generated by the longitudinal spin-acceleration component through the term $(p_yu^1-p_xu^2)u^3a_z$ multiplied by $p^x\d\Sigma_x + p^y\d\Sigma_y$. Notably, the transverse volume element~\eqref{eq:dSigmaPxY} lacks the factor $m_T$, and hence, $\langle P_z\sin(2\phi)\rangle$ is not expected to increase as much as the in-plane polarization. This explains the behavior in Fig.~\ref{fig:Pz_4d}.

Increased precision in measurements of $\langle P_z\sin(2\phi)\rangle$ at high $p_T$ would be instrumental in discriminating between different theoretical frameworks. Specifically, it could distinguish whether spin polarization is induced by vorticity and shear or by a genuine spin potential, as these contributions are expected to exhibit distinct momentum dependencies~\cite{Buzzegoli:2021wlg}. Recent measurements of the second-harmonic longitudinal spin polarization in isobar collisions (Ru$+$Ru and Zr$+$Zr) at $\sqrt{s_{\rm NN}}=200$ GeV demonstrate this improved precision, showing a clear decrease with $p_T$ and becoming compatible with zero near $p_T\sim 4$ GeV$/$c~\cite{STAR:2023eck}. These data further indicate that the third-order Fourier sine coefficient follows a similar qualitative trend as the second-order one. In contrast, as we are neglecting geometry fluctuations by assuming homogeneity of the spin potential in the transverse direction, the current model predicts a vanishing third harmonic at mid-rapidity, i.e., $\langle P_z\sin(3\phi)\rangle=0$, a result that follows directly from the explicit expressions derived in Appendix~\ref{sec:GeneralFormOmegaP} and the transverse flow~\eqref{eq:uperp}.

The centrality dependence shown in Fig.~\ref{fig:Pz_centrality_4d} is derived by assuming a linear scaling of $a_{z0}$ with centrality, with the slope determined by fitting the data points in the $35\%$ and $45\%$ centrality classes. Within uncertainties, it reproduces the data. However, the underestimation observed in ultra-peripheral collisions may stem from several factors: the transverse flow was not explicitly tuned to the hadronic spectra (see Fig.~\ref{fig:hadronspectra_4d}), $a_{z0}$ may exhibit a super-linear dependence on centrality, or additional spin polarization mechanisms may become relevant in this regime.

\subsection{Sensitivity to initial conditions and transverse flow}
\label{subsec:VarParam}
\begin{figure*}[ht!]
    \centering
    \includegraphics[width=8.9cm]{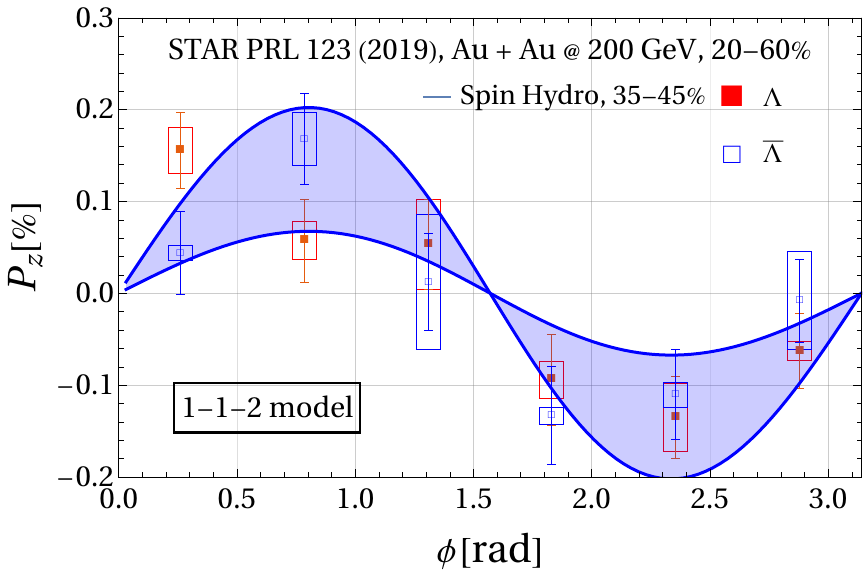}
    \includegraphics[width=8.9cm]{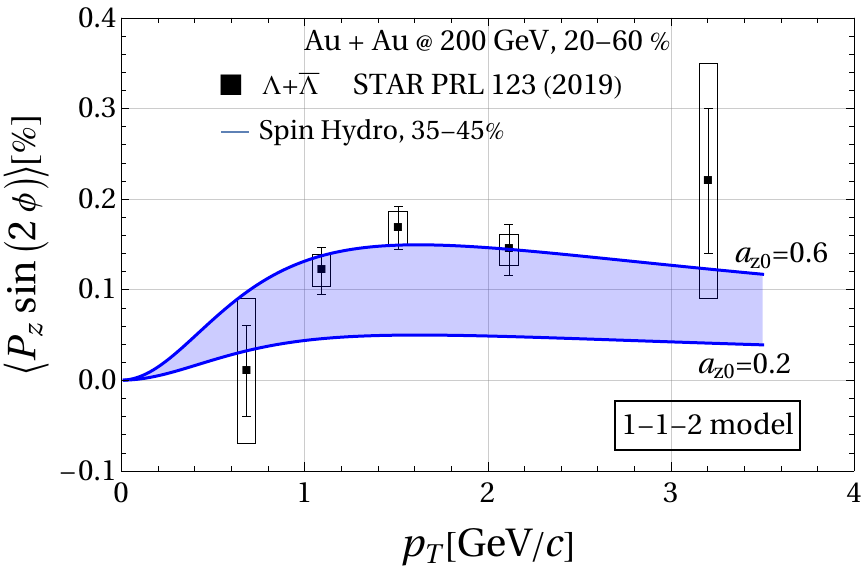}
    \caption{The local longitudinal spin polarization $P_z$ as a function of momentum azimuthal angle $\phi$ (left) and its second Fourier coefficient with the transverse momentum $p_T$ (right) for Au$+$Au collision at $\sqrt{s_{\rm NN}}=200$ GeV within centrality class $35-45$\% for a $1+1+2$D freeze-out, varying $a_{z0}$ from $0.2$ to $0.6$.}
\label{fig:Pz_4d_var_az0}
\end{figure*}

Having discussed the primary results, we conclude by evaluating the sensitivity of the model predictions to our main parameters and assumptions.
First, we examine the dependence of the longitudinal local spin polarization $P_z$ on the initialization parameter $a_{z0}$, denoting the initial longitudinal spin potential acceleration intensity, see Eq.~\eqref{eq:MinimalLongitudinalSpinIC1+3}. Figure~\ref{fig:Pz_4d_var_az0} shows $P_z$ as $a_{z0}$ is varied from $0.2$ to $0.6$. Since $P_z$ is directly proportional to $a_z$  (and thus to $a_{z0}$), altering $a_{z0}$ merely rescales the magnitude of the predicted polarization while preserving its qualitative trend. The overall profile of $P_z$ (similarly for $P_x$ and $P_J$) across transverse momentum, rapidity, and azimuthal angle is primarily governed by the underlying hydrodynamic flow and the freeze-out hypersurface, both of which are constrained by independent observables as described above. Variations in the functional form of the spin potential in the $\tau$--$\eta$ plane at the initial time $\tau_s$ (Eqs.~\eqref{eq:MinimalLongitudinalSpinIC1+3} and \eqref{eq:MinimalSpinIC1+3}) exert negligible influence, as the potential is strongly constrained by symmetry. However, the relaxation of our assumption of transverse homogeneity for the spin potential ($x$--$y$ plane) may introduce nontrivial effects, which we reserve for future investigation.

\begin{figure}[ht!]
\centering
    \includegraphics[width=1\linewidth]{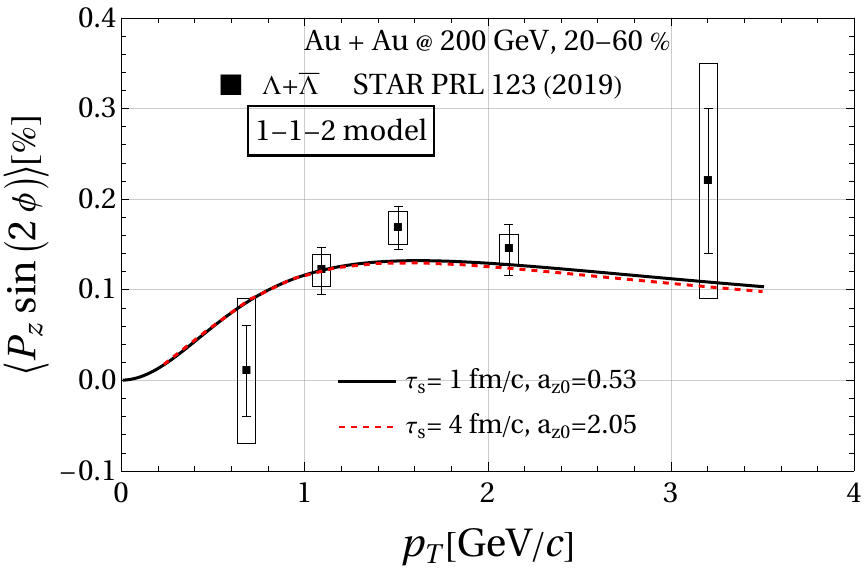}
\caption{The sensitivity of the second Fourier coefficient of the longitudinal local spin polarization to the spin evolution time $\tau_s$.
}
\label{fig:Pz_4d_var_taus}
\end{figure}

\begin{figure}[ht!]
\centering
    \includegraphics[width=1\linewidth]{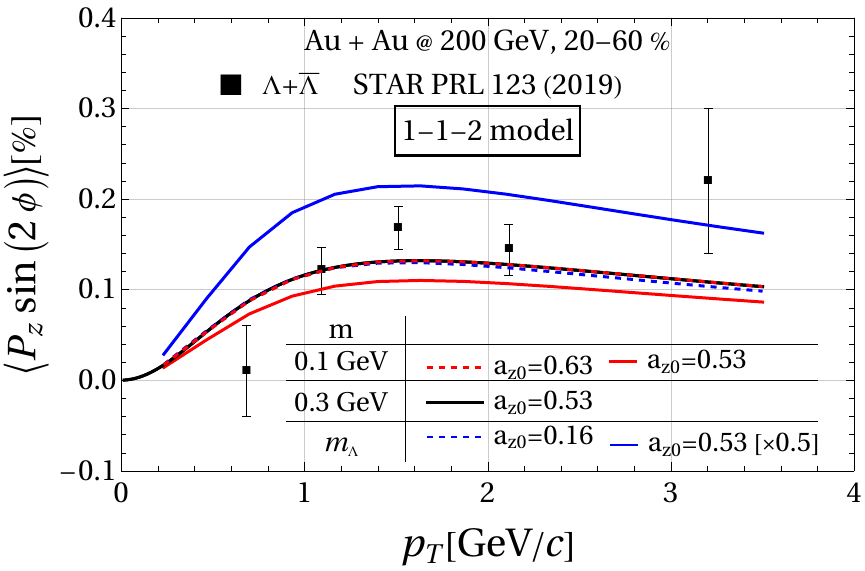}
\caption{The sensitivity of the second Fourier coefficient of the longitudinal local spin polarization to the effective mass $m$. Red, black and blue lines are respectively for $m=100$ MeV, $m=300$ MeV and $m=m_\Lambda$. Solid lines use $a_{z0}=0.53$ (the $m_\Lambda$ one has been divided by 2), and dashed lines are rescaled to fit the data: $a_{z0}=0.63$ for $m=100$ MeV, and $a_{z0}=0.16$ for $m=m_\Lambda$.}
\label{fig:Pz_4d_var_mass}
\end{figure}

We also studied the dependence of $P_z$ on the starting spin evolution time $\tau_s$ in Fig.~\ref{fig:Pz_4d_var_taus}. Interestingly, choosing $\tau_s$ higher than what we used in the main analysis $(\tau_s = 1\,{\rm fm}/c)$ barely changes the final results. However, the fitted parameter needed to reproduce the experimental data is now $a_{z0}>1$. For such values, our main physical assumption of small spin polarization may break down. That is the reason we used the early spin evolution time $\tau_s = 1\,{\rm fm}/c$. It must be noted, however, that changing the initialization of $a_{z}$ in Eq.~\eqref{eq:MinimalLongitudinalSpinIC1+3} with a function that has a large derivative at $\eta_s=0$ would result in a larger longitudinal spin polarization and the possibility of choosing a later spin evolution time.
In accordance with the spin potential evolution in Figs.~\ref{fig:SpinPot_var_mass_no_rescale} and \ref{fig:SpinPot_az_var_mass_no_rescale} and with ref.~\cite{Singh:2024cub}, Fig.~\ref{fig:Pz_4d_var_mass} shows that the spin polarization increase with the effective mass in the spin conservation equations, while the qualitative trend remains unaffected. As the spin potential initialization parameters $b_{y0}$ and $a_{z0}$ are fitted to the data, different effective masses results in the same spin polarization predictions but give different estimates for the initial spin potential intensity.

\begin{figure*}[ht!]
    \centering
    \includegraphics[width=8.9cm]{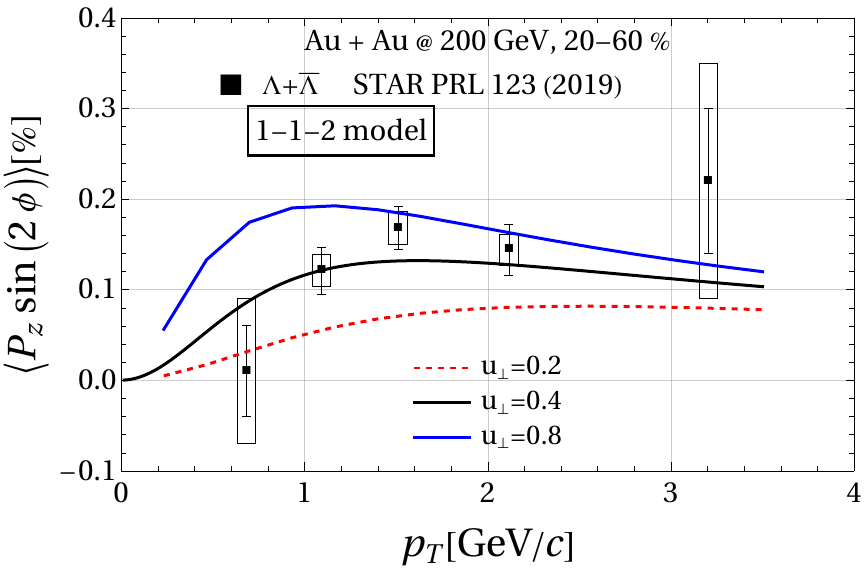}
    \includegraphics[width=8.9cm]{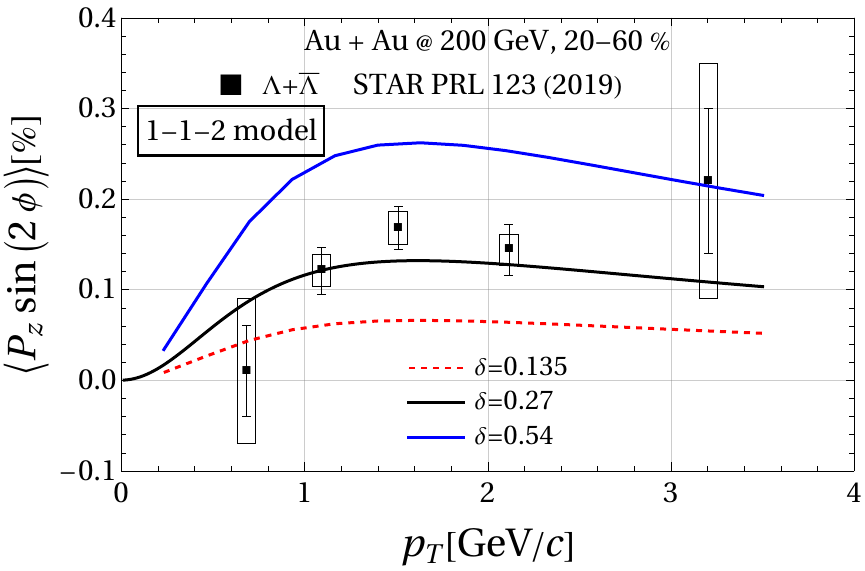}
    \caption{The sensitivity of the second Fourier coefficient of the longitudinal local spin polarization to the transverse flow strength $u_\perp$ (left) and the elliptic shape $\delta$ (right).}
\label{fig:Pz_4d_var_uperp_delta}
\end{figure*}

\begin{figure*}[ht!]
    \centering
    \includegraphics[width=8.9cm]{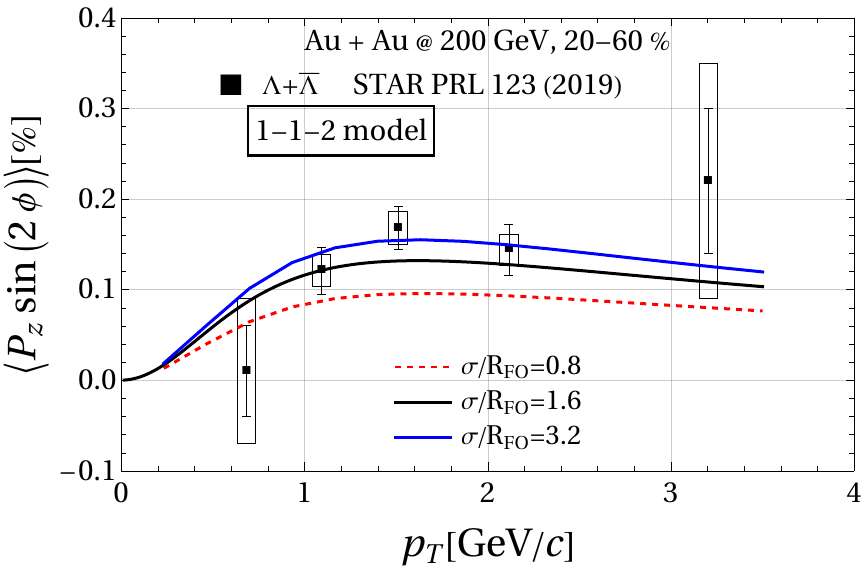}
    \includegraphics[width=8.9cm]{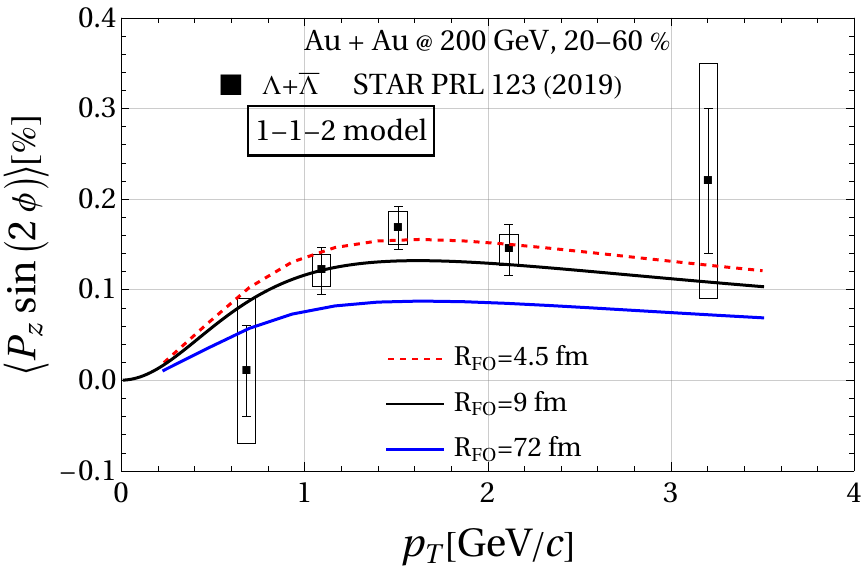}
    \caption{The sensitivity of the second Fourier coefficient of the longitudinal local spin polarization to the temperature transverse width $\sigma/R_{\rm FO}$ (left) and the freeze-out transverse radius $R_{\rm FO}$ (right).}
\label{fig:Pz_4d_var_sigmabar_Rfo}
\end{figure*}

Lastly, Figures~\ref{fig:Pz_4d_var_uperp_delta} and \ref{fig:Pz_4d_var_sigmabar_Rfo} show the sensitivity of local spin polarization to our $1-1-2$ model parameters $u_\perp$, $\delta$, $\sigma$, and $R_{\rm FO}$. In order to distinguish the differences, unlike the study of the variation of $\tau_s$, we didn't fit $a_{z0}$ for each new parameter; rather, we kept the value $a_{z0}=0.53$. 
We observe that $P_z$ is sensitive to the parameters $u_\perp$ (strength of transverse flow) and $\delta$ (elliptic deformation). As can be shown analytically from Eqs.~\eqref{eq:uperp} and \eqref{eq:Pz-even-1-1-2}, reducing $\delta$ or $u_\perp$ to zero makes the second Fourier coefficient of the longitudinal local spin polarization vanish.
We also find that $P_z$ does not vary strongly with the $\sigma$ and $R_{\rm FO}$, and the qualitative behavior remains intact.

\section{Summary and outlook}
\label{sec:summary}
In this work, we have investigated the dynamics of spin polarization in relativistic heavy-ion collisions within the framework of ideal relativistic spin hydrodynamics, focusing on longitudinally expanding systems beyond boost invariance. Employing the GLW formulation in the regime of small spin polarization, we have treated the spin potential perturbatively and studied its evolution on top of a hydrodynamic background described by non-boost-invariant analytic solutions~\cite{Shi:2022iyb}.

We first analyzed the evolution of spin degrees of freedom in $(1+1)$D setup, where the system is assumed to be homogeneous in the transverse plane but exhibits a non-trivial rapidity structure. In this framework, we derived and solved the coupled equations governing the spin potential components and identified their symmetry properties under rapidity reflections. We showed that a minimal set of spin initial conditions, constrained by symmetry considerations and the angular momentum structure of non-central collisions, is sufficient to generate non-trivial spin dynamics. The resulting spin polarization observables, including local and global polarization of $\Lambda$ hyperons, exhibit qualitative and, after fitting the spin potential initial intensity, quantitative agreement with experimental data. However, the $(1+1)$D description is intrinsically limited, particularly in its inability to reproduce azimuthal structures such as the quadrupole pattern of longitudinal polarization.

To overcome this limitation, we constructed a novel $(1+1+2)$D extension in which transverse flow and spatial anisotropy are incorporated phenomenologically at the level of the freeze-out hypersurface while retaining the longitudinal dynamics of the exact solutions. 
Although our present model is less realistic than a full $(3+1)\mathrm{D}$ spin hydrodynamic simulation (such as Ref.~\cite{Singh:2024cub}), where a proper transfer of spin and orbital angular momentum remains to be developed, this extended framework demonstrates that including a longitudinal spin acceleration component alongside transverse flow induces a quadrupole structure in the longitudinal spin polarization. This clearly indicates that spin acceleration plays an important role in shaping polarization observables.

Furthermore, our results show that the combined effect of longitudinal dynamics, transverse expansion, and appropriately chosen spin initial conditions can reproduce several key features observed experimentally, including the azimuthal dependence and transverse-momentum dependence of spin polarization. In particular, we identify the high $p_T$ behavior of the second sine Fourier coefficient of the longitudinal local spin polarization as a crucial observable to discriminate between different models.
This analysis validates ideal non-boost invariant spin hydrodynamics in the GLW pseudogauge as a viable path to understanding and describing the spin polarization in high energy Au$+$Au collisions with good qualitative agreement.
By fitting the spin potential intensity to the data, we achieve good quantitative description and a reasonable estimate of the spin potential: $\omega_y\sim 0.007$, $a_z\sim 0.5$.

Although the initialization of the spin potential components in this work was constrained by expected physical properties rather than external bias, the subsequent evolution of the spin potential remains highly sensitive to its initial shape. In the absence of a specific physical mechanism to uniquely determine the spin potential initialization, the model retains a significant degree of freedom. Nevertheless, this analysis clarifies the functional form and magnitude that the spin potential must adopt to remain consistent with experimental data.

The present study points to several natural directions for future work. For instance, as mentioned, a systematic study of the sensitivity of polarization observables to different initialization scenarios for the spin potential would help constrain the microscopic mechanisms responsible for spin generation and equilibration in the early stages of the collision.

Furthermore, it would be of interest to explore the applicability and extendibility of the present framework to smaller systems and lower collision energies, where gradients are larger and non-equilibrium effects are expected to be more pronounced~\cite{Noronha:2024dtq,Grosse-Oetringhaus:2024bwr,Buzzegoli:2025zud,Alice:2026ch}. In these regimes, spin polarization may provide a sensitive probe of the limits of hydrodynamic descriptions and the interplay between microscopic dynamics, macroscopic evolution, and stochastic fluctuations.

In this regard, analyzing spin dynamics on top of a dissipative hydrodynamic background could be important. The simplest setting for such an analysis is the recently developed first-order dissipative hydrodynamics in the density frame~\cite{Basar:2024qxd,Bhambure:2024axa,Bhambure:2024gnf}. However, a more realistic direction would be to incorporate spin degrees of freedom directly into the density-frame formulation, thereby developing a first-order dissipative spin hydrodynamic theory. Such a framework would allow one to describe relaxation effects associated with the spin degrees of freedom and to systematically modify the coupling between spin polarization and flow gradients. Including these effects may improve quantitative comparisons with experimental data and lead to a more realistic description of the late-time evolution of the system.

Overall, the present work represents a significant step toward a more comprehensive understanding of spin polarization in relativistic heavy-ion collisions by integrating analytically controlled hydrodynamic backgrounds with a consistent treatment of spin dynamics. Our findings underscore the necessity of moving beyond highly symmetric setups and highlight the role of both vorticity and acceleration in determining polarization patterns. Furthermore, this study points toward the requirement for more exhaustive frameworks that fully incorporate dissipative effects and non-equilibrium dynamics.
\begin{acknowledgments}
The work of M.B. and A.G. is supported by the European Union - NextGenerationEU through grant No.\ 760079/23.05.2023, funded by the Romanian Ministry of Research, Innovation and Digitization through Romania's National Recovery and Resilience Plan, call no.\ PNRR-III-C9-2022-I8. R.S. was supported by a postdoctoral fellowship from West University of Timișoara, Romania and acknowledges the support of the Humboldt research fellowship for postdocs from the Alexander von Humboldt Foundation.

R.S. thanks Charles Gale, Sangyong Jeon, and Shuzhe Shi for the initial fruitful discussions and collaboration. R.S. also acknowledges the CERN Theory Department, where part of this work has been carried out. We are grateful to Sourav Kundu, Andrea Palermo, and Himanshu Sharma for their clarifying comments. 
\end{acknowledgments}

\noindent \textbf{Data availability statement}: The data created in our analysis is publicly available in the GitHub repository; see~\cite{ThisWorkDataRepository}.
\appendix
\section{Analysis on spin initialization and general form of spin polarization components}
\label{sec:GeneralFormOmegaP}
In this appendix, we provide the explicit expressions used in our analysis for the spin polarization \eqref{meanspin} in the rest frame of the $\Lambda$ hyperon, obtained with the boost \eqref{meanspinBoost}, in terms of the spin potential components~\eqref{eq:spin_potential_a} and \eqref{eq:spin_potential_omega}.
We used these expressions to choose how to initialize the spin potential components in order to reproduce the known features of spin polarization.

The numerator of Eqs.~\eqref{meanspin} and \eqref{meanspinBoost} shows that the decompositions of local spin polarization in terms of the spin potential components are given by the dual spin potential contracted with the momentum in the rest frame of the particle, i.e., $(\tilde{\omega}_{\mu\nu} p^\nu)^*$. To write it in terms of the spin potential components, we first take the dual of the decomposition \eqref{eq:spin_potential} and contract it with the momentum $p$ expressed as in Eq.~\eqref{eq:Momentum_p_rapidity}. Denoting with $m_T=\sqrt{m_\Lambda^2+p_T^2}$, we obtain
\begin{align}
    \tilde{\omega}_{0\nu}p^\nu =&
    m_T \sinh y_p \left[-u^2 a_x+u^1 a_y+\omega_z \left((u^0)^2-(u^3)^2\right)\right]\\
    &+p_y a_x u^3-p_x a_y u^3+p_x u^2 a_z u^0-p_y u^1 a_z u^0\nn\\
    &+p_x \omega_x u^0+p_y \omega_y u^0-p_x u^1 \omega_z u^3-p_y u^2 \omega_z u^3,\nn
\end{align}
\begin{align}
    \tilde{\omega}_{1\nu}p^\nu =& 
    m_T \sinh y_p \left[-u^0 (a_y+u^1 \omega_z)-u^3 (u^2 a_z+\omega_x)\right]\\
    &+m_T \cosh y_p \left[u^0 (u^2 a_z+\omega_x)-u^3 (a_y+u^1 \omega_z)\right]\nn\\
    &+p_y \left[a_z \left((u^3)^2-(u^0)^2\right)-u^2 \omega_x+u^1 \omega_y\right],\nn
\end{align}
\begin{align}
    \tilde{\omega}_{2\nu}p^\nu =&
    -m_T\sinh y_p \left[u^0 (a_x-u^2 \omega_z)+u^3 (\omega_y-u^1 a_z)\right]\\
    &+ m_T \cosh y_p\left[u^3 (a_x-u^2 \omega_z)+u^0 (\omega_y-u^1 a_z)\right]\nn\\
    &+p_x \left[a_z \left((u^0)^2-(u^3)^2\right)+u^2 \omega_x+u^1 (-\omega_y)\right],\nn
\end{align}
\begin{align}
\tilde{\omega}_{3\nu}p^\nu =& a_x (p_y u^0-u^2 m_T\cosh y_p)\\
    &+a_y (u^1 m_T\cosh y_p-p_x u^0) \nn\\
    &+u^3 \left[a_z (p_x u^2-p_y u^1)+p_x \omega_x+p_y \omega_y\right]\nn\\
    &+\omega_z u^0 (\cosh y_p m_T u^0-p_x u^1-p_y u^2)\nn\\
    &-m_T \cosh y_p \omega_z (u^3)^2.\nn
\end{align}
Noting that $\tilde{\omega}_{\mu\nu}p^\nu$ is a Lorentz four-vector, the decomposition of the local spin polarization in the rest frame of the $\Lambda$ is obtained by boosting $\tilde{\omega}_{\mu\nu}p^\nu$ in the same fashion as Eq.~\eqref{meanspinBoost}. At mid-rapidity ($y_p=0$), we obtain:
\beq
&&\left(\tilde{\omega}_{0\nu} p^\nu\right)^* = 0 ,\\
&& \left(\tilde{\omega}_{1\nu} p^\nu\right)^* = -\f{1}{m_\Lambda}\Bigg[p_x p_y u^3 a_x \left(-1+ \ap\right)\nn\\
&& \quad  +\, u^3 a_y \left(p_x^2 (1 - \ap) -\mT \mL  \right)\nn\\
&& \quad  +\, a_z \Big[ \mL p_y((u^3)^2 - (u^0)^2) + \mT\mL u^0 u^2  \nn\\
&& \quad  +\, p_x u^0 (p_y u^1 - p_x u^2 ) (1-\ap) \Big]\\
&& \quad  +\, \omega_x \left(\mT \mL u^0  - \mL p_y u^2 + p_x^2 u^0  (\ap-1)\right)\nn\\
&& \quad  +\, \omega_y p_y ( \mL u^1 + p_x u^0 (\ap-1))\nn\\
&& \quad  +\, \omega_z u^3 \Big(p_x (p_x u^1 + p_y u^2) (1-\ap)-\mT \mL u^1 
\Big)
\Bigg],\nn
\eeq
\beq
&& \left(\tilde{\omega}_{2\nu} p^\nu\right)^* = -\f{1}{m_\Lambda}\Bigg[u^3 a_x \left(\mT \mL - p_y^2 (1 -  \ap) \right)\nn\\ 
&& \quad  +\, p_x p_y u^3 a_y \left(1- \ap\right)\nn\\
&& \quad  +\, a_z \Big[ \mL p_x ((u^0)^2 - (u^3)^2)-\mT\mL u^0 u^1  \nn\\
&& \quad  +\,   p_y u^0 (p_y u^1-p_x u^2) (1-\ap) \Big]\nn\\
&& \quad  +\, \omega_x p_x \left( \mL u^2  -  p_y u^0 (1-\ap)\right)\\
&& \quad  +\, \omega_y (\mT \mL u^0 -p_y^2 u^0 (1-\ap) - \mL p_x u^1 )\nn\\
&& \quad +\, \omega_z u^3 \Big(p_y (p_x u^1 + p_y u^2) (1+\ap) - \mT \mL u^2 
\Big)
\Bigg],\nn\\
&& \left(\tilde{\omega}_{3\nu} p^\nu\right)^* = \left(\mT u^2 -p_y u^0\right) a_x + \left(p_x u^0 - \mT u^1\right) a_y\nn\\
&& \quad +\,  a_z u^3 \left(p_y u^1 - p_x u^2\right) - u^3 p_x \omega_x - u^3 p_y \omega_y \\
&& \quad +\, \omega_z \left(p_x u^0 u^1 + p_y u^0 u^2 + \mT \left((u^3)^2 - (u^0)^2\right)\right)\nn.
\eeq
where $\ap = \f{\mT}{\mL+ E_p}$. These terms give, respectively, the momentum dependent spin polarization components $P_x$, $P_y$ and $P_z$ obtained in the experiments and shown in the figures of this analysis.
Reminding that i) $u^0$, $u^1$, $u^2$, $\omega_y$, and $a_y$ are $\eta_s$-even functions, ii) $u^3$, $\omega_x$, $\omega_z$, $a_x$, and $a_z$ are $\eta_s$-odd functions,
the terms that survive the integration over an $\eta_s$-even freeze-out hypersurface are the $\eta_s$-even terms, which are
\begin{align}
\left.(\tilde{\omega}_{1\nu}p^\nu)^*\right|_{\rm even} =& 
    \f{p_x p_y}{\mL} (1 - \ap) u^3 a_x \label{eq:Px_112_model}\\
    &-\, p_y [ u^1 +  \f{p_x}{\mL} u^0 (\ap-1)] \omega_y \nn\\
     -\, \Big[\f{p_x}{\mL}& (p_x u^1 + p_y u^2)(1 - \ap)-\mT u^1 \Big] u^3 \omega_z,\nn\\
\left.(\tilde{\omega}_{2\nu}p^\nu)^*\right|_{\rm even} =&
    \Big[ \f{p_y^2}{\mL} (1 -  \ap) -\mT  \Big] u^3 a_x\\
    - \Big[\mT u^0& -\f{p_y^2}{\mL} u^0 (1-\ap) - p_x u^1 \Big]\omega_y\nn\\
    - \Big[\f{p_y}{\mL}& (p_x u^1 + p_y u^2) (1+\ap) - \mT u^2\Big]u^3 \omega_z,\nn\\
\left.(\tilde{\omega}_{3\nu}p^\nu)^*\right|_{\rm even} =& 
    (p_x u^0 - \mT u^1) a_y \label{eq:Pz-even-1-1-2}\\
    & + (p_y u^1 - p_x u^2) u^3 a_z - p_x u^3 \omega_x.\nn
\end{align}
We note that, as discussed in the text, $a_x,\,\omega_y$ and $\omega_z$ contribute to $P_x$ and $P_y$, while $a_y,\,a_z,\,\omega_x$ contributes only to $P_z$. We also observe that $u^1$ and $u^2$ in Eq.~\eqref{eq:uperp} are odd functions of $x$ and $y$ respectively, and we see that in $(\tilde{\omega}_{3\nu}p^\nu)^*$, which gives $P_z$, $a_z$ is always multiplied by $u^1$ or $u^2$, while $a_y$ and $\omega_x$ are not. As a consequence, when coupled with the transverse volume element $p^x\d\Sigma_x + p^y\d\Sigma_y\propto p_x\, x/\sigma_x^2 + p_y\, y/\sigma_y^2$ (see Appendix~\ref{sec:FOIntegral}), only $a_z$ survives the integration, giving terms proportional to $p_x p_y$. Therefore, $a_z$ is the only spin potential component that can create the quadrupole structure. We also verified numerically in the $1-1-2$ model that $a_y$ and $\omega_x$ yield a vanishing $\langle P_z \sin(2\phi)\rangle$. On the contrary, $(\tilde{\omega}_{1\nu}p^\nu)^*$ directly contains terms proportional to $p_x p_y$ without the need for the transverse volume element. These terms generate the quadrupole structure of $P_x$ in both the $(1+1)$D and the $1-1-2$ models, see Figs.~\ref{fig:density_plot_2d} and~\ref{fig:density_plot_4d}, which means that transverse flow components are not necessary for the emergence of quadrupole pattern in $P_x$. This analysis motivated our choice for the initialization of the spin potential components.

In the $(1+1)$D model, the transverse velocity is neglected ($u^1=u^2=0$) and the expressions reduce to
\begin{align}
\left.(\tilde{\omega}_{1\nu}p^\nu)^*\right|^{u_\perp=0}_{\rm even} =& 
    \f{p_x p_y}{\mL} (1 - \ap) (u^3 a_x + u^0 \omega_y),\label{eq:PxExp11}\\
\left.(\tilde{\omega}_{2\nu}p^\nu)^*\right|^{u_\perp=0}_{\rm even} =&
    \Big[ \f{p_y^2}{\mL} (1 -  \ap) -\mT  \Big] (u^3 a_x+u^0\omega_y)\label{eq:PyExp11}\\
\label{eq:PzExp11}
\left.(\tilde{\omega}_{3\nu}p^\nu)^*\right|^{u_\perp=0}_{\rm even} =& 
    p_x (u^0 a_y - u^3 \omega_x).
\end{align}
In this case, there is no contribution from $a_z$ and $\omega_z$.
These expressions motivated the spin initialization discussed in Sec.~\ref{sec:Spin_1_plus_1}, where $a_z$ and $\omega_z$ are absent, and we kept $a_x,\,a_y,\,\omega_x$ and $\omega_y$.
It is clear from Eq.~\eqref{eq:PzExp11} that $P_z(y_p=0,\bm{p}_T)\propto p_x$. as shown in Fig.~\ref{fig:density_plot_2d}, and that $\langle P_z \sin(2\phi)\rangle=0$.
\section{The freeze-out integral in the \texorpdfstring{$1+1+2$}{1+1+2}D model}
\label{sec:FOIntegral}
In the $1-1-2$ model, we introduced the transverse direction through the transverse fluid velocity in Eqs.~\eqref{eq:u_1_plus_3} and~\eqref{eq:uperp}, and through the elliptic deformation of temperature given in Eq.~\eqref{eq:EllipticTemp}. In this appendix, we compute the explicit expression of the freeze-out element $\d^3\Sigma_\mu$ resulting from this model, and we report the expressions used to obtain the hadron spectra in Fig.~\ref{fig:hadronspectra_4d}.

In order to compute the observables through the Cooper-Frye procedure \eqref{eq:Cooper-Frye-formula}, we first need to obtain the coordinates and the volume element for the freeze-out hypersurface, which is done following Ref.~\cite{Shi:2022iyb} with the inclusion of transverse flow components and the elliptic shape.
The freeze-out hypersurface is defined as the 3D volume with constant temperature $T_{\rm FO}$. Setting $T(\tau,\, x,\, y,\, \eta_s)=T_{\rm FO}$ and taking the logarithm of Eq.~(\ref{eq:EllipticTemp}), we obtain
\beq
\frac{3}{2}\ln\left(\frac{T_0}{T_{\rm FO}}\right)^4 &=&  \left(4 - a^{-2} - a^2\right) \ln\left(\frac{t_0}{\tau_0}\right)\nn\\
    &+& \left(2-a^{-2}\right) \ln\left(1 + \frac{a\tau}{t_0}\E^{\eta_s}  \right)\\
    &+& \left(2-a^2\right) \ln\left(1 + \frac{\tau}{t_0\,a}\E^{-\eta_s}  \right)
    + \frac{3r^2}{\sigma^2},\nn
\eeq
where we used the elliptic coordinates
\begin{equation}
\label{eq:EllipticCoordApp}
x = \frac{r}{\sqrt{1+\delta}}\cos\varphi\,,\quad
y = \frac{r}{\sqrt{1-\delta}}\sin\varphi\,.
\end{equation}
Defining the constant
\begin{equation}
C_f = \frac{3}{2}\ln\left(\frac{T_0}{T_{\rm FO}}\right)^4
    + \left(4 - a^{-2} - a^2\right) \ln\left(\frac{\tau_0}{t_0}\right),
\end{equation}
and using the definitions~\eqref{eq:q1-q2},
the constant temperature equation is written as
\begin{equation}
\label{eq:q1q1ConstantT}
C_f = \left(2-a^{-2}\right) q_1 + \left(2-a^2\right) q_2 + \frac{3r^2}{\sigma^2}.
\end{equation}
The variables $\tau$ and $\eta_s$ are obtained from $q_1$ and $q_2$ as
\beq
\tau &=& t_0 \sqrt{(\E^{q_1}-1)(\E^{q_2}-1)}\,,\\
\eta_s &=& -\ln a + \frac{1}{2}\ln\left(\frac{\E^{q_1}-1}{\E^{q_2}-1} \right).
\eeq
Introducing a new variable $\zeta$,
\begin{equation}
\zeta = \frac{q_1-q_2}{2},
\end{equation}
provides the values of $q_1$ and $q_2$ on the constant temperature hypersurface (\ref{eq:q1q1ConstantT}) as
\begin{align}
q_1 = & \frac{C_f + 2 \left(2-a^2 \right)\zeta -\Delta_{\rm T}}{4-a^2-a^{-2}}\,,\\
q_2 = & \frac{C_f -2 \left(2-a^{-2}\right)\zeta -\Delta_{\rm T}}{4-a^2-a^{-2}}\,,
\end{align}
where
\begin{equation}
\Delta_{\rm T} =  \frac{3r^2}{\sigma^2}=\frac{3x^2}{\sigma_x^2}+\frac{3y^2}{\sigma_y^2}\,.
\end{equation}
Now that we have the freeze-out hypersurface coordinates, we can obtain the surface volume element using Eq.~\eqref{eq:VolumeElement}.
Straightforward calculations give:
\beq\d\Sigma_\tau &=& \tau\D\zeta \D x \D y
    \left(\frac{\pd\eta}{\pd q_1}\frac{\pd q_1}{\pd\zeta} + \frac{\pd\eta}{\pd q_2}\frac{\pd q_2}{\pd\zeta}\right)\\
&=& \frac{\tau\D\zeta \D x \D y}{4-a^2-a^{-2}}
    \left(\frac{2-a^2}{1-\E^{-q_1}} + \frac{2-a^{-2}}{1-\E^{-q_2}}\right),\nn
\eeq
\beq
\d\Sigma_\eta &=& -\tau\D\zeta \D x \D y
    \left(\frac{\pd\tau}{\pd q_1}\frac{\pd q_1}{\pd\zeta} + \frac{\pd\tau}{\pd q_2}\frac{\pd q_2}{\pd\zeta}\right)\\
&=& \frac{\tau^2\D\zeta \D x \D y}{4-a^2-a^{-2}}
    \left(\frac{2-a^{-2}}{1-\E^{-q_2}}-\frac{2-a^2}{1-\E^{-q_1}}\right),\nn
\eeq
\beq
&&\d\Sigma_x =  -\tau\D\zeta \D x \D y
        \left(\frac{\pd\tau}{\pd x}\frac{\pd\eta}{\pd\zeta} - \frac{\pd\tau}{\pd\zeta}\frac{\pd\eta}{\pd x}\right)= -\tau\D\zeta \D x \D y \nn\\
 && \quad \times\left[
\left(\frac{\pd\tau}{\pd q_1}\frac{\pd q_1}{\pd x} + \frac{\pd\tau}{\pd q_2}\frac{\pd q_2}{\pd x}\right)\left(\frac{\pd\eta}{\pd q_1}\frac{\pd q_1}{\pd\zeta} + \frac{\pd\eta}{\pd q_2}\frac{\pd q_2}{\pd\zeta}\right)\right.\nn\\
&& \quad -\left.\left(\frac{\pd\tau}{\pd q_1}\frac{\pd q_1}{\pd\zeta} + \frac{\pd\tau}{\pd q_2}\frac{\pd q_2}{\pd\zeta}\right)\left(\frac{\pd\eta}{\pd q_1}\frac{\pd q_1}{\pd x} + \frac{\pd\eta}{\pd q_2}\frac{\pd q_2}{\pd x}\right)\right],\nn\\
&& \quad = \frac{\tau^2\D\zeta \D x \D y}{4-a^2-a^{-2}}\\
&& \times\left[ 
\left(\frac{3 x/\sigma_x^2}{1-\E^{-q_1}} + \frac{3 x/\sigma_x^2}{1-\E^{-q_2}}\right)\left(\frac{2-a^2}{1-\E^{-q_1}} + \frac{2-a^{-2}}{1-\E^{-q_2}}\right)\right.\nn\\
 &&-\left.\left( \frac{2-a^{-2}}{1-\E^{-q_2}}-\frac{2-a^2}{1-\E^{-q_1}}\right)\left(\frac{3 x/\sigma_x^2}{1-\E^{-q_2}}-\frac{3 x/\sigma_x^2}{1-\E^{-q_1}} \right)\right],\nn\\
&&  \quad = \frac{\tau^2\D\zeta \D x \D y}{4-a^2-a^{-2}} \frac{6 x}{\sigma_x^2}\frac{1}{1-\E^{-q_1}}\frac{1}{1-\E^{-q_2}}\nn\,,
\eeq
and similarly
\beq
\d\Sigma_y =  \frac{\tau^2\D\zeta \D x \D y}{4-a^2-a^{-2}} \frac{6 y}{\sigma_y^2}\frac{1}{1-\E^{-q_1}}\frac{1}{1-\E^{-q_2}}\,.
\eeq
The Cooper-Fry formula~\eqref{eq:Cooper-Frye-formula} and the spin polarization averages~\eqref{meanspin}-\eqref{eq:momentum-integrated-polarization} depend on the scalar product of this generalized volume element with the momentum: $\d^3\Sigma\cdot p$.
Using the results above and the momentum Eq.~\eqref{eq:Momentum_p_rapidity} for a hadron $h$ with mass $m_h$, we obtain
\begin{equation}\label{eq:dSigmaP112}
\d^3\Sigma\cdot p = p^\tau\d\Sigma_\tau + p^\eta\d\Sigma_\eta
    + p^x\d\Sigma_x + p^y\d\Sigma_y,
\end{equation}
where
\begin{multline}\label{eq:dSigmaPtauEta}
p^\tau\d\Sigma_\tau + p^\eta\d\Sigma_\eta = 
    \frac{2 m_T \tau \D\zeta \D x \D y}{4-a^2-a^{-2}}\times\\
    \sqrt{\frac{(2-a^2)(2-a^{-2})}{(1-\E^{-q_1})(1-\E^{-q_2})}}\cosh\left[y_p - \zeta + \frac{1}{2}\ln\frac{2a^2 - 1}{2-a^2} \right]\\
=  2 t_0 m_T \D\zeta \D x \D y \frac{\sqrt{(2-a^2)(2-a^{-2})}}{4-a^2-a^{-2}}\E^{\frac{q_1 + q_2}{2}}\\
\times \cosh\left[y_p - \zeta + \frac{1}{2}\ln\frac{2a^2 - 1}{2-a^2} \right],    
\end{multline}
and
\begin{equation}\label{eq:dSigmaPxY}
\begin{split}
p^x\d\Sigma_x + p^y\d\Sigma_y &= 
    \frac{\tau^2 \D\zeta \D x \D y}{4-a^2-a^{-2}}\frac{6\left(\frac{p_x x}{\sigma_x^2} + \frac{p_y y}{\sigma_y^2} \right)}{(1-\E^{-q_1})(1-\E^{-q_2})}\\
&=  \frac{6 t_0^2 \D\zeta \D x \D y}{4-a^2-a^{-2}}\E^{q_1 + q_2}\left(\frac{p_x x}{\sigma_x^2} + \frac{p_y y}{\sigma_y^2} \right),
\end{split}
\end{equation}
with $m_T=\sqrt{m_h^2 + p_T^2}$ being the transverse mass of the particle, and
\begin{equation}
\E^{\frac{q_1 + q_2}{2}} = \exp\left[
    \frac{C_f + \zeta \left(a^{-2}-a^2\right)\zeta - \Delta_{\rm T}}{4 - a^2 - a^{-2}}
\right].
\end{equation}
Furthermore, in the integration over the freeze-out hypersurface, it is useful to use the elliptic coordinates (\ref{eq:EllipticCoordApp}) from which follows that
\begin{equation}
\D x \, \D y = \frac{r\,\D r \D\varphi}{\sqrt{1-\delta^2}}.
\end{equation}
Using this volume element, the hadron spectra in Fig.~\ref{fig:hadronspectra_4d} are obtained as follow.
Summing the contributions of pions $(\pi^\pm)$, kaons $(K^\pm)$, and protons $p(\bar{p})$, the pseudo-rapidity distribution of charged particle multiplicity is obtained with
\begin{equation}
\f{\d N_h}{\d\eta} = \!\int_{\Sigma_{\rm FO}}\!\!\!\f{\d^3\Sigma\cdot p}{(2\pi)^3}
   \int\f{p_T^2\cosh\eta \,\d p_T}{\sqrt{m_h^2+p_T^2\cosh^2\eta}}
    \f{\Theta(\tau-\tau_i)}{\E^{p\cdot u/T_{\rm FO}} \pm 1},
    \label{eq:charged particle multiplicity-1-1-2}
\end{equation}
where the subscript $h$ denotes the contribution of a single hadron, $\pm$ is $+$ for fermions and $-$ for bosons, the momentum is on-shell $p^2=m_h^2$, and $p_x=p_T\cos\phi$, $p_y=p_T\sin\phi$.

Similarly, the $p_T$ dependence of elliptic flow $v_2$ of the charged particles is obtained with
\begin{equation}
v_2^h =
 \int_0^{2\pi}\!\!\!\d \phi\int_{\Sigma_{\rm FO}}\!\!\!\f{\d^3\Sigma\cdot p}{(2\pi)^4}
   \f{\cos(2\phi)}{\langle N_{\rm ch}\rangle}
    \f{\Theta(\tau-\tau_i)}{\E^{p\cdot u/T_{\rm FO}} \pm 1},
    \label{eq:v2-1-1-2}
\end{equation}
with
\begin{equation}
\langle N_{\rm ch}\rangle  = \sum_h
 \int_0^{2\pi}\!\!\!\d \phi\int_{\Sigma_{\rm FO}}\!\!\!\f{\d^3\Sigma\cdot p}{(2\pi)^4}
    \f{\Theta(\tau-\tau_i)}{\E^{p\cdot u/T_{\rm FO}} \pm 1},
\end{equation}
where both expressions are evaluated at mid-rapidity $y_p=0$.
Averaging the charged particle contributions as \mbox{$(h^+ + h^-)/2$}, and including pions, kaons, and protons, the transverse momentum spectrum of charged particles is obtained at mid-rapidity $y_p=0$ with
\begin{align}
\f{1}{2\pi p_T} \left. \f{\d^2 N_{\rm ch}}{\d p_T\d\eta}\right|_{\eta=0} =&
 \int_0^{2\pi}\!\!\!\d \phi\int_{\Sigma_{\rm FO}}\!\!\!\f{\d^3\Sigma\cdot p}{(2\pi)^4}
   \f{p_T}{\sqrt{m_h^2+p_T^2}}\nn\\
    &\times\f{\Theta(\tau-\tau_i)}{\E^{p\cdot u/T_{\rm FO}} \pm 1}.
    \label{eq:transverse momentum spectrum-1-1-2}
\end{align}
\bibliographystyle{utphys}
\bibliography{pv_ref}

\providecommand{\href}[2]{#2}\begingroup\raggedright\begin{thebibliography}{10}

\bibitem{Heinz:2013th}
U.~Heinz and R.~Snellings, ``{Collective flow and viscosity in relativistic heavy-ion collisions},'' \href{http://dx.doi.org/10.1146/annurev-nucl-102212-170540}{{\em Ann. Rev. Nucl. Part. Sci.} {\bfseries 63} (2013) 123--151}, \href{http://arxiv.org/abs/1301.2826}{{\ttfamily arXiv:1301.2826 [nucl-th]}}.

\bibitem{Gale:2013da}
C.~Gale, S.~Jeon, and B.~Schenke, ``{Hydrodynamic Modeling of Heavy-Ion Collisions},'' \href{http://dx.doi.org/10.1142/S0217751X13400113}{{\em Int. J. Mod. Phys. A} {\bfseries 28} (2013) 1340011}, \href{http://arxiv.org/abs/1301.5893}{{\ttfamily arXiv:1301.5893 [nucl-th]}}.

\bibitem{Braun-Munzinger:2015hba}
P.~Braun-Munzinger, V.~Koch, T.~Sch{\"a}fer, and J.~Stachel, ``{Properties of hot and dense matter from relativistic heavy ion collisions},'' \href{http://dx.doi.org/10.1016/j.physrep.2015.12.003}{{\em Phys. Rept.} {\bfseries 621} (2016) 76--126}, \href{http://arxiv.org/abs/1510.00442}{{\ttfamily arXiv:1510.00442 [nucl-th]}}.

\bibitem{Arslandok:2023utm}
M.~Arslandok {\em et~al.}, ``{Hot QCD White Paper},'' \href{http://dx.doi.org/0.48550/arXiv.2303.17254}{{\em pre-print} (3, 2023) 1--190}, \href{http://arxiv.org/abs/2303.17254}{{\ttfamily arXiv:2303.17254 [nucl-ex]}}.

\bibitem{Teaney:2009qa}
D.~A. Teaney, \href{http://dx.doi.org/10.1142/9789814293297_0004}{``{Viscous Hydrodynamics and the Quark Gluon Plasma},''} in {\em {Quark-gluon plasma 4}}, R.~C. Hwa and X.-N. Wang, eds., pp.~207--266.
\newblock World Scientific, 2010.
\newblock \href{http://arxiv.org/abs/0905.2433}{{\ttfamily arXiv:0905.2433 [nucl-th]}}.

\bibitem{Schafer:2009dj}
T.~Sch{\"a}fer and D.~Teaney, ``{Nearly Perfect Fluidity: From Cold Atomic Gases to Hot Quark Gluon Plasmas},'' \href{http://dx.doi.org/10.1088/0034-4885/72/12/126001}{{\em Rept. Prog. Phys.} {\bfseries 72} (2009) 126001}, \href{http://arxiv.org/abs/0904.3107}{{\ttfamily arXiv:0904.3107 [hep-ph]}}.

\bibitem{Martinoia:2024hip}
L.~Martinoia and R.~Singh, ``{A Note on the Canonical Approach to Hydrodynamics and Linear Response Theory},'' \href{http://dx.doi.org/10.5506/APhysPolB.56.1-A4}{{\em Acta Phys. Polon. B} {\bfseries 56} no.~1, (2025) 1--A4}, \href{http://arxiv.org/abs/2408.10698}{{\ttfamily arXiv:2408.10698 [hep-th]}}.

\bibitem{Becattini:2007sr}
F.~Becattini, F.~Piccinini, and J.~Rizzo, ``{Angular momentum conservation in heavy ion collisions at very high energy},'' \href{http://dx.doi.org/10.1103/PhysRevC.77.024906}{{\em Phys. Rev. C} {\bfseries 77} (2008) 024906}, \href{http://arxiv.org/abs/0711.1253}{{\ttfamily arXiv:0711.1253 [nucl-th]}}.

\bibitem{Becattini:2024uha}
F.~Becattini, M.~Buzzegoli, T.~Niida, S.~Pu, A.-H. Tang, and Q.~Wang, ``{Spin polarization in relativistic heavy-ion collisions},'' \href{http://dx.doi.org/10.1142/9789811294679_0005}{{\em Int. J. Mod. Phys. E} {\bfseries 33} no.~06, (2024) 2430006}, \href{http://arxiv.org/abs/2402.04540}{{\ttfamily arXiv:2402.04540 [nucl-th]}}.

\bibitem{Niida:2024ntm}
T.~Niida and S.~A. Voloshin, ``{Polarization phenomenon in heavy-ion collisions},'' \href{http://dx.doi.org/10.1142/S0218301324300108}{{\em Int. J. Mod. Phys. E} {\bfseries 33} no.~09, (2024) 2430010}, \href{http://arxiv.org/abs/2404.11042}{{\ttfamily arXiv:2404.11042 [nucl-ex]}}.

\bibitem{STAR:2017ckg}
{\bfseries STAR} Collaboration, L.~Adamczyk {\em et~al.}, ``{Global $\Lambda$ hyperon polarization in nuclear collisions: evidence for the most vortical fluid},'' \href{http://dx.doi.org/10.1038/nature23004}{{\em Nature} {\bfseries 548} (2017) 62--65}, \href{http://arxiv.org/abs/1701.06657}{{\ttfamily arXiv:1701.06657 [nucl-ex]}}.

\bibitem{Liang:2004ph}
Z.-T. Liang and X.-N. Wang, ``{Globally polarized quark-gluon plasma in non-central A+A collisions},'' \href{http://dx.doi.org/10.1103/PhysRevLett.94.102301}{{\em Phys. Rev. Lett.} {\bfseries 94} (2005) 102301}, \href{http://arxiv.org/abs/nucl-th/0410079}{{\ttfamily arXiv:nucl-th/0410079}}. [Erratum: Phys.Rev.Lett. 96, 039901 (2006)].

\bibitem{STAR:2019erd}
{\bfseries STAR} Collaboration, J.~Adam {\em et~al.}, ``{Polarization of $\Lambda$ ($\bar{\Lambda}$) hyperons along the beam direction in Au+Au collisions at $\sqrt{s_{_{NN}}}$ = 200 GeV},'' \href{http://dx.doi.org/10.1103/PhysRevLett.123.132301}{{\em Phys. Rev. Lett.} {\bfseries 123} no.~13, (2019) 132301}, \href{http://arxiv.org/abs/1905.11917}{{\ttfamily arXiv:1905.11917 [nucl-ex]}}.

\bibitem{ALICE:2019onw}
{\bfseries ALICE} Collaboration, S.~Acharya {\em et~al.}, ``{Global polarization of $\Lambda \bar \Lambda$ hyperons in Pb-Pb collisions at $\sqrt {s_{NN}}$ = 2.76 and 5.02 TeV},'' \href{http://dx.doi.org/10.1103/PhysRevC.101.044611}{{\em Phys. Rev. C} {\bfseries 101} no.~4, (2020) 044611}, \href{http://arxiv.org/abs/1909.01281}{{\ttfamily arXiv:1909.01281 [nucl-ex]}}. [Erratum: Phys.Rev.C 105, 029902 (2022)].

\bibitem{ALICE:2021pzu}
{\bfseries ALICE} Collaboration, S.~Acharya {\em et~al.}, ``{Polarization of $\Lambda$ and $\bar \Lambda$ Hyperons along the Beam Direction in Pb-Pb Collisions at $\sqrt {s_{NN}}$=5.02{\,}{\,}TeV},'' \href{http://dx.doi.org/10.1103/PhysRevLett.128.172005}{{\em Phys. Rev. Lett.} {\bfseries 128} no.~17, (2022) 172005}, \href{http://arxiv.org/abs/2107.11183}{{\ttfamily arXiv:2107.11183 [nucl-ex]}}.

\bibitem{CMS:2025nqr}
{\bfseries CMS} Collaboration, A.~Hayrapetyan {\em et~al.}, ``{Observation of {\ensuremath{\Lambda}} Hyperon Local Polarization in p-Pb Collisions at sNN=8.16{\,}{\,}TeV},'' \href{http://dx.doi.org/10.1103/6ywq-gm61}{{\em Phys. Rev. Lett.} {\bfseries 135} no.~13, (2025) 132301}, \href{http://arxiv.org/abs/2502.07898}{{\ttfamily arXiv:2502.07898 [nucl-ex]}}.

\bibitem{Becattini:2011zz}
F.~Becattini, ``{Hydrodynamics of fluids with spin},'' \href{http://dx.doi.org/10.1134/S1547477111080036}{{\em Phys. Part. Nucl. Lett.} {\bfseries 8} (2011) 801--804}.

\bibitem{Florkowski:2017ruc}
W.~Florkowski, B.~Friman, A.~Jaiswal, and E.~Speranza, ``{Relativistic fluid dynamics with spin},'' \href{http://dx.doi.org/10.1103/PhysRevC.97.041901}{{\em Phys. Rev.} {\bfseries C97} no.~4, (2018) 041901},
\href{http://arxiv.org/abs/1705.00587}{{\ttfamily arXiv:1705.00587 [nucl-th]}}.

\bibitem{Florkowski:2018fap}
W.~Florkowski, R.~Ryblewski, and A.~Kumar, ``{Relativistic hydrodynamics for spin-polarized fluids},'' \href{http://dx.doi.org/10.1016/j.ppnp.2019.07.001}{{\em Prog. Part. Nucl. Phys.} {\bfseries 108} (2019) 103709}, \href{http://arxiv.org/abs/1811.04409}{{\ttfamily arXiv:1811.04409 [nucl-th]}}.

\bibitem{Montenegro:2017rbu}
D.~Montenegro, L.~Tinti, and G.~Torrieri, ``{The ideal relativistic fluid limit for a medium with polarization},'' \href{http://dx.doi.org/10.1103/PhysRevD.96.056012}{{\em Phys. Rev.} {\bfseries D96} no.~5, (2017) 056012},
\href{http://arxiv.org/abs/1701.08263}{{\ttfamily arXiv:1701.08263 [hep-th]}}.

\bibitem{Hattori:2019lfp}
K.~Hattori, M.~Hongo, X.-G. Huang, M.~Matsuo, and H.~Taya, ``{Fate of spin polarization in a relativistic fluid: An entropy-current analysis},'' \href{http://dx.doi.org/10.1016/j.physletb.2019.05.040}{{\em Phys. Lett. B} {\bfseries 795} (2019) 100--106}, \href{http://arxiv.org/abs/1901.06615}{{\ttfamily arXiv:1901.06615 [hep-th]}}.

\bibitem{Hongo:2021ona}
M.~Hongo, X.-G. Huang, M.~Kaminski, M.~Stephanov, and H.-U. Yee, ``{Relativistic spin hydrodynamics with torsion and linear response theory for spin relaxation},'' \href{http://dx.doi.org/10.1007/JHEP11(2021)150}{{\em JHEP} {\bfseries 11} (2021) 150}, \href{http://arxiv.org/abs/2107.14231}{{\ttfamily arXiv:2107.14231 [hep-th]}}.

\bibitem{Gallegos:2022jow}
A.~D. Gallegos, U.~Gursoy, and A.~Yarom, ``{Hydrodynamics, spin currents and torsion},'' \href{http://dx.doi.org/10.1007/JHEP05(2023)139}{{\em JHEP} {\bfseries 05} (2023) 139}, \href{http://arxiv.org/abs/2203.05044}{{\ttfamily arXiv:2203.05044 [hep-th]}}.

\bibitem{Das:2022azr}
A.~Das, W.~Florkowski, A.~Kumar, R.~Ryblewski, and R.~Singh, ``{Semi-classical Kinetic Theory for Massive Spin-half Fermions with Leading-order Spin Effects},'' \href{http://dx.doi.org/10.5506/APhysPolB.54.8-A4}{{\em Acta Phys. Polon. B} {\bfseries 54} no.~8, (2023) 8--A4}, \href{http://arxiv.org/abs/2203.15562}{{\ttfamily arXiv:2203.15562 [hep-th]}}.

\bibitem{Singh:2025hnb}
R.~Singh, ``{Guiding center hydrodynamics with spin},'' \href{http://dx.doi.org/10.1016/j.jspc.2025.100255}{{\em J. Subatomic Part. Cosmol.} {\bfseries 4} (2025) 100255}, \href{http://arxiv.org/abs/2503.23061}{{\ttfamily arXiv:2503.23061 [nucl-th]}}.

\bibitem{Cao:2022aku}
Z.~Cao, K.~Hattori, M.~Hongo, X.-G. Huang, and H.~Taya, ``{Gyrohydrodynamics: Relativistic spinful fluid with strong vorticity},'' \href{http://dx.doi.org/10.1093/ptep/ptac091}{{\em PTEP} {\bfseries 2022} no.~7, (2022) 071D01}, \href{http://arxiv.org/abs/2205.08051}{{\ttfamily arXiv:2205.08051 [hep-th]}}.

\bibitem{Weickgenannt:2022zxs}
N.~Weickgenannt, D.~Wagner, E.~Speranza, and D.~H. Rischke, ``{Relativistic second-order dissipative spin hydrodynamics from the method of moments},'' \href{http://dx.doi.org/10.1103/PhysRevD.106.096014}{{\em Phys. Rev. D} {\bfseries 106} no.~9, (2022) 096014}, \href{http://arxiv.org/abs/2203.04766}{{\ttfamily arXiv:2203.04766 [nucl-th]}}.

\bibitem{Weickgenannt:2022qvh}
N.~Weickgenannt, D.~Wagner, E.~Speranza, and D.~H. Rischke, ``{Relativistic dissipative spin hydrodynamics from kinetic theory with a nonlocal collision term},'' \href{http://dx.doi.org/10.1103/PhysRevD.106.L091901}{{\em Phys. Rev. D} {\bfseries 106} no.~9, (2022) L091901}, \href{http://arxiv.org/abs/2208.01955}{{\ttfamily arXiv:2208.01955 [nucl-th]}}.

\bibitem{Becattini:2023ouz}
F.~Becattini, A.~Daher, and X.-L. Sheng, ``{Entropy current and entropy production in relativistic spin hydrodynamics},'' \href{http://dx.doi.org/10.1016/j.physletb.2024.138533}{{\em Phys. Lett. B} {\bfseries 850} (2024) 138533}, \href{http://arxiv.org/abs/2309.05789}{{\ttfamily arXiv:2309.05789 [nucl-th]}}.

\bibitem{Drogosz:2024gzv}
Z.~Drogosz, W.~Florkowski, and M.~Hontarenko, ``{Hybrid approach to perfect and dissipative spin hydrodynamics},'' \href{http://dx.doi.org/10.1103/PhysRevD.110.096018}{{\em Phys. Rev. D} {\bfseries 110} no.~9, (2024) 096018}, \href{http://arxiv.org/abs/2408.03106}{{\ttfamily arXiv:2408.03106 [hep-ph]}}.

\bibitem{Daher:2025pfq}
A.~Daher, X.-L. Sheng, D.~Wagner, and F.~Becattini, ``{Dissipative currents and transport coefficients in relativistic spin hydrodynamics},'' \href{http://dx.doi.org/10.1103/zttq-cs4l}{{\em Phys. Rev. D} {\bfseries 112} no.~9, (2025) 094020}, \href{http://arxiv.org/abs/2503.03713}{{\ttfamily arXiv:2503.03713 [nucl-th]}}.

\bibitem{Abboud:2025shb}
N.~Abboud, L.~Gavassino, R.~Singh, and E.~Speranza, ``{Perfect spinfluid: A divergence-type approach},'' \href{http://dx.doi.org/10.1103/bngt-lbdv}{{\em Phys. Rev. D} {\bfseries 112} no.~9, (2025) 094043}, \href{http://arxiv.org/abs/2506.19786}{{\ttfamily arXiv:2506.19786 [nucl-th]}}.

\bibitem{Bhadury:2025wuh}
S.~Bhadury, Z.~Drogosz, W.~Florkowski, S.~K. Kar, and V.~Mykhaylova, ``{Nonlinear causality and stability of perfect spin hydrodynamics and its nonperturbative character},'' \href{http://dx.doi.org/10.1103/ly9v-nw35}{{\em Phys. Rev. D} {\bfseries 113} no.~3, (2026) 036017}, \href{http://arxiv.org/abs/2511.19295}{{\ttfamily arXiv:2511.19295 [hep-ph]}}.

\bibitem{Sapna:2025yss}
Sapna, S.~K. Singh, and D.~Wagner, ``{Spin polarization of {\ensuremath{\Lambda}} hyperons from dissipative spin hydrodynamics },'' \href{http://dx.doi.org/10.1103/1s6g-fs8w}{{\em Phys. Rev. C} {\bfseries 112} no.~5, (2025) 054902}, \href{http://arxiv.org/abs/2503.22552}{{\ttfamily arXiv:2503.22552 [hep-ph]}}.

\bibitem{Matsuda:2025whj}
H.~Matsuda, K.~Hattori, and K.~Murase, ``{Achieving angular-momentum conservation with physics-informed neural networks in computational relativistic spin hydrodynamics},'' \href{http://arxiv.org/abs/2512.17971}{{\ttfamily arXiv:2512.17971 [physics.flu-dyn]}}.

\bibitem{Hehl:1976vr}
F.~W. Hehl, ``{On the Energy Tensor of Spinning Massive Matter in Classical Field Theory and General Relativity},'' \href{http://dx.doi.org/10.1016/0034-4877(76)90016-1}{{\em Rept. Math. Phys.} {\bfseries 9} (1976) 55--82}.

\bibitem{Buzzegoli:2024mra}
M.~Buzzegoli and A.~Palermo, ``{Emergent Canonical Spin Tensor in the Chiral-Symmetric Hot QCD},'' \href{http://dx.doi.org/10.1103/PhysRevLett.133.262301}{{\em Phys. Rev. Lett.} {\bfseries 133} no.~26, (2024) 262301}, \href{http://arxiv.org/abs/2407.14345}{{\ttfamily arXiv:2407.14345 [hep-ph]}}.

\bibitem{Becattini:2025oyi}
F.~Becattini and R.~Singh, ``{On the local thermodynamic relations in relativistic spin hydrodynamics},'' \href{http://dx.doi.org/10.1140/epjc/s10052-025-15071-3}{{\em Eur. Phys. J. C} {\bfseries 85} no.~11, (2025) 1338}, \href{http://arxiv.org/abs/2506.20681}{{\ttfamily arXiv:2506.20681 [nucl-th]}}.

\bibitem{DeGroot:1980dk}
S.~De~Groot, W.~Van~Leeuwen, and C.~Van~Weert, {\em {Relativistic Kinetic Theory. Principles and Applications}}.
\newblock North Holland, Amsterdam, 1, 1980.

\bibitem{Florkowski:2018ahw}
W.~Florkowski, A.~Kumar, and R.~Ryblewski, ``{Thermodynamic versus kinetic approach to polarization-vorticity coupling},'' \href{http://dx.doi.org/10.1103/PhysRevC.98.044906}{{\em Phys. Rev.} {\bfseries C98} no.~4, (2018) 044906},
\href{http://arxiv.org/abs/1806.02616}{{\ttfamily arXiv:1806.02616 [hep-ph]}}.

\bibitem{Buzzegoli:2021wlg}
M.~Buzzegoli, ``{Pseudogauge dependence of the spin polarization and of the axial vortical effect},'' \href{http://dx.doi.org/10.1103/PhysRevC.105.044907}{{\em Phys. Rev. C} {\bfseries 105} no.~4, (2022) 044907}, \href{http://arxiv.org/abs/2109.12084}{{\ttfamily arXiv:2109.12084 [nucl-th]}}.

\bibitem{Florkowski:2019qdp}
W.~Florkowski, A.~Kumar, R.~Ryblewski, and R.~Singh, ``{Spin polarization evolution in a boost invariant hydrodynamical background},'' \href{http://dx.doi.org/10.1103/PhysRevC.99.044910}{{\em Phys. Rev. C} {\bfseries 99} no.~4, (2019) 044910}, \href{http://arxiv.org/abs/1901.09655}{{\ttfamily arXiv:1901.09655 [hep-ph]}}.

\bibitem{Florkowski:2021wvk}
W.~Florkowski, R.~Ryblewski, R.~Singh, and G.~Sophys, ``{Spin polarization dynamics in the non-boost-invariant background},'' \href{http://dx.doi.org/10.1103/PhysRevD.105.054007}{{\em Phys. Rev. D} {\bfseries 105} no.~5, (2022) 054007}, \href{http://arxiv.org/abs/2112.01856}{{\ttfamily arXiv:2112.01856 [hep-ph]}}.

\bibitem{Singh:2024cub}
S.~K. Singh, R.~Ryblewski, and W.~Florkowski, ``{Spin dynamics with realistic hydrodynamic background for relativistic heavy-ion collisions},'' \href{http://dx.doi.org/10.1103/PhysRevC.111.024907}{{\em Phys. Rev. C} {\bfseries 111} no.~2, (2025) 024907}, \href{http://arxiv.org/abs/2411.08223}{{\ttfamily arXiv:2411.08223 [hep-ph]}}.

\bibitem{Singh:2020rht}
R.~Singh, G.~Sophys, and R.~Ryblewski, ``{Spin polarization dynamics in the Gubser-expanding background},'' \href{http://dx.doi.org/10.1103/PhysRevD.103.074024}{{\em Phys. Rev. D} {\bfseries 103} no.~7, (2021) 074024}, \href{http://arxiv.org/abs/2011.14907}{{\ttfamily arXiv:2011.14907 [hep-ph]}}.

\bibitem{Singh:2026wvf}
R.~Singh and A.~Soloviev, ``{Spin hydrodynamics on a hyperbolic expanding background},'' \href{http://arxiv.org/abs/2603.02296}{{\ttfamily arXiv:2603.02296 [nucl-th]}}.

\bibitem{Karpenko:2016jyx}
I.~Karpenko and F.~Becattini, ``{Study of $\Lambda $ polarization in relativistic nuclear collisions at $\sqrt{s_\mathrm {NN}}=7.7$ {\textendash}200 GeV},'' \href{http://dx.doi.org/10.1140/epjc/s10052-017-4765-1}{{\em Eur. Phys. J. C} {\bfseries 77} no.~4, (2017) 213}, \href{http://arxiv.org/abs/1610.04717}{{\ttfamily arXiv:1610.04717 [nucl-th]}}.

\bibitem{Xie:2017upb}
Y.~Xie, D.~Wang, and L.~P. Csernai, ``{Global {\ensuremath{\Lambda}} polarization in high energy collisions},'' \href{http://dx.doi.org/10.1103/PhysRevC.95.031901}{{\em Phys. Rev. C} {\bfseries 95} no.~3, (2017) 031901}, \href{http://arxiv.org/abs/1703.03770}{{\ttfamily arXiv:1703.03770 [nucl-th]}}.

\bibitem{Wu:2019eyi}
H.-Z. Wu, L.-G. Pang, X.-G. Huang, and Q.~Wang, ``{Local spin polarization in high energy heavy ion collisions},'' \href{http://dx.doi.org/10.1103/PhysRevResearch.1.033058}{{\em Phys. Rev. Research.} {\bfseries 1} (2019) 033058}, \href{http://arxiv.org/abs/1906.09385}{{\ttfamily arXiv:1906.09385 [nucl-th]}}.

\bibitem{Ivanov:2020udj}
Y.~B. Ivanov, ``{Global $\Lambda$ polarization in moderately relativistic nuclear collisions},'' \href{http://dx.doi.org/10.1103/PhysRevC.103.L031903}{{\em Phys. Rev. C} {\bfseries 103} no.~3, (2021) L031903}, \href{http://arxiv.org/abs/2012.07597}{{\ttfamily arXiv:2012.07597 [nucl-th]}}.

\bibitem{Fu:2020oxj}
B.~Fu, K.~Xu, X.-G. Huang, and H.~Song, ``{Hydrodynamic study of hyperon spin polarization in relativistic heavy ion collisions},'' \href{http://dx.doi.org/10.1103/PhysRevC.103.024903}{{\em Phys. Rev. C} {\bfseries 103} no.~2, (2021) 024903}, \href{http://arxiv.org/abs/2011.03740}{{\ttfamily arXiv:2011.03740 [nucl-th]}}.

\bibitem{Li:2017slc}
H.~Li, L.-G. Pang, Q.~Wang, and X.-L. Xia, ``{Global $\Lambda$ polarization in heavy-ion collisions from a transport model},'' \href{http://dx.doi.org/10.1103/PhysRevC.96.054908}{{\em Phys. Rev. C} {\bfseries 96} no.~5, (2017) 054908}, \href{http://arxiv.org/abs/1704.01507}{{\ttfamily arXiv:1704.01507 [nucl-th]}}.

\bibitem{Wei:2018zfb}
D.-X. Wei, W.-T. Deng, and X.-G. Huang, ``{Thermal vorticity and spin polarization in heavy-ion collisions},'' \href{http://dx.doi.org/10.1103/PhysRevC.99.014905}{{\em Phys. Rev. C} {\bfseries 99} no.~1, (2019) 014905}, \href{http://arxiv.org/abs/1810.00151}{{\ttfamily arXiv:1810.00151 [nucl-th]}}.

\bibitem{Becattini:2021suc}
F.~Becattini, M.~Buzzegoli, and A.~Palermo, ``{Spin-thermal shear coupling in a relativistic fluid},'' \href{http://dx.doi.org/10.1016/j.physletb.2021.136519}{{\em Phys. Lett. B} {\bfseries 820} (2021) 136519}, \href{http://arxiv.org/abs/2103.10917}{{\ttfamily arXiv:2103.10917 [nucl-th]}}.

\bibitem{Becattini:2021iol}
F.~Becattini, M.~Buzzegoli, G.~Inghirami, I.~Karpenko, and A.~Palermo, ``{Local Polarization and Isothermal Local Equilibrium in Relativistic Heavy Ion Collisions},'' \href{http://dx.doi.org/10.1103/PhysRevLett.127.272302}{{\em Phys. Rev. Lett.} {\bfseries 127} no.~27, (2021) 272302}, \href{http://arxiv.org/abs/2103.14621}{{\ttfamily arXiv:2103.14621 [nucl-th]}}.

\bibitem{Liu:2021uhn}
S.~Y.~F. Liu and Y.~Yin, ``{Spin polarization induced by the hydrodynamic gradients},'' \href{http://dx.doi.org/10.1007/JHEP07(2021)188}{{\em JHEP} {\bfseries 07} (2021) 188}, \href{http://arxiv.org/abs/2103.09200}{{\ttfamily arXiv:2103.09200 [hep-ph]}}.

\bibitem{Fu:2021pok}
B.~Fu, S.~Y.~F. Liu, L.~Pang, H.~Song, and Y.~Yin, ``{Shear-Induced Spin Polarization in Heavy-Ion Collisions},'' \href{http://dx.doi.org/10.1103/PhysRevLett.127.142301}{{\em Phys. Rev. Lett.} {\bfseries 127} no.~14, (2021) 142301}, \href{http://arxiv.org/abs/2103.10403}{{\ttfamily arXiv:2103.10403 [hep-ph]}}.

\bibitem{Yi:2021ryh}
C.~Yi, S.~Pu, and D.-L. Yang, ``{Reexamination of local spin polarization beyond global equilibrium in relativistic heavy ion collisions},'' \href{http://dx.doi.org/10.1103/PhysRevC.104.064901}{{\em Phys. Rev. C} {\bfseries 104} no.~6, (2021) 064901}, \href{http://arxiv.org/abs/2106.00238}{{\ttfamily arXiv:2106.00238 [hep-ph]}}.

\bibitem{Palermo:2024tza}
A.~Palermo, E.~Grossi, I.~Karpenko, and F.~Becattini, ``{$\Lambda $ polarization in very high energy heavy ion collisions as a probe of the quark{\textendash}gluon plasma formation and properties},'' \href{http://dx.doi.org/10.1140/epjc/s10052-024-13229-z}{{\em Eur. Phys. J. C} {\bfseries 84} no.~9, (2024) 920}, \href{http://arxiv.org/abs/2404.14295}{{\ttfamily arXiv:2404.14295 [nucl-th]}}.

\bibitem{Arslan:2025tan}
A.~Arslan, W.-B. Dong, C.~Gale, S.~Jeon, Q.~Wang, and X.-Y. Wu, ``{In-plane transverse polarization in heavy-ion collisions},'' \href{http://dx.doi.org/10.1103/47bs-l3wk}{{\em Phys. Rev. C} {\bfseries 113} no.~3, (2026) 034920}, \href{http://arxiv.org/abs/2509.00796}{{\ttfamily arXiv:2509.00796 [nucl-th]}}.

\bibitem{Singh:2022uyy}
R.~Singh, ``{Collective dynamics of polarized spin-half fermions in relativistic heavy-ion collisions},'' \href{http://dx.doi.org/10.1142/S0217751X23300119}{{\em Int. J. Mod. Phys. A} {\bfseries 38} no.~20, (2023) 2330011}, \href{http://arxiv.org/abs/2212.06569}{{\ttfamily arXiv:2212.06569 [hep-ph]}}.

\bibitem{Shi:2022iyb}
S.~Shi, S.~Jeon, and C.~Gale, ``{Family of new exact solutions for longitudinally expanding ideal fluids},'' \href{http://dx.doi.org/10.1103/PhysRevC.105.L021902}{{\em Phys. Rev. C} {\bfseries 105} no.~2, (2022) L021902}, \href{http://arxiv.org/abs/2201.06670}{{\ttfamily arXiv:2201.06670 [hep-ph]}}.

\bibitem{Chen:2023vrk}
S.~Chen and S.~Shi, ``{Exact solution of Boltzmann equation~in a longitudinal expanding system},'' \href{http://dx.doi.org/10.1103/PhysRevC.109.L051901}{{\em Phys. Rev. C} {\bfseries 109} no.~5, (2024) L051901}, \href{http://arxiv.org/abs/2311.09575}{{\ttfamily arXiv:2311.09575 [nucl-th]}}.

\bibitem{Chen:2024pez}
S.~Chen and S.~Shi, ``{Attractor for (1+1)D viscous hydrodynamics with general rapidity distribution},'' \href{http://dx.doi.org/10.1103/PhysRevC.111.L021902}{{\em Phys. Rev. C} {\bfseries 111} no.~2, (2025) L021902}, \href{http://arxiv.org/abs/2407.15209}{{\ttfamily arXiv:2407.15209 [hep-ph]}}.

\bibitem{Chen:2024grb}
S.~Chen and S.~Shi, ``{Anisotropic hydrodynamics with a boost-noninvariant expansion},'' \href{http://dx.doi.org/10.1103/PhysRevD.111.014001}{{\em Phys. Rev. D} {\bfseries 111} no.~1, (2025) 014001}, \href{http://arxiv.org/abs/2409.19897}{{\ttfamily arXiv:2409.19897 [nucl-th]}}.

\bibitem{Shao:2025ygy}
H.~Shao, S.~Chen, and S.~Shi, ``{Onset of Bjorken Flow in Quantum Evolution of the Massive Schwinger Model},'' \href{http://arxiv.org/abs/2509.10855}{{\ttfamily arXiv:2509.10855 [hep-ph]}}.

\bibitem{Florkowski:2011jg}
W.~Florkowski and R.~Ryblewski, ``{Projection method for boost-invariant and cylindrically symmetric dissipative hydrodynamics},'' \href{http://dx.doi.org/10.1103/PhysRevC.85.044902}{{\em Phys. Rev. C} {\bfseries 85} (2012) 044902}, \href{http://arxiv.org/abs/1111.5997}{{\ttfamily arXiv:1111.5997 [nucl-th]}}.

\bibitem{ParticleDataGroup:2020ssz}
{\bfseries Particle Data Group} Collaboration, P.~A. Zyla {\em et~al.}, ``{Review of Particle Physics},'' \href{http://dx.doi.org/10.1093/ptep/ptaa104}{{\em PTEP} {\bfseries 2020} no.~8, (2020) 083C01}.

\bibitem{Becattini:2017gcx}
F.~Becattini and I.~Karpenko, ``{Collective Longitudinal Polarization in Relativistic Heavy-Ion Collisions at Very High Energy},'' \href{http://dx.doi.org/10.1103/PhysRevLett.120.012302}{{\em Phys. Rev. Lett.} {\bfseries 120} no.~1, (2018) 012302}, \href{http://arxiv.org/abs/1707.07984}{{\ttfamily arXiv:1707.07984 [nucl-th]}}.

\bibitem{Back:2002wb}
B.~B. Back {\em et~al.}, ``{The Significance of the fragmentation region in ultrarelativistic heavy ion collisions},'' \href{http://dx.doi.org/10.1103/PhysRevLett.91.052303}{{\em Phys. Rev. Lett.} {\bfseries 91} (2003) 052303}, \href{http://arxiv.org/abs/nucl-ex/0210015}{{\ttfamily arXiv:nucl-ex/0210015}}.

\bibitem{STAR:2018gyt}
{\bfseries STAR} Collaboration, J.~Adam {\em et~al.}, ``{Global polarization of $\Lambda$ hyperons in Au+Au collisions at $\sqrt{s_{_{NN}}}$ = 200 GeV},'' \href{http://dx.doi.org/10.1103/PhysRevC.98.014910}{{\em Phys. Rev. C} {\bfseries 98} (2018) 014910}, \href{http://arxiv.org/abs/1805.04400}{{\ttfamily arXiv:1805.04400 [nucl-ex]}}.

\bibitem{STAR:2021beb}
{\bfseries STAR} Collaboration, M.~S. Abdallah {\em et~al.}, ``{Global $\Lambda$-hyperon polarization in Au+Au collisions at $\sqrt {s_{NN}}$=3~GeV},'' \href{http://dx.doi.org/10.1103/PhysRevC.104.L061901}{{\em Phys. Rev. C} {\bfseries 104} no.~6, (2021) L061901}, \href{http://arxiv.org/abs/2108.00044}{{\ttfamily arXiv:2108.00044 [nucl-ex]}}.

\bibitem{STAR:2023eck}
{\bfseries STAR} Collaboration, M.~Abdulhamid {\em et~al.}, ``{Hyperon Polarization along the Beam Direction Relative to the Second and Third Harmonic Event Planes in Isobar Collisions at sNN=200{\,}{\,}GeV},'' \href{http://dx.doi.org/10.1103/PhysRevLett.131.202301}{{\em Phys. Rev. Lett.} {\bfseries 131} no.~20, (2023) 202301}, \href{http://arxiv.org/abs/2303.09074}{{\ttfamily arXiv:2303.09074 [nucl-ex]}}.

\bibitem{Voloshin:2017kqp}
S.~A. Voloshin, ``{Vorticity and particle polarization in heavy ion collisions (experimental perspective)},'' \href{http://dx.doi.org/10.1051/epjconf/201817107002}{{\em EPJ Web Conf.} {\bfseries 171} (2018) 07002}, \href{http://arxiv.org/abs/1710.08934}{{\ttfamily arXiv:1710.08934 [nucl-ex]}}.

\bibitem{Niida:2018hfw}
{\bfseries STAR} Collaboration, T.~Niida, ``{Global and local polarization of $\Lambda$ hyperons in Au+Au collisions at 200 GeV from STAR},'' \href{http://dx.doi.org/10.1016/j.nuclphysa.2018.08.034}{{\em Nucl. Phys. A} {\bfseries 982} (2019) 511--514}, \href{http://arxiv.org/abs/1808.10482}{{\ttfamily arXiv:1808.10482 [nucl-ex]}}.

\bibitem{HADES:2014ttv}
{\bfseries HADES} Collaboration, G.~Agakishiev {\em et~al.}, ``{Lambda hyperon production and polarization in collisions of p(3.5 GeV)+Nb},'' \href{http://dx.doi.org/10.1140/epja/i2014-14081-2}{{\em Eur. Phys. J. A} {\bfseries 50} (2014) 81}, \href{http://arxiv.org/abs/1404.3014}{{\ttfamily arXiv:1404.3014 [nucl-ex]}}.

\bibitem{HADES:2022enx}
{\bfseries HADES} Collaboration, R.~Abou~Yassine {\em et~al.}, ``{Measurement of global polarization of {\ensuremath{\Lambda}} hyperons in few-GeV heavy-ion collisions},'' \href{http://dx.doi.org/10.1016/j.physletb.2022.137506}{{\em Phys. Lett. B} {\bfseries 835} (2022) 137506}, \href{http://arxiv.org/abs/2207.05160}{{\ttfamily arXiv:2207.05160 [nucl-ex]}}.

\bibitem{Heiselberg:1998es}
H.~Heiselberg and A.-M. Levy, ``{Elliptic flow and HBT in noncentral nuclear collisions},'' \href{http://dx.doi.org/10.1103/PhysRevC.59.2716}{{\em Phys. Rev. C} {\bfseries 59} (1999) 2716--2727}, \href{http://arxiv.org/abs/nucl-th/9812034}{{\ttfamily arXiv:nucl-th/9812034}}.

\bibitem{STAR:2004jwm}
{\bfseries STAR} Collaboration, J.~Adams {\em et~al.}, ``{Azimuthal anisotropy in Au+Au collisions at s(NN)**(1/2) = 200-GeV},'' \href{http://dx.doi.org/10.1103/PhysRevC.72.014904}{{\em Phys. Rev. C} {\bfseries 72} (2005) 014904}, \href{http://arxiv.org/abs/nucl-ex/0409033}{{\ttfamily arXiv:nucl-ex/0409033}}.

\bibitem{STAR:2003fka}
{\bfseries STAR} Collaboration, J.~Adams {\em et~al.}, ``{Transverse momentum and collision energy dependence of high p(T) hadron suppression in Au+Au collisions at ultrarelativistic energies},'' \href{http://dx.doi.org/10.1103/PhysRevLett.91.172302}{{\em Phys. Rev. Lett.} {\bfseries 91} (2003) 172302}, \href{http://arxiv.org/abs/nucl-ex/0305015}{{\ttfamily arXiv:nucl-ex/0305015}}.

\bibitem{Xia:2018tes}
X.-L. Xia, H.~Li, Z.-B. Tang, and Q.~Wang, ``{Probing vorticity structure in heavy-ion collisions by local $\Lambda$ polarization},'' \href{http://dx.doi.org/10.1103/PhysRevC.98.024905}{{\em Phys. Rev. C} {\bfseries 98} (2018) 024905}, \href{http://arxiv.org/abs/1803.00867}{{\ttfamily arXiv:1803.00867 [nucl-th]}}.

\bibitem{Sun:2021nsg}
Y.~Sun, Z.~Zhang, C.~M. Ko, and W.~Zhao, ``{Evolution of {\ensuremath{\Lambda}} polarization in the hadronic phase of heavy-ion collisions},'' \href{http://dx.doi.org/10.1103/PhysRevC.105.034911}{{\em Phys. Rev. C} {\bfseries 105} no.~3, (2022) 034911}, \href{http://arxiv.org/abs/2112.14410}{{\ttfamily arXiv:2112.14410 [nucl-th]}}.

\bibitem{Noronha:2024dtq}
J.~Noronha, B.~Schenke, C.~Shen, and W.~Zhao, ``{Progress and challenges in small systems},'' \href{http://dx.doi.org/10.1142/9789811294679_0004}{{\em Int. J. Mod. Phys. E} {\bfseries 33} no.~06, (2024) 2430005}, \href{http://arxiv.org/abs/2401.09208}{{\ttfamily arXiv:2401.09208 [nucl-th]}}.

\bibitem{Grosse-Oetringhaus:2024bwr}
J.~F. Grosse-Oetringhaus and U.~A. Wiedemann, ``{A Decade of Collectivity in Small Systems},'' \href{http://arxiv.org/abs/2407.07484}{{\ttfamily arXiv:2407.07484 [hep-ex]}}.

\bibitem{Buzzegoli:2025zud}
M.~Buzzegoli, ``{Kubo formulas for spin polarization in dissipative relativistic spin hydrodynamics: a first-order gradient expansion approach},'' \href{http://dx.doi.org/10.1007/JHEP07(2025)255}{{\em JHEP} {\bfseries 07} (2025) 255}, \href{http://arxiv.org/abs/2502.15520}{{\ttfamily arXiv:2502.15520 [nucl-th]}}.

\bibitem{Alice:2026ch}
C.~Jauch, ``{Local Lambda polarization in light-ion with ALICE at LHC}.'' {Strangness in Quark Matter 2026}.

\bibitem{Basar:2024qxd}
G.~Ba{\c{s}}ar, J.~Bhambure, R.~Singh, and D.~Teaney, ``{Stochastic relativistic advection diffusion equation~from the Metropolis algorithm},'' \href{http://dx.doi.org/10.1103/PhysRevC.110.044903}{{\em Phys. Rev. C} {\bfseries 110} no.~4, (2024) 044903}, \href{http://arxiv.org/abs/2403.04185}{{\ttfamily arXiv:2403.04185 [nucl-th]}}.

\bibitem{Bhambure:2024axa}
J.~Bhambure, A.~Mazeliauskas, J.-F. Paquet, R.~Singh, M.~Singh, D.~Teaney, and F.~Zhou, ``{Relativistic viscous hydrodynamics in the density frame: Numerical tests and comparisons},'' \href{http://dx.doi.org/10.1103/PhysRevC.111.064910}{{\em Phys. Rev. C} {\bfseries 111} no.~6, (2025) 064910}, \href{http://arxiv.org/abs/2412.10303}{{\ttfamily arXiv:2412.10303 [nucl-th]}}.

\bibitem{Bhambure:2024gnf}
J.~Bhambure, R.~Singh, and D.~Teaney, ``{Stochastic relativistic viscous hydrodynamics from the Metropolis algorithm},'' \href{http://dx.doi.org/10.1103/15y6-6gmz}{{\em Phys. Rev. C} {\bfseries 111} no.~6, (2025) 064909}, \href{http://arxiv.org/abs/2412.10306}{{\ttfamily arXiv:2412.10306 [nucl-th]}}.

\bibitem{ThisWorkDataRepository}
M.~Buzzegoli, A.~Geci\'{c}, and R.~Singh, ``{Modeling Lambda polarization in Au+Au collisions at 200 GeV using relativistic spin hydrodynamics}.'' \url{https://github.com/rs240291/rs-HYDRO.git}, 2026.

\end{thebibliography}\endgroup
\end{document}